\documentclass[%
 aip,jcp,
 sd,%
 amsmath,amssymb,
 reprint,%
]{revtex4-1}

\usepackage[usenames,dvipsnames]{xcolor}
\usepackage{graphicx}
\usepackage{mathrsfs}
\usepackage{float}
\usepackage{wasysym}
\DeclareMathAlphabet{\mathpzc}{OT1}{pzc}{m}{it}
\usepackage{dcolumn}
\usepackage{bm}
\newcommand{\etal}{{\it{et al.}}}

\definecolor{dgreen}{rgb}{0,.5,0}
\newcommand{\manu}[1]{{\textcolor{dgreen}{ Manu: #1 }} }

\begin{document}

\preprint{AIP/123-QED}

\title{Combining extrapolation with ghost interaction correction in range-separated 
ensemble density functional theory for excited states}

\author{Md. Mehboob Alam}
\altaffiliation[Also at ]{Department of Chemistry, Universitetet i Troms\o}
 \email{mehboob.cu@gmail.com}
 \affiliation{ 
Laboratoire de Chimie Quantique, Institut de Chimie, CNRS/Universit\'{e} de Strasbourg,
4 rue Blaise Pascal, 67000 Strasbourg, France
}%
\author{Killian Deur}
 \affiliation{ 
Laboratoire de Chimie Quantique, Institut de Chimie, CNRS/Universit\'{e} de Strasbourg,
4 rue Blaise Pascal, 67000 Strasbourg, France
}%
\author{Stefan Knecht}
\affiliation{%
Laboratory of Physical Chemistry, ETH, Z\"{u}rich, Vladimir-Prelog Weg 2, CH-8093,
Z\"{u}rich, Switzerland
}
\author{Emmanuel Fromager}
 \affiliation{ 
Laboratoire de Chimie Quantique, Institut de Chimie, CNRS/Universit\'{e} de Strasbourg,
4 rue Blaise Pascal, 67000 Strasbourg, France
}

\date{\today}

\begin{abstract}

The extrapolation technique of Savin [J. Chem. Phys. 140, 18A509 (2014)], which was initially applied to
range-separated ground-state-density-functional Hamiltonians, 
is adapted in this work to ghost-interaction-corrected (GIC) range-separated ensemble density-functional
theory (eDFT) for excited states. While standard extrapolations rely on
energies that decay as $\mu^{-2}$ in the large range-separation-parameter
$\mu$ limit, we show analytically that (approximate) range-separated
GIC ensemble energies converge more rapidly (as $\mu^{-3}$) towards  
their pure wavefunction theory values ($\mu\rightarrow+\infty$ limit),
thus requiring a different extrapolation correction. The purpose of such
a correction is to further improve
on the convergence and, consequently, to obtain more accurate excitation
energies for a finite (and, in practice, relatively small) $\mu$ value. 
As a proof of concept, we apply the extrapolation method 
to He and small molecular systems (viz.
H$_{2}$, HeH$^{+}$ and LiH), thus considering different types of
excitations like
Rydberg, charge transfer and double excitations. Potential energy 
profiles of the first three and four singlet $\Sigma^+$ excitation energies in
HeH$^{+}$ and H$_{2}$, respectively, are studied with a particular
focus on avoided crossings for the latter. Finally, the
extraction of individual state energies from the ensemble energy is discussed in the context of
range-separated eDFT, as a perspective.  
%
\end{abstract}

\pacs{Valid PACS appear here}
\keywords{Ensemble density functional theory, Range-separation, Ghost-interaction correction, 
Excited states, Excitation energy}
\maketitle

\section{\label{sec:level1}Introduction\protect\\
}
Electronic excitation energies in atoms, molecules and solids
can in principle be obtained from
density-functional theory (DFT)~\cite{hktheo,KS} by using either time-dependent~\cite{TDDFT_Gross,Casida_tddft_review_2012,Casida_eqs,marques2004time} 
or
time-independent\cite{GUNNARSSON:1976p1781,DederichsPRL1984,Ziegler1977,BarthPRA1979,JPC79_Theophilou_equi-ensembles,PRA_GOK_RRprinc,PRA_GOK_EKSDFT,GOK3,Nagy_enseXpot,TasnadiJPB2003}
approaches. The Hohenberg--Kohn (HK) theorem states that the 
(time-independent) 
ground-state 
density carries all the information about 
the system and hence excitation energies can, in principle, be extracted 
from it. While the ground-state energy can be obtained variationally
from the ground-state density, the HK variational principle has no
trivial extension to the excited states. This is one of the reason why
the standard approach for computing  
excitation energies is nowadays linear response {time-dependent} DFT
(TD-DFT). Despite its success and the significant efforts put into its
development over the last two decades, TD-DFT still suffers from some
limitations {\it e.g.} the 
inability to account for multiconfigurational
effects~\cite{gritsenko2000excitation} (which is inherited from
standard Kohn--Sham DFT), the 
poor description\cite{CasidaJCP1998chargetransfer,DreuwJCP2003chargetransfer} 
of charge-transfer and Rydberg states with semi-local functionals and
the absence of double excitations\cite{maitra2004double} 
from its spectra. These limitations are associated with mainly three
aspects 
of the theory: its single-determinantal nature, the wrong asymptotic 
behavior of approximate density-functional exchange-correlation
potentials and the adiabatic approximation 
({\it i.e.} the use of a frequency-independent exchange-correlation kernel in
the response equations).
\cite{Casida_tddft_review_2012} Some of these limitations can be overcome 
by the use of Savin's idea of range separation\cite{savin1988combined,savinstoll,savinbook} 
and double-hybrid
kernels.\cite{HuixRotllantCP2011hybridkernel}\\
The need for time-independent alternatives to TD-DFT for 
modeling excited states has attracted increasing attention over the
years~\cite{filatov2015ensemble,krykunov2013jctc,ziegler2009jcp,levy2016computation,ayers2009pra,
yang2017prl,PRA13_Pernal_srEDFT,franck2014generalised,PRB17_Deur_eDFT_Hubbard_dimer}.
In this work we focus on one of them, namely ensemble DFT (eDFT), which is an
in-principle-exact approach for the calculation of excitation
energies. It was first 
formulated by Theophilou {\etal}\cite{JPC79_Theophilou_equi-ensembles,theophilou_book} and then developed further by Gross, Oliveira 
and Kohn.\cite{PRA_GOK_RRprinc,PRA_GOK_EKSDFT,GOK3} Since, in eDFT, the
basic variable is the ensemble density ({\it i.e.} the weighted sum of
ground- and excited-state densities), the approach is in principle well suited for  
modelling multiconfigurational problems (like bond dissociations) or
multiple excitations.\cite{filatov2015ensemble} 
A quasi-local-density approximation and an ensemble exchange potential 
was developed respectively by Kohn\cite{KohnPRA1986} and Nagy,
\cite{Nagy_functional,nagy2001} but no further attempt was taken to 
develop density-functional approximations for ensembles. An ensemble is
characterized by weights that are attributed to the states belonging to
the ensemble. The exchange-correlation ensemble-density-functional
energy depends in principle on these weights. It still remains a
challenge to model this weight-dependence which actually  
plays a crucial role in the calculation of excitation
energies~\cite{PRA_GOK_EKSDFT,franck2014generalised,yang2014exact,Burke_ensemble,PRB17_Deur_eDFT_Hubbard_dimer,yang2017prl}. 
\\
The recent resurgence of eDFT in the literature is partly due to the
fact that, when combined with wavefunction theory
by means of range separation for
example~\cite{PRA13_Pernal_srEDFT,franck2014generalised}, it leads to a rigorous
state-averaged multiconfigurational DFT. Like in conventional Kohn--Sham-eDFT,
the weight-dependence of the complementary short-range
exchange-correlation density functional should be
modelled~\cite{franck2014generalised}, which is of
course not trivial. A standard approximation consists in using
(weight-independent) ground-state short-range functionals~\cite{PRA13_Pernal_srEDFT}, thus leading to
weight-dependent excitation energies~\cite{SenjeanPRA2015}. This problem
can be fixed either by using the ensemble weights as
parameters~\cite{PRA13_Pernal_srEDFT} or by performing a linear interpolation between
equiensemble energies~\cite{SenjeanPRA2015}. Using ground-state
functionals induces also so-called {\it ghost-interaction
errors}~\cite{ensemble_ghost_interaction,Nagy_functional,TasnadiJPB2003,pastorczak2014ensemble}.
The latter are induced by the (short-range) Hartree energy that is
quadratic in the ensemble density, hence the unphysical coupling terms
between two different state densities. In the context of range-separated
eDFT, this error can be removed either by constructing individual state
energies~\cite{PRA13_Pernal_srEDFT,pernal2016ghost} or by introducing an
alternative separation of ensemble exchange and correlation short-range
energies~\cite{alam_gic}. The latter approach has the advantage of
making approximate range-separated ensemble energies essentially linear with
respect to the ensemble weights. It also gave very encouraging results
for the description of charge-transfer and double
excitations~\cite{alam_gic}. Finally, while using a relatively small
range separation parameter $\mu$ value (typically
$\mu=0.4$-$0.5$~\cite{FromagerJCP2007,nancycalib} up to
1.0 a.u.~\cite{PRA13_Pernal_srEDFT}) is preferable in terms of computational
cost, since a significant part of the two-electron repulsion (including
the Coulomb hole~\cite{JCP15_Odile_basis_convergence_srDFT}) is modelled by a density functional in this case,
excitation energies might be underestimated, essentially because
(weight-independent) short-range local or semi-local density functional
approximations are used~\cite{extrapol_edft}. As initially shown by
Savin~\cite{savin2014towards} in the context of ground-state
range-separated DFT, the Taylor expansion of the energy around the pure
wavefunction theory limit ({\it i.e.} $\mu\rightarrow+\infty$) can be
used for improving the energy at a given (finite) $\mu$ value. This
approach, known as extrapolation technique, has been extended to
excited states by considering the ($\mu$-dependent) individual excited-state energies of a
ground-state-density-functional long-range-interacting
Hamiltonian~\cite{rebolini2015pra}. An extension to range-separated eDFT
has been recently proposed by Senjean~{\it et
al.}~\cite{extrapol_edft} One drawback of the latter approach is that it does not incorporate ghost interaction
corrections. The purpose of this work is to show how these corrections
can be combined with extrapolation techniques in order to obtain
accurate excitation energies.\\ 
The paper is organized 
as follows. After an introduction to 
range-separated eDFT (Sec.~\ref{subsec:rs_edft})
and the ghost-interaction error (Sec.~\ref{subsec:gi_gic_edft}), the
calculation of excitation energies by linear interpolation
(Sec.~\ref{subsec:interpolation}) will be briefly reviewed. 
The central result of this paper, which is the combination of extrapolation
techniques with ghost-interaction corrections, is presented in
Sec.~\ref{subsec:extrapolation}. Higher-order extrapolation corrections
will also be introduced in Sec.~\ref{subsec:higher_order_extrapo}. 
Sec.~\ref{sec:computational_details} contains the computational 
details. The results are discussed in Sec.~\ref{sec:results} followed by
a perspective section (Sec.~\ref{sec:individual_energies}) on the construction of individual
state energies in range-separated eDFT. Conclusions are given in Sec.~\ref{sec:conclusion}.
\section{Theory}\label{sec:theory}

\subsection{Range-separated ensemble DFT for excited states}\label{subsec:rs_edft}

Let $\hat{H} = \hat{T} + \hat{W}_{\rm{ee}} + 
\int d{\bf r}\,v_{\rm{ne}}({\bf r})\hat{n}({\bf r})$ be the electronic
Hamiltonian with nuclear potential $v_{\rm{ne}}({\bf r})$ where $\hat{T}$,
$\hat{W}_{\rm{ee}}$, and $\hat{n}({\bf r})$ are the kinetic energy,
two-electron repulsion and density operators, respectively. In the
following, we consider the ensemble
$\left\{ \Psi_{k} \right\}_{0 \leq k \leq M-1}$ of eigenfunctions
associated to the $M$ lowest eigenvalues $E_{0}\leq
E_{1}\leq \dots\leq E_{M-1}$ of $\hat{H}$ with ensemble 
weights ${\bf w} \equiv (w_{0},w_{1},\dots,w_{M-1})$ where 
$w_{0}\geq w_{1}\geq\dots\geq w_{M-1}\geq 0$ and 
\begin{eqnarray}\label{eq:normalization_weights}
\sum^{M-1}_{k=0}w_k=1.
\end{eqnarray}  
The ensemble energy 
\begin{eqnarray}
\begin{array}{l}\label{eq:exact_ens_en}
 E^{\bf w} = \sum\limits_{k=0}^{M-1}w_{k}E_{k},
\end{array}
\end{eqnarray}
which is the weighted sum of ground- and excited-state energies, 
is a functional of the ensemble density~\cite{PRA_GOK_EKSDFT,PRA_GOK_RRprinc,GOK3}
\begin{eqnarray}
\begin{array}{l}\label{eq:exact_ens_dens}
n_{\hat{\Gamma}^{\bf w}}({\bf r}) = \sum\limits_{k=0}^{M-1}w_{k}n_{\Psi_{k}}({\bf r}) = \text{Tr}\left[\hat{\Gamma}^{\bf w}\hat{n}({\bf r})\right]
,
\end{array}
\end{eqnarray}
where Tr denotes the trace and $\hat{\Gamma}^{\bf
w}=\sum^{M-1}_{k=0}w_k\vert\Psi_k\rangle\langle\Psi_k\vert$, and it 
can be obtained variationally as follows,
\begin{eqnarray}
\begin{array}{l}\label{eq:gok_ens_en}
\begin{split}
E^{\bf w} & = \min\limits_{n} \left\{ F^{\bf w}[n] 
+ \int d{\bf r}v_{\rm{ne}}({\bf r})n({\bf r}) \right\}\\
 & = F^{\bf w}[n_{\hat{\Gamma}^{\bf w}}] 
 + \int d{\bf r}v_{\rm{ne}}({\bf r})n_{\hat{\Gamma}^{\bf w}}({\bf r}),
\end{split}
\end{array}
\end{eqnarray}
where 
\begin{eqnarray}
\begin{array}{l}\label{eq:ens_ll_func}
\begin{split}
F^{\bf w}[n] = \min\limits_{\hat{\gamma}^{\bf w}\rightarrow n} 
\text{Tr}\left[ \hat{\gamma}^{\bf w} \left( \hat{T} + \hat{W}_{\rm{ee}} \right) \right]
\end{split}
\end{array}
\end{eqnarray}
is the ensemble Levy--Lieb (LL) functional. Note that the minimization
in Eq.~(\ref{eq:ens_ll_func}) is restricted to trial ensemble density matrix
operators $\hat{\gamma}^{\bf w}=\sum_{k=0}^{M-1}w_{k}\vert
\tilde{\Psi}_{k} \rangle \langle \tilde{\Psi}_{k} \vert $ with density $n$:
\begin{eqnarray}
\begin{array}{l}
\displaystyle{
n_{\hat{\gamma}^{\bf w}}({\bf r}) 
=\text{Tr}\left[\hat{\gamma}^{\bf w}\hat{n}(\bf r)\right] 
= \sum\limits_{k=0}^{M-1}w_{k}n_{\tilde{\Psi}_{k}}({\bf r})
= n({\bf r})
}.
\end{array}
\end{eqnarray}
A rigorous combination of wavefunction-based and eDFT methods can be
obtained from the separation of the two-electron interaction into long-
and short-range
parts~\cite{savin1988combined,savinstoll,savinbook,PRA13_Pernal_srEDFT,franck2014generalised},
\begin{eqnarray}
\begin{array}{l}\label{eq:gs_rangeseparation}
\begin{split}
\displaystyle
& \hat{W}_{\rm{ee}} = \hat{W}_{\rm{ee}}^{\rm{lr,\mu}} + \hat{W}_{\rm{ee}}^{\rm{sr,\mu}} 
\equiv \sum\limits_{i < j} \left\{w_{\rm{ee}}^{\rm{lr,\mu}}(r_{ij}) + w_{\rm{ee}}^{\rm{sr,\mu}}(r_{ij})\right\} \\
& w_{\rm{ee}}^{\rm{lr,\mu}}(r) = \frac{\text{erf}(\mu r)}{r}, 
w_{\rm{ee}}^{\rm{sr,\mu}}(r) = \frac{\text{erfc}(\mu r)}{r}
,
\end{split}
\end{array}
\end{eqnarray}
where $\text{erf}$ is the error function, 
$\text{erfc}(\mu r) = 1 - \text{erf}(\mu r)$ and $\mu$ is the 
range-separation parameter in $[0,+\infty[$.
As a consequence of Eq.~(\ref{eq:gs_rangeseparation}),
the ensemble LL functional in 
Eq.~(\ref{eq:ens_ll_func}) can be rewritten as follows, 
\begin{eqnarray}
\begin{array}{l}\label{eq:rs_of_ll_func}
F^{\bf w}[n] = F^{\rm{lr,\mu},\bf w}[n] + E_{\rm{Hxc}}^{\rm{sr,\mu},\bf w}[n],
\end{array}
\end{eqnarray}
where
\begin{eqnarray}
\begin{array}{l}\label{eq:gok_longrangeLLfunctional}
\begin{split}
F^{\rm{lr,\mu,{\bf w}}}[n] & = \min\limits_{\hat{\gamma}^{\bf w} \rightarrow n}
\text{Tr} \left[ \hat{\gamma}^{\bf w}\left(\hat{T} + \hat{W}_{\rm{ee}}^{\rm{lr,\mu}}\right)\right],\\
& = \text{Tr} \left[ \hat{\Gamma}^{\rm{\mu,{\bf w}}}[n]\left(\hat{T} + \hat{W}_{\rm{ee}}^{\rm{lr,\mu}}\right)\right]
\end{split}
\end{array}
\end{eqnarray}
is the long-range ensemble LL functional and, by definition, 
$E_{\rm{Hxc}}^{\rm{sr,\mu,{\bf w}}}[n]=F^{\bf w}[n]-F^{\rm{lr,\mu},\bf w}[n]$ is the complementary 
short-range ensemble Hartree-exchange-correlation (Hxc) functional, which
is both ${\bf w}$ and $\mu$ dependent. Note that the minimizing 
ensemble density matrix operator {{$\hat{\Gamma}^{\rm{\mu,{\bf
w}}}[n]$ in Eq.~(\ref{eq:gok_longrangeLLfunctional}) is 
the long-range-interacting one with density $n$.}} The short-range
ensemble Hxc functional is usually decomposed
as follows~\cite{PRA13_Pernal_srEDFT,franck2014generalised},
\begin{eqnarray}
\begin{array}{l}\label{eq:sr_hxc_ks_decomp}
E_{\rm{Hxc}}^{\rm{sr,\mu,{\bf w}}}[n] = 
E_{\rm{H}}^{\rm{sr,\mu}}[n] 
+ E_{\rm{xc}}^{\rm{sr,\mu,{\bf w}}}[n],
\end{array}
\end{eqnarray}
where
\begin{eqnarray}
\label{eq:ensemble_hartree}
E_{\rm{H}}^{\rm{sr,\mu}}[n] = \frac{1}{2}\int\int d{\bf r}d{\bf r^{\prime}}n({\bf r})n({\bf r}^{\prime})
{w_{\rm{ee}}^{\rm{sr,\mu}}}({\vert {\bf r} - {\bf r}^{\prime} \vert})
\end{eqnarray}
is the ${\bf w}$-independent but $\mu$-dependent short-range Hartree
functional
and $E_{\rm{xc}}^{\rm{sr,\mu,{\bf w}}}[n]$ is the complementary 
ensemble short-range xc functional. 
Since, according to the first line
of Eq.~(\ref{eq:gok_longrangeLLfunctional}), the following inequality is
fulfilled for {\it any} trial
ensemble density matrix operator $\hat{\gamma}^{\rm{\bf w}}$,
\begin{eqnarray}
\text{Tr}\left[ \hat{\gamma}^{\rm\bf w}\left( \hat{T} + \hat{W}_{\rm ee}^{\rm lr, \mu} \right) \right] \geq F^{\rm lr,\mu,{\bf w}}[n_{\hat{\gamma}^{\rm \bf w}}]
,
\end{eqnarray}
thus leading to
\begin{eqnarray}
&&\text{Tr}\left[ \hat{\gamma}^{\rm\bf w}\left( \hat{T} + \hat{W}_{\rm
ee}^{\rm
lr, \mu} \right) \right] + E_{\rm{Hxc}}^{\rm{sr,\mu,{\bf
w}}}[n_{\hat{\gamma}^{\bf w}}] + \int d{\bf r}v_{\rm{ne}}({\bf r})n_{\hat{\gamma}^{\bf w}}({\bf r})
\nonumber\\
&&\geq F^{\rm lr,\mu,{\bf w}}[n_{\hat{\gamma}^{\rm \bf w}}] +
E_{\rm{Hxc}}^{\rm{sr,\mu,{\bf w}}}[n_{\hat{\gamma}^{\bf w}}] + \int
d{\bf r}v_{\rm{ne}}({\bf r})n_{\hat{\gamma}^{\bf w}}({\bf r}),
\end{eqnarray}
or, equivalently, according to Eqs.~(\ref{eq:gok_ens_en}) and (\ref{eq:rs_of_ll_func}),
\begin{eqnarray}
&&\text{Tr}\left[ \hat{\gamma}^{\rm\bf w}\left( \hat{T} + \hat{W}_{\rm
ee}^{\rm
lr, \mu}+\hat{V}_{\rm ne} \right) \right] + E_{\rm{Hxc}}^{\rm{sr,\mu,{\bf
w}}}[n_{\hat{\gamma}^{\bf w}}] 
\nonumber\\
&&\geq F^{\rm \bf w}[n_{\hat{\gamma}^{\rm \bf w}}]+ \int
d{\bf r}v_{\rm{ne}}({\bf r})n_{\hat{\gamma}^{\bf w}}({\bf r})
\geq E^{\rm \bf w},
\end{eqnarray}
where $\hat{V}_{\rm{ne}}=\int d{\bf r}\,v_{\rm{ne}}({\bf r})\hat{n}({\bf
r})$, we finally obtain an {\it
exact} range-separated variational expression for the ensemble energy
where the minimization is performed over all possible density matrix
operators ({\it without} any density constraint),
\begin{multline}\label{eq:gok_rs_ens_en}
E^{\bf w}
= \min\limits_{\hat{\gamma}^{\bf w}}\left\{\text{Tr}
\left[ \hat{\gamma}^{\bf w} (\hat{T} + \hat{W}_{\rm{ee}}^{\rm{lr,\mu}} 
+ \hat{V}_{\rm{ne}})\right]
 + E_{\rm{Hxc}}^{\rm{sr,\mu,{\bf w}}}[n_{\hat{\gamma}^{\bf w}}] \right\}\\
= \text{Tr}\left[ \hat{\Gamma}^{\rm{\mu,{\bf w}}} (\hat{T} 
+ \hat{W}_{\rm{ee}}^{\rm{lr,\mu}} + \hat{V}_{\rm{ne}})\right] 
+ E_{\rm{Hxc}}^{\rm{sr,\mu,{\bf w}}}[n_{\hat{\Gamma}^{\rm{\mu,{\bf
w}}}}].
\end{multline}
Note that the minimizing long-range-interacting ensemble density matrix
operator 
\begin{eqnarray}\label{eq:lrDM}
\hat{\Gamma}^{\rm{\mu,{\bf w}}} = 
\sum_{k=0}^{M-1}w_{k}\vert \Psi_{k}^{\rm{\mu,{\bf w}}} \rangle \langle \Psi_{k}^{\rm{\mu,{\bf w}}} \vert
\end{eqnarray}
in Eq.~(\ref{eq:gok_rs_ens_en}) reproduces the exact ensemble density of
the physical (fully interacting) system, 
\begin{eqnarray}
\begin{array}{l}\label{eq:gok_rs_ens_dens}
\text{Tr}\left[ \hat{\Gamma}^{\rm{\mu,{\bf w}}} \hat{n}({\bf r}) \right] 
= n_{\hat{\Gamma}^{\rm{\mu,{\bf w}}}}({\bf r}) = n_{\hat{\Gamma}^{\bf w}}({\bf r}),
\end{array}
\end{eqnarray}
and that the (multideterminantal) wavefunctions $\{ \Psi_{k}^{\rm{\mu,{\bf w}}} \}_{0 \leq k \leq M-1}$ 
are solutions of the self-consistent equation~\cite{SenjeanPRA2015,PRA13_Pernal_srEDFT}
\begin{multline}\label{eq:gok_rs_scf}
\left( \hat{T} + \hat{W}_{\rm{ee}}^{\rm{lr,\mu}} 
+ \hat{V}_{\rm{ne}} + \int d{\bf r}
\frac{\delta E_{\rm{Hxc}}^{\rm{sr,\mu,{\bf w}}}[n_{\hat{\Gamma}^{\rm{\mu,{\bf w}}}}]}
{\delta n({\bf r})}\hat{n}({\bf r}) \right) \vert \Psi_{k}^{\rm{\mu,{\bf w}}} \rangle \\
 = \mathcal{E}_{k}^{\rm{\mu,{\bf w}}} \vert \Psi_{k}^{\rm{\mu,{\bf w}}} \rangle,
\hspace{0.2cm}0 \leq k \leq M-1,
\end{multline}
from which the standard Schr\"{o}dinger and Kohn--Sham
(KS) eDFT~\cite{PRA_GOK_EKSDFT} equations  
are recovered in the $\mu\rightarrow+\infty$ and $\mu \rightarrow 0$
limits, respectively.\\

In practice, long-range-interacting wavefunctions are usually computed
(self-consistently) at the {\it configuration interaction} (CI)
level~\cite{SenjeanPRA2015,PRA13_Pernal_srEDFT,pernal2016ghost} within
the \emph{weight-independent
density functional approximation} (WIDFA), which simply consists in
substituting in Eqs.~(\ref{eq:gok_rs_ens_en}) and (\ref{eq:gok_rs_scf}) the ground-state ($w_0=1$) short-range xc functional
$E_{\rm{xc}}^{\rm{sr,\mu}}[n]$ (which is approximated by a local or
semi-local functional~\cite{srDFT,Goll2005PCCP}) for the ensemble one.
The (approximate) WIDFA range-separated ensemble energy reads
\begin{multline}\label{eq:widfa_ens_en}
\tilde{E}^{\rm{\mu,{\bf w}}} = \min\limits_{\hat{\gamma}^{\bf w}}
\{ \text{Tr}\left[ \hat{\gamma}^{\bf w} (\hat{T} + \hat{W}_{\rm ee}^{\rm{lr,\mu}} 
+ \hat{V}_{\rm ne}) \right] + E_{\rm{Hxc}}^{\rm{sr,\mu}}[n_{\hat{\gamma}^{\bf w}}] \}\\
=\text{Tr}\left[ \hat{\gamma}^{\mu,\bf w} (\hat{T} + \hat{W}_{\rm ee}^{\rm{lr,\mu}} 
+ \hat{V}_{\rm ne}) \right] + E_{\rm{Hxc}}^{\rm{sr,\mu}}[n_{\hat{\gamma}^{\mu,\bf w}}],
\end{multline}
where $E_{\rm{Hxc}}^{\rm{sr,\mu}}[n] = E_{\rm{H}}^{\rm{sr,\mu}}[n] + E_{\rm{xc}}^{\rm{sr,\mu}}[n]$.
The minimizing ensemble density matrix operator 
$\hat{\gamma}^{\rm{\mu,{\bf w}}}=
\sum_{k=0}^{M-1}w_{k}\vert \tilde{\Psi}_{k}^{\rm{\mu,{\bf w}}} \rangle
 \langle \tilde{\Psi}_{k}^{\rm{\mu,{\bf w}}} \vert$, which is an
approximation to $\hat{\Gamma}^{\rm{\mu,{\bf w}}}$, fulfills the
following self-consistent equation~\cite{SenjeanPRA2015},
\begin{multline}\label{eq:widfa_scf}
\left( \hat{T} + \hat{W}_{\rm{ee}}^{\rm{lr,\mu}} 
+ \hat{V}_{\rm{ne}} + \int d{\bf r}
\frac{\delta E_{\rm{Hxc}}^{\rm{sr,\mu}}[n_{\hat{\gamma}^{\rm{\mu,{\bf w}}}}]}
{\delta n({\bf r})}\hat{n}({\bf r}) \right) \vert \tilde{\Psi}_{k}^{\rm{\mu,{\bf w}}} \rangle \\
 = \tilde{\mathcal{E}}_{k}^{\rm{\mu,{\bf w}}} \vert \tilde{\Psi}_{k}^{\rm{\mu,{\bf w}}} \rangle,
\hspace{0.2cm}0 \leq k \leq M-1.
\end{multline}
Let us
finally mention that, in case of near degeneracies, a state-averaged multiconfigurational
self-consistent field is preferable to CI for the description of
long-range correlation effects. Work is currently in progress in this
direction. 
\subsection{Ghost interaction correction}\label{subsec:gi_gic_edft}

As readily seen from Eqs.~(\ref{eq:sr_hxc_ks_decomp}), (\ref{eq:ensemble_hartree}),
(\ref{eq:gok_rs_ens_en}), (\ref{eq:lrDM}), and
(\ref{eq:gok_rs_ens_dens}), the short-range Hartree density-functional
contribution to the exact range-separated ensemble energy can be written
as follows,
\begin{eqnarray}
\label{eq:gi_error1}
&& E_{\rm{H}}^{\rm{sr,\mu}}[n_{\hat{\Gamma}^{\rm{\mu,{\bf w}}}}] =
 \sum\limits_{k=0}^{M-1} w_{k}^{2} E_{\rm{H}}^{\rm{sr,\mu}}[n_{{{\Psi}}_{k}^{\rm{\mu,{\bf w}}}}]
+  \frac{1}{2}\sum\limits_{j\neq k}^{M-1} w_{j}w_{k}
\nonumber\\
&&\times \int\int d{\bf r} d{\bf r}^{{\prime}}
n_{{{\Psi}}_{j}^{\rm{\mu,{\bf w}}}}({\bf r})n_{{{\Psi}}_{k}^{\rm{\mu,{\bf w}}}}({\bf r}^{\prime})w_{\rm{ee}}^{\rm{sr,\mu}}(\vert {\bf r} 
- {\bf{r}}{^{\prime}} \vert),
\end{eqnarray}
where the individual-state contributions (first term on the right-hand
side of Eq.~(\ref{eq:gi_error1}))
are {\it quadratic} in the ensemble weights, while the exact total ensemble
energy is of course {\it linear} (see Eq.~(\ref{eq:exact_ens_en})), and the second term describes the {\it
unphysical} interaction, known as ghost interaction
(GI)~\cite{ensemble_ghost_interaction}, between two different states
belonging to the ensemble. Both errors should of course be compensated
by the exact ensemble short-range xc functional but, in practice ({\it
i.e.} when WIDFA is applied), this is not the
case~\cite{SenjeanPRA2015,extrapol_edft,alam_gic}. In order to correct
for GI errors, one can either construct individual state
energies~\cite{PRA13_Pernal_srEDFT,pernal2016ghost} or use an
alternative separation of short-range ensemble exchange and correlation
energies, as proposed recently by some of the authors~\cite{alam_gic}.
The first approach will be discussed further in
Sec.~\ref{sec:individual_energies}. For now we focus on the second one
which relies on the following {\it exact} decomposition of
the ensemble short-range xc functional, 
\begin{eqnarray}
\begin{array}{l}\label{eq:gic_dAC_xc}
E_{\rm{xc}}^{\rm{sr,\mu,{\bf w}}}[n] = E_{\rm{x,md}}^{\rm{sr,\mu,{\bf w}}}[n] 
+ E_{\rm{c,md}}^{\rm{sr,\mu,{\bf w}}}[n],
\end{array}
\end{eqnarray}
where 
\begin{eqnarray}
\begin{array}{l}\label{eq:md_x}
E_{\rm{x,md}}^{\rm{sr,\mu,{\bf w}}}[n] 
= \text{Tr}\left[\hat{\Gamma}^{\rm{\mu,{\bf
w}}}[n]\hat{W}_{\rm{ee}}^{\rm{sr,\mu}}\right] 
- E_{\rm{H}}^{\rm{sr,\mu}}[n]
\end{array}
\end{eqnarray}
is the analog for ensembles of the ground-state {\it
multi-determinantal} [hence the subscript 'md' in the functionals of
Eq.~(\ref{eq:gic_dAC_xc})] short-range
exact exchange functional introduced by Toulouse
{\etal}~\cite{Toulouse2005TCA}, and 
$E_{\rm{c,md}}^{\rm{sr,\mu,{\bf w}}}[n]$ is the 
complementary short-range ensemble correlation functional adapted to
the multi-determinantal definition of the short-range ensemble exchange energy.
Combining Eqs.~(\ref{eq:gok_rs_ens_en}), ~(\ref{eq:gic_dAC_xc})
 and ~(\ref{eq:md_x}) leads to the following exact ensemble energy
expression~\cite{alam_gic},
\begin{eqnarray}
\begin{array}{l}\label{eq:gic_ens_en}
E^{\bf w} = \text{Tr}\left[ \hat{\Gamma}^{\rm{\mu,{\bf w}}}\hat{H} \right] 
+ E_{\rm{c,md}}^{\rm{sr,\mu,{\bf w}}}[n_{\hat{\Gamma}^{\rm{\mu,{\bf w}}}}],
\end{array}
\end{eqnarray}
where, as readily seen, the GI error arising from the short-range Hartree energy has been
removed. 
 Note that the energy expression in Eq.~(\ref{eq:gic_ens_en}) is
{\it not} variational with respect to the
ensemble density matrix operator. A straight minimization over all possible
density matrix operators would lead to a fully interacting solution and
therefore to double counting problems since the exact solution
$\hat{\Gamma}^{\rm{\mu,{\bf w}}}$ is long-range-interacting only. In
practice, we use the WIDFA solution $\hat{\gamma}^{\rm{\mu,{\bf w}}}$
introduced in Eq.~(\ref{eq:widfa_ens_en}) in conjunction with the
complementary ground-state functional $E_{\rm{c,md}}^{\rm{sr,\mu}}[n]$,
for which a local density-functional approximation has been developed by
Paziani {\etal}~\cite{Paziani2006PRB}, thus leading to the approximate GI corrected
(GIC) range-separated energy~\cite{alam_gic},
\begin{eqnarray}
\begin{array}{l}\label{eq:gok_widfa_gic_ens_en}
\begin{split}
\tilde{E}^{\rm{\mu,{\bf w}}}_{\rm{GIC}} & = 
\text{Tr}\left[ \hat{\gamma}^{\rm{\mu,{\bf w}}}\hat{H} \right] 
+ E_{\rm{c,md}}^{\rm{sr,\mu}}[n_{\hat{\gamma}^{\rm{\mu,{\bf w}}}}]\\
& = \sum\limits_{k=0}^{M-1}w_{k}
\langle \tilde{\Psi}_{k}^{\rm{\mu,{\bf w}}} \vert \hat{H} \vert 
\tilde{\Psi}_{k}^{\rm{\mu,{\bf w}}} \rangle 
+ E_{\rm{c,md}}^{\rm{sr,\mu}}[n_{\hat{\gamma}^{\rm{\mu,{\bf w}}}}],
\end{split}
\end{array}
\end{eqnarray}
which is in principle both $\mu$ and weight dependent.
Note that $\tilde{E}^{\rm{\mu,{\bf w}}}_{\rm{GIC}}$ converges towards the
exact ensemble energy when $\mu\rightarrow+\infty$.

\subsection{Extraction of excitation energies by linear interpolation}\label{subsec:interpolation}

Following the seminal work of Gross \etal~\cite{PRA_GOK_EKSDFT}, we
consider an ensemble consisting of the $(I+1)$ lowest multiplets in
energy. This ensemble contains $M_I$ states in total (degeneracy is included) and is characterized by the following weights, 
\begin{eqnarray}
\begin{array}{l}\label{eq:gok_generalized_LIM_weights}
   w_{k} = \left\{
                    \begin{array}{ll}
                       \dfrac{1-w\mathrm{g}_{I}}{M_{I-1}} \ \ \ \ \  0 \leq k \leq M_{I-1}-1\\
                       w \ \ \ \ \ \ \ \ \ \ M_{I-1} \leq k \leq M_{I}-1,
                     \end{array}
\right.
\end{array}
\end{eqnarray}
where ${\mathrm{g}_{K}}$ is the degeneracy of the $K$th multiplet with
energy $E_K$ and $M_{K} = \sum_{L=0}^{K}\mathrm{g}_{L}$ is the total
number of states with energies lower or equal to $E_K$. 
Note that all the weights are controlled by a single weight $w$ in the
range $0 \leq w \leq \frac{1}{M_{I}}$. Consequently, the ensemble energy
becomes a function of $w$ and reads, according to
Eq.~(\ref{eq:exact_ens_en}),
\begin{eqnarray}
E_I^w=\dfrac{1-w\mathrm{g}_{I}}{M_{I-1}}\left(\sum^{I-1}_{K=0}\mathrm{g}_{K}E_K\right)+w\mathrm{g}_{I}E_I.
\end{eqnarray}
The $I$th excitation energy $\omega_I=E_I-E_0$ can be extracted from
$E_I^w$ and the lower excitation energies as
follows~\cite{PRA_GOK_EKSDFT,SenjeanPRA2015},  
\begin{eqnarray}\label{eq:XE_deriv}
\omega_I=\dfrac{1}{\mathrm{g}_{I}}\dfrac{dE_I^w}{d
w}+\dfrac{1}{M_{I-1}}\sum^{I-1}_{K=1}\mathrm{g}_K\omega_K,
\end{eqnarray}
or, alternatively,
\begin{eqnarray}\label{eq:XE_lim}
\omega_I&=&\dfrac{M_I}{\mathrm{g}_{I}}
\left(E^{1/M_I}_I-E^{1/M_{I-1}}_{I-1}\right)+\dfrac{1}{M_{I-1}}\sum^{I-1}_{K=1}\mathrm{g}_K\omega_K,
\nonumber\\
&=&\dfrac{1}{\mathrm{g}_{I}}\left(M_IE^{1/M_I}_I-M_{I-1}E^{1/M_{I-1}}_{I-1}\right)-E_0,
\end{eqnarray}
where we used the linearity of the ensemble energy in $w$ and the
equality $E^{w=0}_I=E^{1/M_{I-1}}_{I-1}$ on the first
line and, on the second line, the fact
that $E^{1/M_0}_0$ equals the ground-state energy $E_0$. While the first expression in
Eq.~(\ref{eq:XE_deriv}) involves the derivative of the ensemble energy, the
second expression in Eq.~(\ref{eq:XE_lim}) uses the linear interpolation
between equiensemble energies. In the exact theory, both expressions are
of course equivalent. However, as soon as approximate wavefunctions and
functionals are used, this is not the case anymore. For example, at the
WIDFA level, approximate range-separated ensemble energies exhibit
curvature with respect to the ensemble weight~\cite{SenjeanPRA2015}. Consequently,
Eq.~(\ref{eq:XE_deriv}) will provide weight-dependent excitation
energies which means that the ensemble weight must be used as a
parameter, in addition
to the range separation one. On the other hand, the linear interpolation
method (LIM) sketched in
Eq.~(\ref{eq:XE_lim}) gives, by construction, weight-independent
approximate excitation energies, which is preferable. Let us stress
that, even when approximate ensemble energies are used, the two
expressions in
Eq.~(\ref{eq:XE_lim}) remain equivalent. Note also that  
LIM applies to any approximate ensemble energies (WIDFA or GIC, with
or without range separation~\cite{alam_gic}). Let us
finally mention that other approaches can be used for extracting
excitation energies from an ensemble, in particular by making another
choice for the ensemble
weights and by using derivatives of the ensemble
energy for a direct
extraction~\cite{yang2017prl}. In the latter
case, weight dependence of the xc functional must be introduced, which
is not trivial. Even though LIM is not direct, in a sense that two
equiensemble calculations are necessary to obtain the excitation energy
of interest (in addition to the ground-state calculation),
standard (weight-independent) ground-state density-functional
approximations can be used~\cite{SenjeanPRA2015}.  
In Sec.~\ref{sec:results}, LIM will be applied to non-degenerate states. In the
latter case, the excitation energy expression in Eq.~(\ref{eq:XE_lim}) can be simplified as
follows,
\begin{eqnarray}\label{eq:XE_lim_no_deg}
\omega_I=(I+1)
E^{1/(I+1)}_I-IE^{1/I}_{I-1}-E_0.
\end{eqnarray}

\subsection{Range-separated GIC ensemble energy in the large $\mu$ limit
and extrapolation technique}\label{subsec:extrapolation}

\begin{figure}[!]
\begin{center}
\includegraphics[scale=0.95]{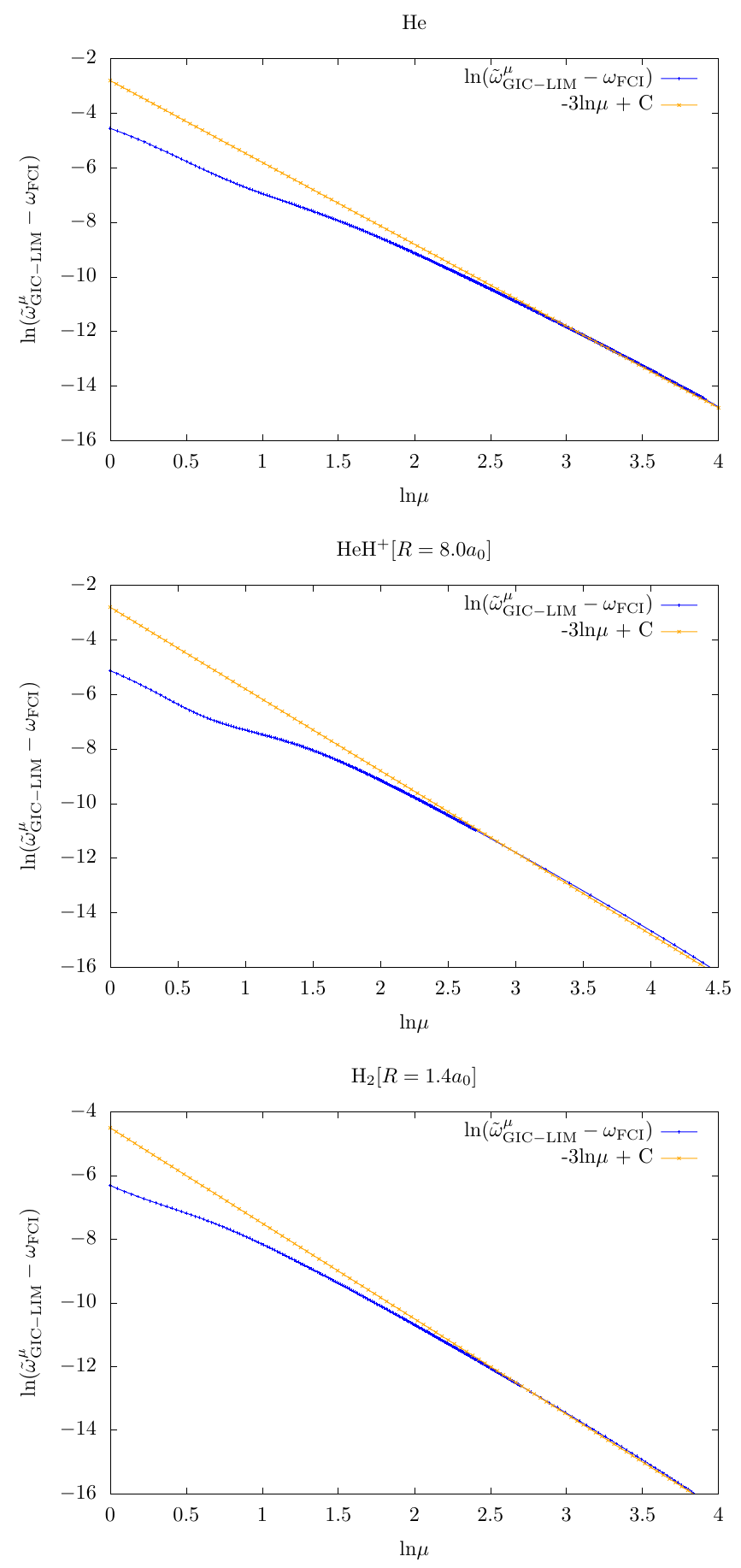}
\end{center}
\caption{Asymptotic behaviour of the GIC-LIM excitation 
energy in the large $\mu$ limit for He,
the stretched HeH$^{+}$ molecule and H$_{2}$ at equilibrium.
The excitations considered are: $\{ 1^{1}{\rm S} \rightarrow 
2^{1}{\rm S}\}$ in He, $\{ 1^{1}\Sigma^{+} \rightarrow 
2^{1}\Sigma^{+} \}$ in HeH$^+$ and $\{ 1^{1}\Sigma^{+}_{g} \rightarrow 
2^{1}\Sigma^{+}_{g} \}$ in H$_{2}$. See text for
further details.
}\label{fig:aysmptotic_giclim_exen}
\end{figure}

Following the seminal work of Savin~\cite{savin2014towards}, Senjean
{\etal}~\cite{extrapol_edft} have shown that, when
$\mu\rightarrow+\infty$, the  
approximate range-separated WIDFA ensemble energy (see
Eq.~(\ref{eq:widfa_ens_en})) converges towards the exact ensemble energy
$E^{{\bf w}}$ as follows, 
\begin{eqnarray}\label{eq:widfa_exp_largemu}
\tilde{E}^{\rm{\mu,{\bf w}}}=E^{{\bf w}}+\dfrac{1}{2}\dfrac{\tilde{E}^{(-2),{{\bf
w}}}}{\mu^2}+\mathcal{O}\left(\dfrac{1}{\mu^3}\right),
\end{eqnarray}
thus leading to
\begin{eqnarray}
\tilde{E}^{\rm{\mu,{\bf w}}}
+\dfrac{\mu}{2}\dfrac{\partial \tilde{E}^{\rm{\mu,{\bf w}}}}{\partial\mu}=E^{{\bf w}}
+\mathcal{O}\left(\dfrac{1}{\mu^3}\right),
\end{eqnarray}
and the corresponding expansion for the combined WIDFA/LIM (just
referred to as LIM in the following) approximate and $\mu$-dependent $I$th excitation energy,
\begin{eqnarray}
\tilde{\omega}_{{\rm{LIM,}}I}^{\mu}
+\dfrac{\mu}{2}\dfrac{\partial
\tilde{\omega}_{{\rm{LIM,}}I}^{\mu}}{\partial\mu}=\omega_I
+\mathcal{O}\left(\dfrac{1}{\mu^3}\right),
\end{eqnarray}
since, according to
Eq.~(\ref{eq:XE_lim}), the latter is a linear combination of WIDFA
(equi-) ensemble energies.
As a result, the deviation of the extrapolated LIM (ELIM) excitation
energy~\cite{extrapol_edft}, 
\begin{eqnarray}
\label{eq:extrapol_lim_exen}
\tilde{\omega}_{{\rm{ELIM,}}I}^{\mu} = \tilde{\omega}_{{\rm{LIM,}}I}^{\mu}  + 
\frac{\mu}{2}\frac{\partial
\tilde{\omega}_{{\rm{LIM,}}I}^{\rm{\mu}}}{\partial \mu},
\end{eqnarray}
from the exact result $\omega_I$ varies as $\mu^{-3}$ while, according
to Eq.~(\ref{eq:widfa_exp_largemu}), it varies as $\mu^{-2}$ in the case
of LIM. Therefore, in practice, the extrapolation correction (second term on the
right-hand side of Eq.~(\ref{eq:extrapol_lim_exen})) will make 
LIM converge faster in $\mu$ towards the pure wavefunction theory ($\mu\rightarrow+\infty$) result~\cite{extrapol_edft}.\\
Let us now consider the approximate range-separated GIC ensemble energy
in Eq.~(\ref{eq:gok_widfa_gic_ens_en})
which is also $\mu$-dependent. As shown in the Appendix, it varies as
follows for large $\mu$ values,    
\begin{eqnarray}\label{eq:gic_exp_largemu}
\tilde{E}^{\rm{\mu,{\bf w}}}_{\rm{GIC}}=E^{{\bf w}}+\dfrac{1}{6}\dfrac{\tilde{E}^{(-3),{{\bf
w}}}_{\rm{GIC}}}{\mu^3}+\mathcal{O}\left(\dfrac{1}{\mu^4}\right),
\end{eqnarray}
and therefore converges faster in $\mu$ than the WIDFA ensemble energy, thus leading to 
\begin{eqnarray}\label{eq:exp_GIC_ens_ener}
\tilde{E}^{\rm{\mu,{\bf w}}}_{\rm{GIC}}
+\dfrac{\mu}{3}\dfrac{\partial \tilde{E}^{\rm{\mu,{\bf w}}}_{\rm{GIC}}}{\partial\mu}=E^{{\bf w}}
+\mathcal{O}\left(\dfrac{1}{\mu^4}\right),
\end{eqnarray}
and, for the combined GIC-LIM approximate excitation energy,  
\begin{eqnarray}\label{eq:exp_XE_GIC-LIM}
\tilde{\omega}_{{\rm{GIC-LIM,}}I}^{\mu}
+\dfrac{\mu}{3}\dfrac{\partial
\tilde{\omega}_{{\rm{GIC-LIM,}}I}^{\mu}}{\partial\mu}=\omega_I
+\mathcal{O}\left(\dfrac{1}{\mu^4}\right).
\end{eqnarray}
Eqs.~(\ref{eq:exp_GIC_ens_ener}) and (\ref{eq:exp_XE_GIC-LIM}) are the central result of the paper. 
As readily seen, the standard extrapolation correction in
Eq.~(\ref{eq:extrapol_lim_exen})
is {\it not} relevant anymore when ghost-interaction errors are removed.
The factor $1/2$ should be replaced by $1/3$, thus leading to the
extrapolated GIC-LIM (EGIC-LIM) excitation energy expression,  
\begin{eqnarray}
\label{eq:extrapol_giclim_exen}
\tilde{\omega}_{{\rm{EGIC-LIM,}}I}^{\mu} = \tilde{\omega}_{{\rm{GIC-LIM,}}I}^{\mu}  
+ \frac{\mu}{3}\frac{\partial
\tilde{\omega}_{{\rm{GIC-LIM,}}I}^{\rm{\mu}}}{\partial \mu},
\end{eqnarray}
which converges as $\mu^{-4}$ towards the pure wavefunction theory
result while GIC-LIM converges as $\mu^{-3}$, as expected from
Eqs.~(\ref{eq:XE_lim}) and (\ref{eq:gic_exp_largemu}), and illustrated in
Fig.~\ref{fig:aysmptotic_giclim_exen} for He, H$_{2}$ ($R=1.4a_{0}$) and
HeH$^{+}$ ($R=8.0a_{0}$).   

\subsection{Higher-order extrapolation
corrections}\label{subsec:higher_order_extrapo}
As pointed out in Ref.~\cite{rebolini2015calculating}, higher-order energy derivatives can be
used in the extrapolation correction in order to further improve on the convergence of ELIM and EGIC-LIM
excitation energies towards the FCI results in the large $\mu$ limit.
From the Taylor expansion of the WIDFA ensemble energy through third 
order 
\begin{eqnarray}
\tilde{E}^{\mu,{\bf w}} = E^{{\bf w}} + \frac{1}{2}\frac{\tilde{E}^{(-2),{\bf w}}}{\mu^{2}} + \frac{1}{6}\frac{\tilde{E}^{(-3),{\bf w}}}{\mu^{3}} + \mathcal{O}\left(\frac{1}{\mu^{4}}\right),
\end{eqnarray}
we obtain
\begin{eqnarray}
\tilde{E}^{\mu,{\bf w}} + \mu\frac{\partial \tilde{E}^{\mu,{\bf w}}}{\partial \mu} + \frac{\mu^{2}}{6}\frac{\partial^{2} \tilde{E}^{\mu,{\bf w}}}{\partial \mu^{2}} = 
E^{{\bf w}} +\mathcal{O}\left(\frac{1}{\mu^{4}}\right),
\end{eqnarray}
thus leading, after linear interpolation, to the following
{\it second-order} ELIM (ELIM2) excitation energy expression,
\begin{eqnarray}\label{eq:elim2_formula}
\tilde{\omega}_{{\rm{ELIM2}},I}^{\mu} = \tilde{\omega}_{{\rm{LIM}},I}^{\mu} + \mu\frac{\partial \tilde{\omega}_{{\rm{LIM}},I}^{\mu}}{\partial \mu} + 
\frac{\mu^{2}}{6}\frac{\partial^{2}
\tilde{\omega}_{{\rm{LIM}},I}^{\mu}}{\partial \mu^{2}},
\end{eqnarray}
which is exact through third order in $1/\mu$, like the EGIC-LIM
excitation energy. 
Similarly, from the Taylor expansion of the GIC ensemble energy through
fourth order (see the Appendix), 
\begin{eqnarray}
\tilde{E}_{\rm{GIC}}^{\mu,{\bf w}} = E^{{\bf w}} + \frac{1}{6}\frac{\tilde{E}_{\rm{GIC}}^{(-3),{\bf w}}}{\mu^{3}} + 
\frac{1}{24}\frac{\tilde{E}_{\rm{GIC}}^{(-4),{\bf w}}}{\mu^{4}} + \mathcal{O}\left(\frac{1}{\mu^{5}}\right)
\end{eqnarray}
it comes
\begin{eqnarray}
&&\tilde{E}_{\rm{GIC}}^{\mu,{\bf w}} + \frac{2}{3}\mu\frac{\partial \tilde{E}_{\rm{GIC}}^{\mu,{\bf w}}}{\partial \mu} + 
\frac{1}{12}\mu^{2}\frac{\partial^{2} \tilde{E}_{\rm{GIC}}^{\mu,{\bf w}}}{\partial \mu^{2}} 
\nonumber\\
&&= E^{\bf w}
+ \mathcal{O}\left(\frac{1}{\mu^{5}}\right),
\end{eqnarray}
thus leading to the {\it second-order} EGIC-LIM (EGIC-LIM2) excitation
energy expression, 
\begin{eqnarray}\label{eq:EGICLIM2_formula}
\tilde{\omega}_{{\rm{EGIC-LIM2}},I}^{\mu} &=& \tilde{\omega}_{{\rm{GIC-LIM}},I}^{\mu} + \frac{2}{3}\mu\frac{\partial \tilde{\omega}_{{\rm{GIC-LIM}},I}^{\mu}}{\partial \mu} \nonumber\\
&&+ \frac{1}{12}\mu^{2}\frac{\partial^{2} \tilde{\omega}_{{\rm{GIC-LIM}},I}^{\mu}}{\partial \mu^{2}},
\end{eqnarray}
which is exact through fourth order in $1/\mu$.
\section{Computational Details}\label{sec:computational_details}
WIDFA (Eq.~(\ref{eq:widfa_ens_en})) and GIC
(Eq.~(\ref{eq:gok_widfa_gic_ens_en})) range-separated ensemble energies as well as 
LIM and GIC-LIM excitation energies (see Eq.~(\ref{eq:XE_lim_no_deg})),
with (Eqs.~(\ref{eq:extrapol_lim_exen}) and
(\ref{eq:extrapol_giclim_exen})) and without extrapolation, have been
computed 
with a development version of the DALTON program package
\cite{DALTON,DALTON2} for a small test set consisting of He, 
H$_{2} (R=1.4a_{0}, 3.7a_{0})$, HeH$^{+}$ 
and LiH. The extrapolated LIM and GIC-LIM excitation energies (ELIM and EGIC-LIM)
have been calculated using finite differences with 
$\Delta\mu = 0.005a_{0}^{-1}$. 
The long-range-interacting wavefunctions have been calculated using full CI (FCI) level of theory in combination with the 
spin-independent ground-state short-range local density approximation 
of Toulouse {\etal}\cite{toulda,srDFT}. The short-range 
multideterminantal correlation functional of Paziani {\etal}
\cite{Paziani2006PRB} has been used for calculating GIC range-separated ensemble
energies and GIC-LIM
excitation energies. For all systems but LiH, aug-cc-pVQZ basis sets
\cite{dunning1989gaussian,woon1994gaussian} have been used.
For LiH, aug-cc-pVTZ basis set with frozen 1s orbital has been used. 
For calculating the first excitation energy a two-state ensemble is considered in all the 
cases whereas for the higher excitation energies larger ensembles 
(three-, four- and five-state ensembles), consisting of singlet 
 states only, are considered. The corresponding two-state ensembles are 
 $\{1^{1}\Sigma^{+}, 2^{1}\Sigma^{+}\}$ for HeH$^{+}$ and LiH,  
 $\{1^{1}\Sigma^{+}_{g}, 2^{1}\Sigma^{+}_{g}\}$ for H$_{2}$ 
 and $\{1^{1}S, 2^{1}S\}$ for He. 
 The larger ensembles have been used for calculating the 
 $\{1^{1}\Sigma^{+}_{g}\rightarrow 3^{1}\Sigma^{+}_{g}\},
  \{1^{1}\Sigma^{+}_{g} \rightarrow 4^{1}\Sigma^{+}_{g}\}$  
  and $\{1^{1}\Sigma^{+}_{g} \rightarrow 5^{1}\Sigma^{+}_{g}\}$ 
  excitation energies in H$_{2}$ and $\{1^{1}\Sigma^{+} \rightarrow 
  3^{1}\Sigma^{+}\}, \{1^{1}\Sigma^{+} \rightarrow 4^{1}\Sigma^{+}\}$  
 excitation energies in HeH$^{+}$.

\section{Results and discussion}\label{sec:results}

\subsection{Basis set convergence in He}
\begin{figure}[h]
\begin{center}
\includegraphics[scale=0.95]{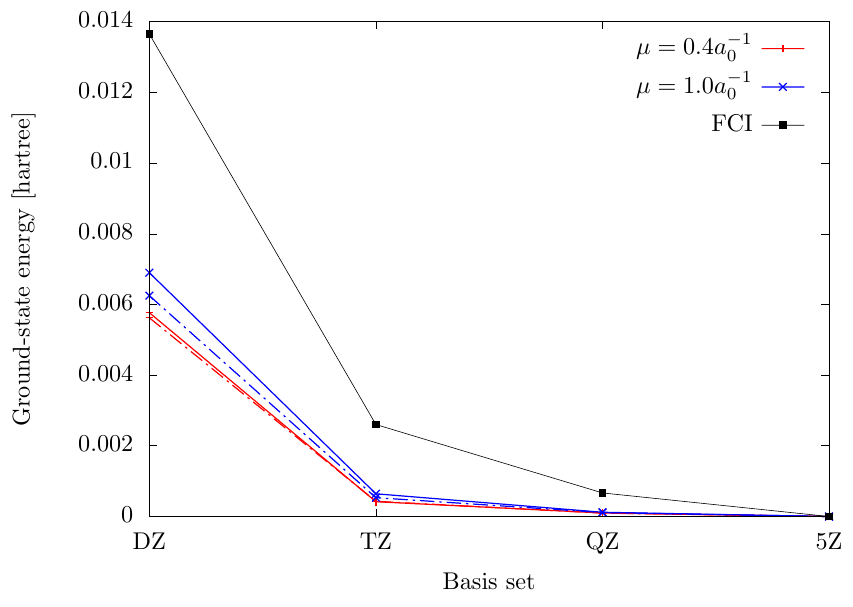}\\
\includegraphics[scale=0.95]{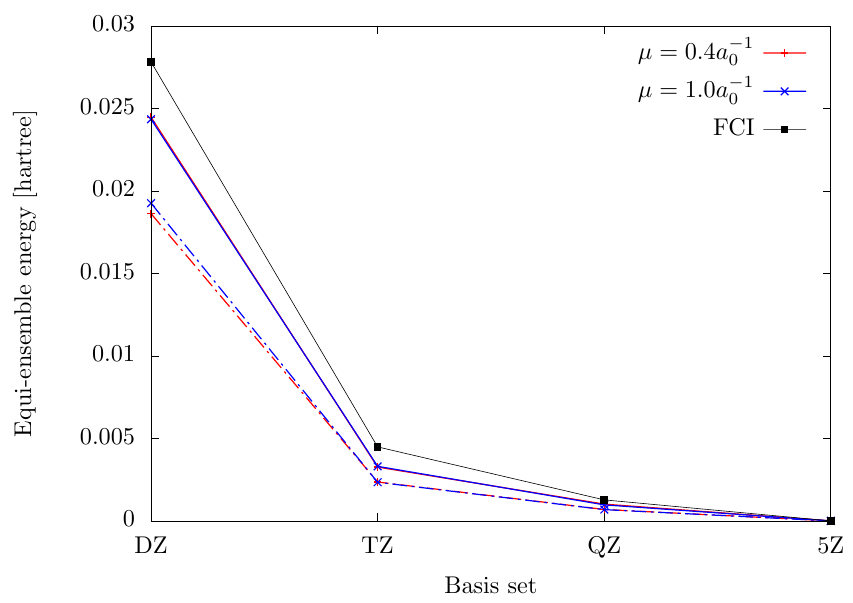}
\end{center}
\caption{Variation of ground-state energy (top panel) and equi-ensemble
energy (bottom panel) with
basis set in He. 
For clarity, 
the energies relative to aug-cc-pV5Z values are plotted.
Basis sets used 
are aug-cc-pVnZ, where n=2-5.
GIC and WIDFA values are represented by solid and dash-dotted lines, respectively. 
Results for $\mu=0.4a_{0}^{-1}$ and $\mu=1.0a_{0}^{-1}$ are represented by the red
and blue colored lines, respectively. For comparison FCI values 
(black curve) are also plotted.}\label{fig:convergence_he}
\end{figure}
The performance of LIM and GIC-LIM (with and without extrapolation) has
already been discussed for He in Ref.~ \cite{alam_gic} The purpose of
this section is to extend the discussion of
Franck~\etal~\cite{JCP15_Odile_basis_convergence_srDFT} on the basis set
convergence of range-separated ground-state energies to ensembles.
In Fig.~\ref{fig:convergence_he}, 
the convergence 
of WIDFA/GIC ground-state ($w$=0) and equiensemble ($w$=1/2)
range-separated energies 
obtained with aug-cc-pVnZ (n=2,3,4,5) basis sets are 
shown for the two-state ensemble $\{1^{1}S, 2^{1}S\}$ in He with the
range-separation parameter set to the
typical~\cite{FromagerJCP2007,PRA13_Pernal_srEDFT}
$\mu=0.4a^{-1}_0$ and $\mu=1.0a_0^{-1}$ values. 
In 
comparison to the FCI values, the WIDFA/GIC range-separated energies
show faster convergence, especially the ground-state ones. The latter 
converge at the same rate with both methods 
 whereas, for the equiensemble energies, WIDFA values converge 
 faster than GIC values (which are actually relatively close to the FCI
ones). The $\mu$ dependence is not the same for the two energies. Equiensemble energies are less sensitive to the 
 range-separation parameter $\mu$ than the ground-state energy. In 
 fact, the GIC equiensemble energies for the two $\mu$ values overlap. 
 The difference between WIDFA and GIC equiensemble energies as well as 
 their slower convergence with the basis set (when comparison is made with the
ground-state energy) are due to the facts that (i) long-range correlation 
 effects are negligible in the ground state but significant in the
equiensemble because of the Rydberg character of the excited state and  
(ii) the GIC energy is less density-dependent than the WIDFA one. Indeed, 
 in the former case, short-range correlation effects only are described by a 
 density functional. 

\subsection{LiH}\label{subsec:gs_es_en_lih}

\begin{figure}[H]
\begin{center}
\includegraphics[scale=0.7]{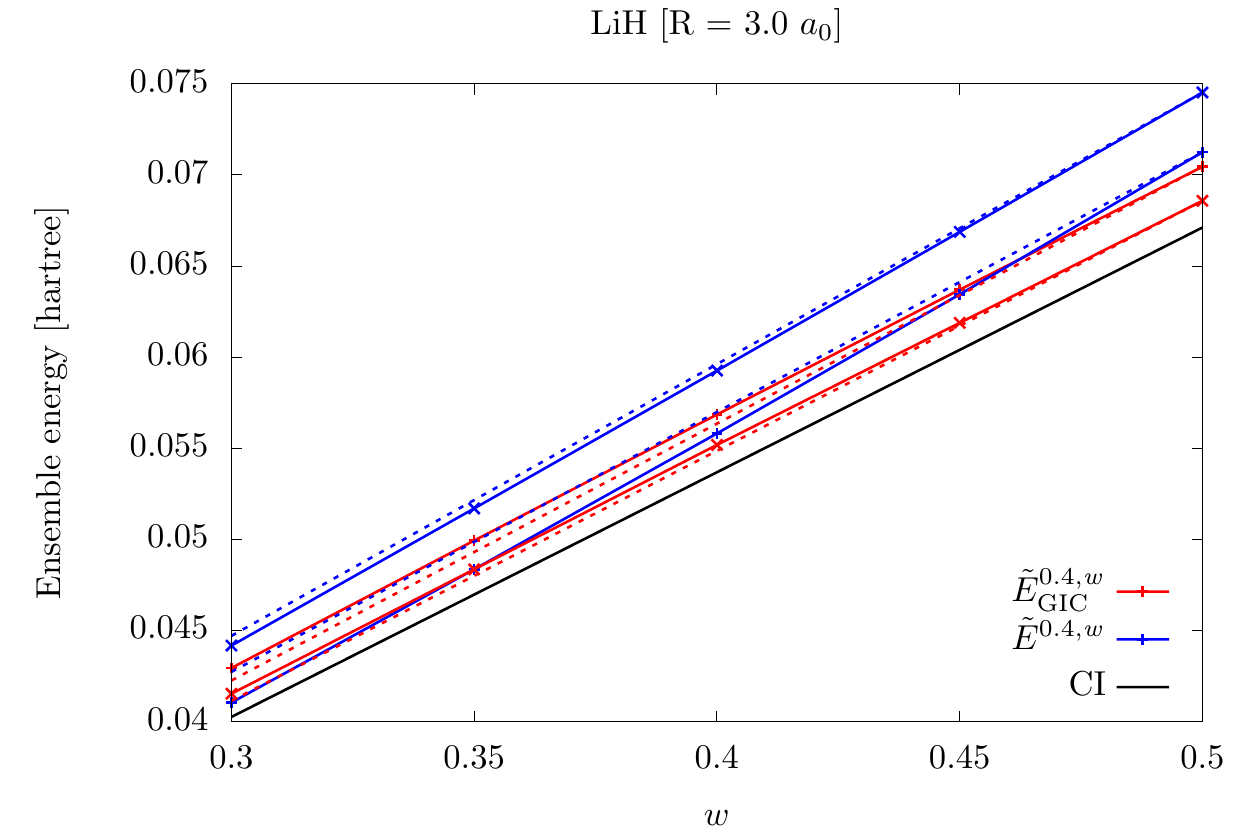}\\
\includegraphics[scale=0.7]{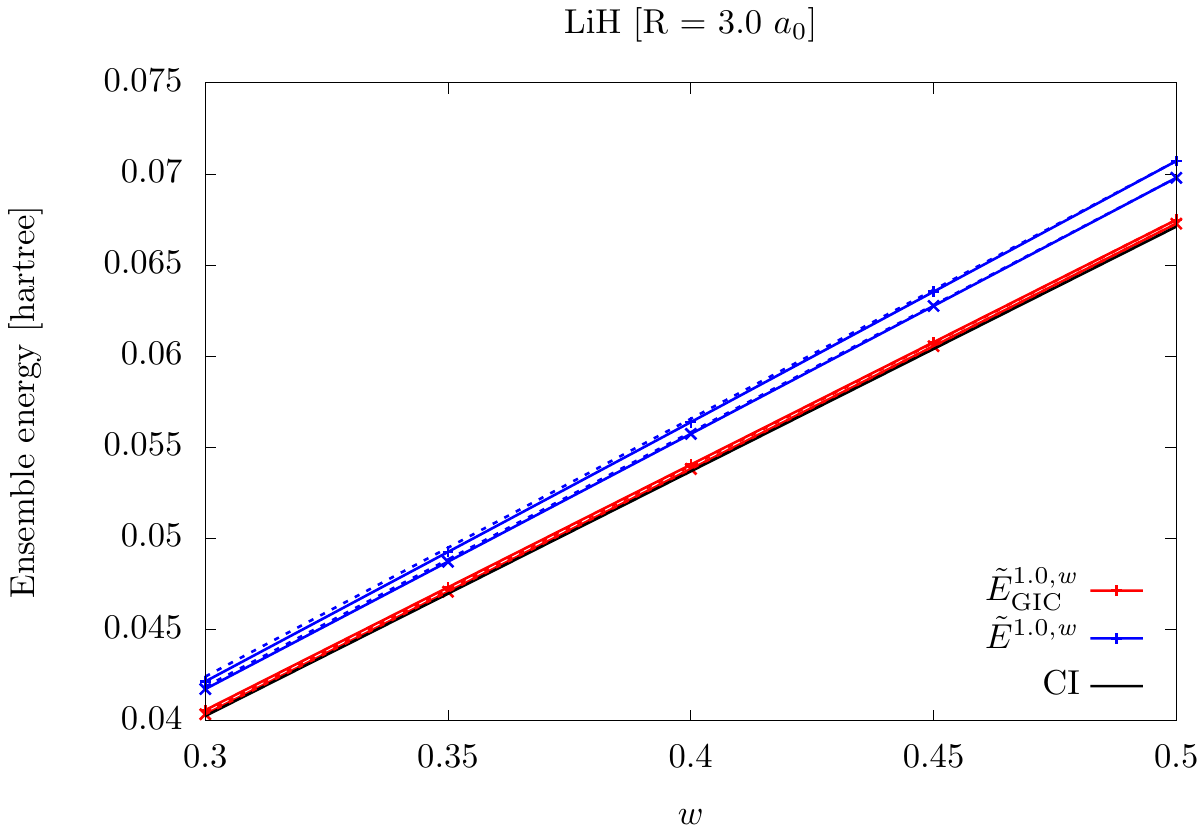}
\end{center}
\caption{Effect of the extrapolation on the range-separated energy of the two-state 
 $\left\{1^{1}\Sigma^{+}, 2^{1}\Sigma^{+}\right\}$ ensemble
with weight $w$ in LiH $(R = 3.0a_{0})$ for $\mu=0.4a_{0}^{-1}$ (top)
 and $1.0a_{0}^{-1}$ (bottom). Red and blue curves with point type 
 ``+" represent the GIC and WIDFA ensemble energies, respectively, whereas the same 
 colored lines with point type ``$\times$" represent the corresponding extrapolated 
values. The dashed lines connecting the two extreme points ({\it i.e.} for $w=0.0$ and 
$w=0.5$) are drawn to show the deviation from linearity.
}
\label{fig:lih_extrapol_ensen}
\end{figure}
The effect of extrapolation on the two-state ensemble energy of LiH in
the large-$w$ region is 
shown in Fig.~\ref{fig:lih_extrapol_ensen} for $\mu=0.4a_{0}^{-1}$
(top panel) and $1.0a_{0}^{-1}$ (bottom panel). 
It is obvious from these plots 
that the WIDFA energy and its extrapolation 
are slightly curved, which is more visible for $\mu=0.4a_{0}^{-1}$, and
that they deviate significantly from the CI straight line 
This is a consequence of using an
approximate (weight-independent) ground-state local short-range xc
functional, as discussed in Sec.~\ref{subsec:gi_gic_edft} and Ref.~\cite{extrapol_edft,SenjeanPRA2015}
On going from $\mu=0.4a_{0}^{-1}$  to $\mu=1.0a_{0}^{-1}$, although 
curvature is reduced, the WIDFA energies (with or without extrapolation)
still differ from the CI result. In a previous work~\cite{alam_gic}, we
have shown that the GIC scheme could almost restore the linearity of the ensemble energy, which is also reflected in Fig.~\ref{fig:lih_extrapol_ensen}. 
Note that, for $\mu=0.4a_{0}^{-1}$, the extrapolation enlarges the
deviation of the WIDFA ensemble energy from the accurate CI result. It
only leads to an improvement when the larger $\mu=1.0a_{0}^{-1}$ value
is used. On
the other hand, GIC energies are always improved after extrapolation.
For $\mu=1.0a_{0}^{-1}$, the extrapolated GIC ensemble energy is almost
on top of the CI one. 

\begin{figure}[H]
\begin{center}
\includegraphics[scale=0.7]{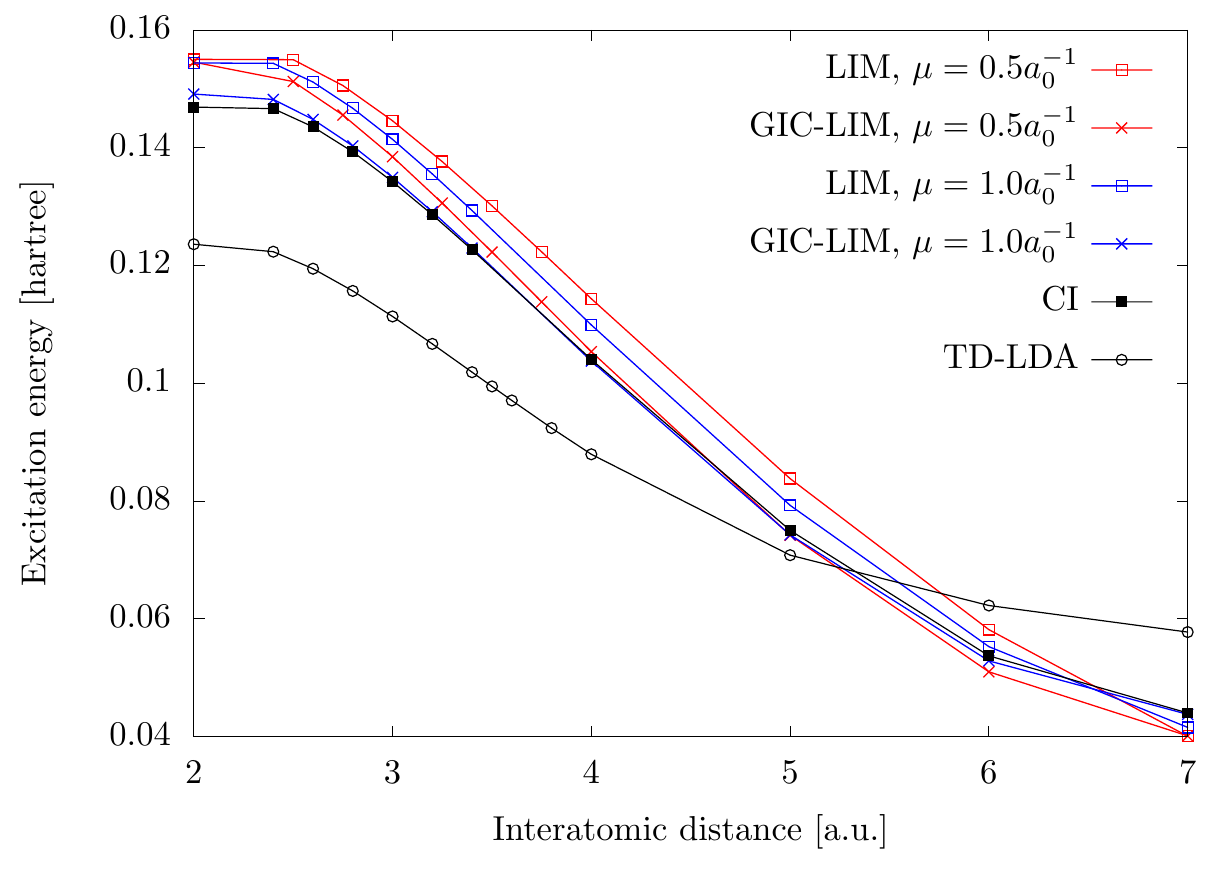}
\end{center}
\caption{Variation of the first $\Sigma^+$ singlet excitation energy with the inter-atomic 
distance in LiH, for $\mu = 0.5a_{0}^{-1}$(red colored lines) 
and $\mu=1.0a_{0}^{-1}$(blue colored lines). LIM 
and GIC-LIM values are represented by $\square$ and $\times$,
respectively. For comparison, TD-LDA ($\fullmoon$) and CI results 
($\blacksquare$) are also plotted.
} \label{fig:lih_gs_es_en}
\end{figure}

In Fig.~\ref{fig:lih_gs_es_en}, we show the variation of the first 
$\Sigma^+$ singlet excitation energy of LiH with the inter-atomic distance, for 
$\mu=0.5$ and $1.0a_{0}^{-1}$. The comparison is made with the CI 
and TD-LDA results. Note that, in contrast to TD-LDA, both LIM and
GIC-LIM reproduce relatively well the shape of the CI curve. As
expected, GIC-LIM is closer to CI than LIM for the two $\mu$ values.
For $\mu=1.0a_{0}^{-1}$, the agreement is actually excellent beyond the
equilibrium distance ($R>3a_0$). At equilibrium ($R=3a_0$),
TD-LDA underestimates the excitation energy by 0.0229 a.u.~(if
comparison is made with the CI result), which was expected since the
$2^1\Sigma^+$ state has a charge-transfer character (from H to Li). On the
other hand, GIC-LIM $(\mu = 0.5 a_{0}^{-1})$ slightly overestimates (by
0.004 a.u.) the excitation energy. For larger bond distances,
the failure of TD-LDA might be related to the multiconfigurational
character of the $2^1\Sigma^+$ state. The multideterminantal treatment of
the long-range interaction in range-separated eDFT enables a proper
description of the excitation energy in the strong correlation
regime. 

\begin{figure}[H]
\begin{center}
\includegraphics[scale=0.65]{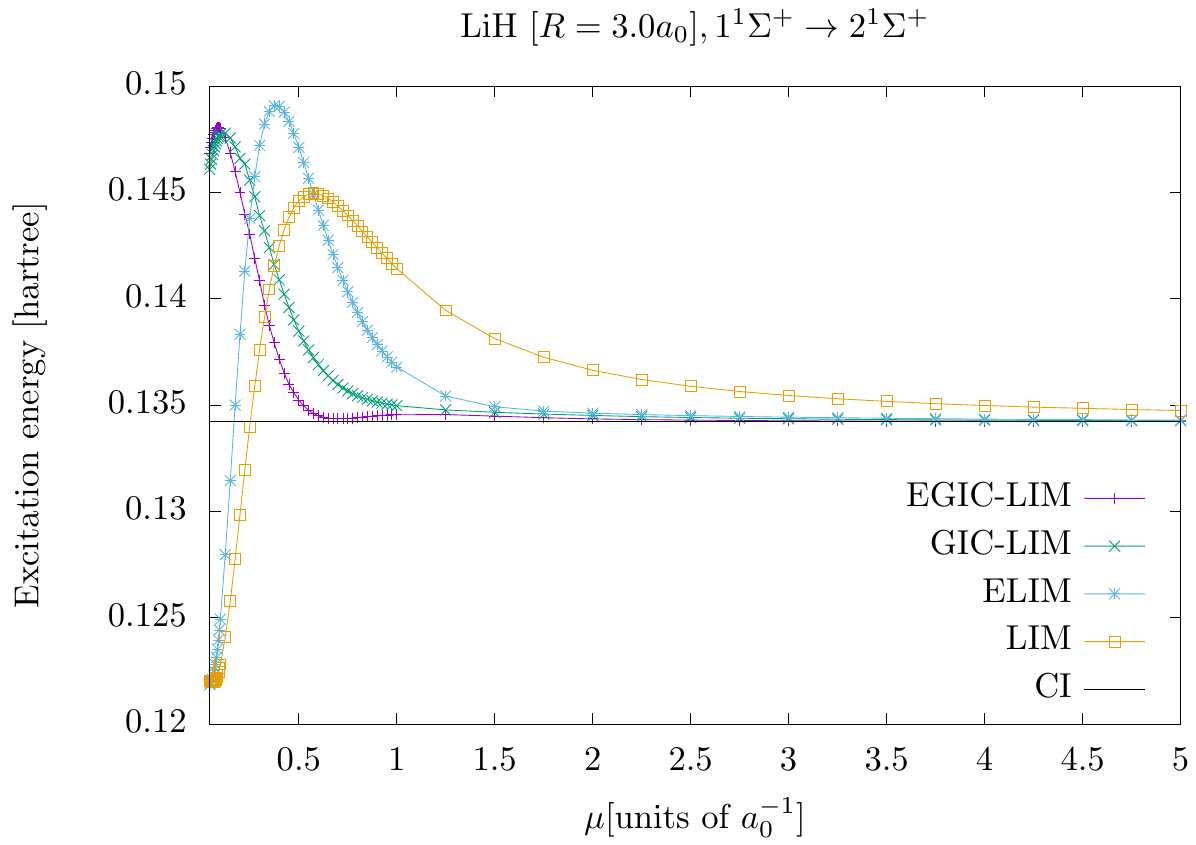}
\end{center}
\caption{Comparison of EGIC-LIM excitation energy with that 
obtained from other schemes in LiH. See text for further
details.
}
\label{fig:gicelim_exen_vs_mu_lih}
\end{figure}
The performance of the extrapolation scheme is now investigated at the
fixed inter-atomic distance $R=3a_0$ when varying the range-separation
parameter. Results are shown in Fig.~\ref{fig:gicelim_exen_vs_mu_lih}.
One can easily see that EGIC-LIM exhibits the fastest convergence in
$\mu$ towards the CI result, as expected. While GIC-LIM and ELIM
excitation energies are almost converged at about $\mu =  2.0
a_{0}^{-1}$, EGIC-LIM reaches the CI result already for the relatively
small $\mu=0.75 a_{0}^{-1}$ value.

\subsection{HeH$^{+}$}\label{subsec:HeH+}
\begin{figure}[H]
\begin{center}
\includegraphics[scale=0.65]{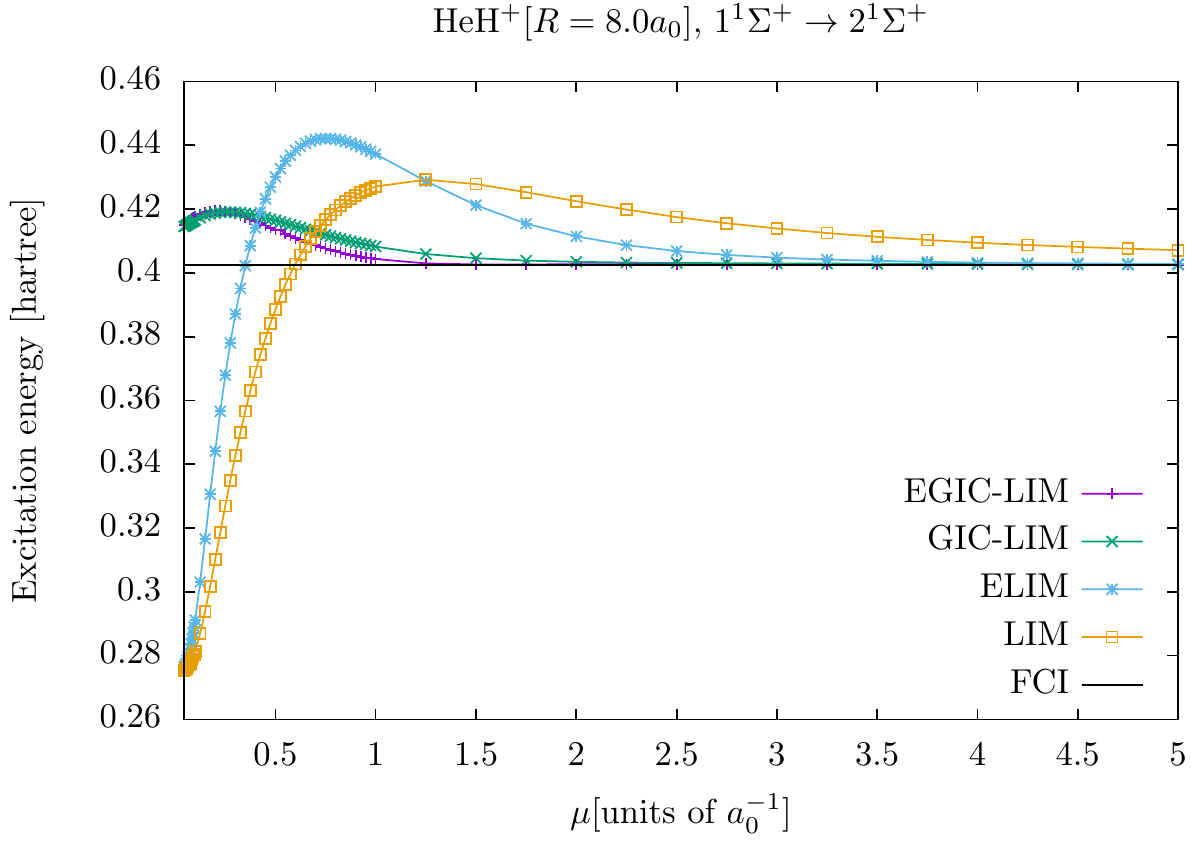}
\end{center}
\caption{Comparison of EGIC-LIM excitation energy with that 
obtained from other schemes in the stretched HeH$^{+}$ molecule. See
text for further details.}
\label{fig:gicelim_exen_vs_mu_heh}
\end{figure}
We show in Fig.~\ref{fig:gicelim_exen_vs_mu_heh} 
the convergence
of the $1^{1}\Sigma^{+}\rightarrow 2^{1}\Sigma^{+}$ charge-transfer
excitation energy with the $\mu$ parameter in the stretched HeH$^{+}$
($R$=8.0$a_{0}^{-1}$) molecule. 
As already observed for LiH, EGIC-LIM 
converges faster (at about $\mu =1.0a_{0}^{-1}$) than the other
methods.\\ 
We also studied the variation of the $1^{1}\Sigma^{+}\rightarrow
n^{1}\Sigma^{+}$ ($n=2,3,4$) excitation energies 
with the bond length for $\mu=0.4$ and $1.0a_{0}^{-1}$ values.
Results are shown in Fig.~{\ref{fig:heh_pes} and comparison is made with
FCI and TD-LDA.
In contrast to TD-LDA, which significantly underestimates the
(charge-transfer) excitation energies, as expected, both LIM and GIC-LIM
(with or without extrapolation) are much closer to FCI for all
interatomic distances. Interestingly, for $\mu=0.4a_{0}^{-1}$, LIM
underestimates the first excitation energy and overestimates the second
and third excitation energies whereas, for $\mu=1.0a_{0}^{-1}$, it
overestimates all the three excitation energies. After extrapolation,
the corresponding ELIM ($\mu=0.4a_{0}^{-1}$) excitation energies
increase, which is an improvement only for the first excited state.  
As expected, GIC-LIM performs better than LIM and ELIM. It slightly overestimates all
excitation energies for both $\mu$ values. The impact of the
extrapolation correction on the curves is hardly visible. Note that, for
$\mu$=$1.0a_{0}^{-1}$, EGIC-LIM and FCI curves are almost on top of each
other. Finally, even for the relatively small range-separation parameter value $\mu=0.4a_{0}^{-1}$, the avoided crossing between the second and
third excited states at about $R=4.0a_{0}$ is 
well reproduced by GIC-LIM (with or without extrapolation).
\begin{figure*}
\begin{center}
\includegraphics[scale=0.9]{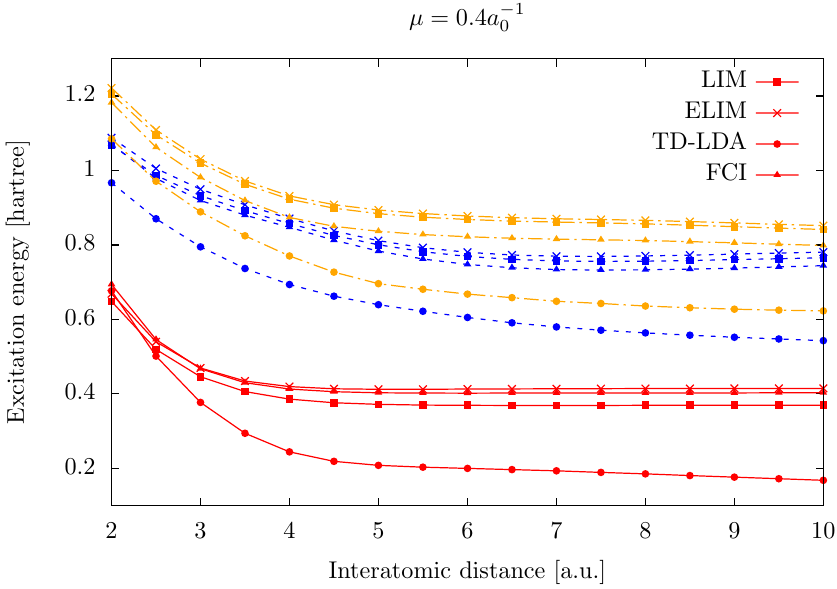}
\includegraphics[scale=0.9]{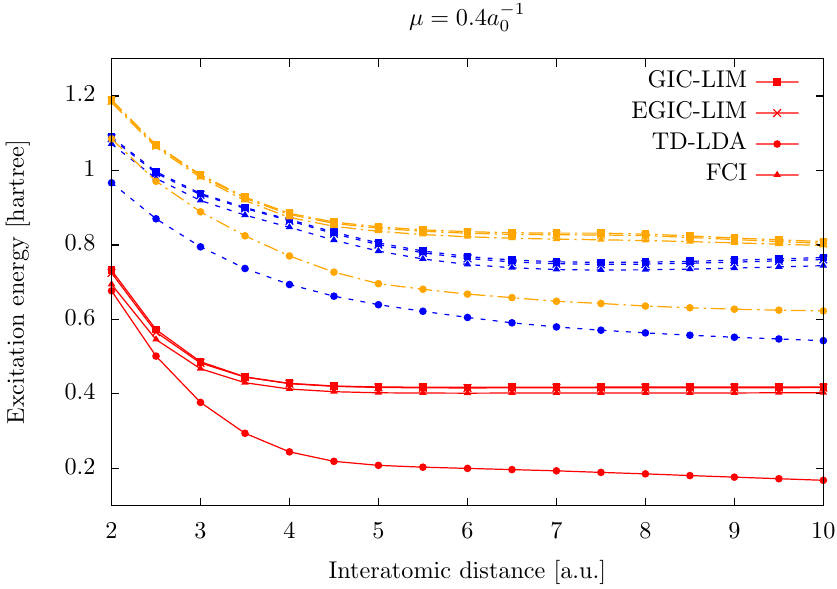}\\
\includegraphics[scale=0.9]{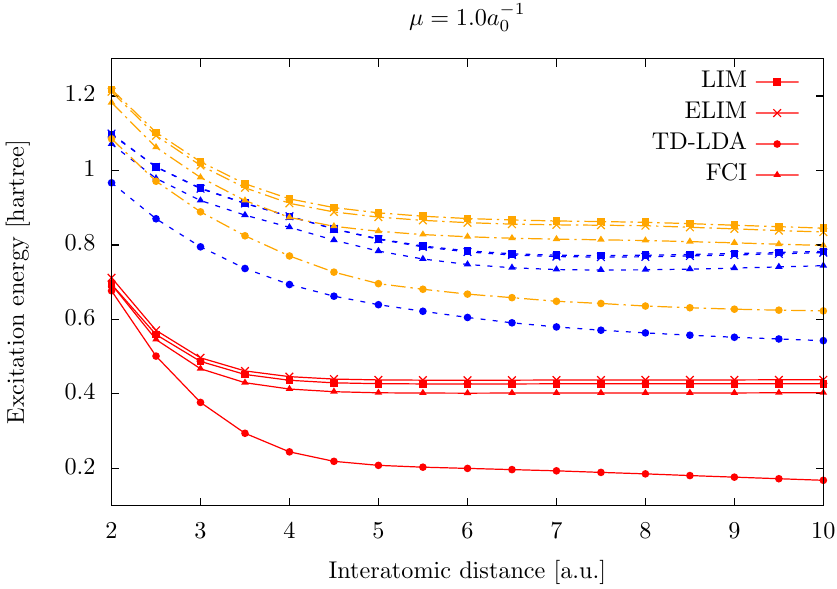}
\includegraphics[scale=0.9]{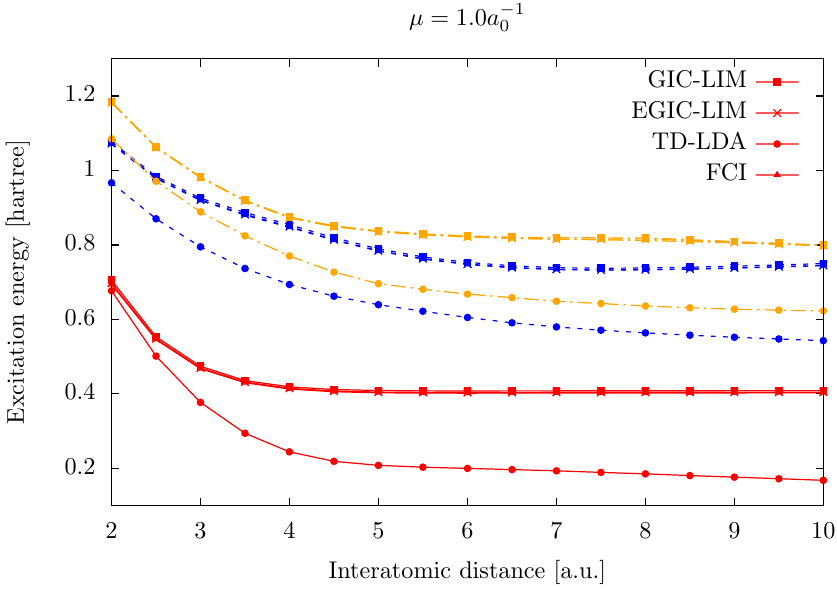}
\end{center}
\caption{Variation of first (red curves; $1^{1}\Sigma^{+} \rightarrow 
2^{1}\Sigma^{+}$), second (blue curves; $1^{1}\Sigma^{+} \rightarrow 
3^{1}\Sigma^{+}$) and 
third (orange curves; $1^{1}\Sigma^{+} \rightarrow 
4^{1}\Sigma^{+}$) excitation energies with the interatomic 
distance in HeH$^{+}$. The following markers are used to 
distinguish results obtained from different methods --
LIM (or GIC-LIM): $\blacksquare$, ELIM (or EGIC-LIM): $\times$, TD-LDA: {$\newmoon$}, and FCI: $\blacktriangle$.}\label{fig:heh_pes}
\end{figure*}

\subsection{Convergence in $\mu$ of higher-order extrapolation schemes}

The variation with $\mu$ of the excitation energies obtained for He and
HeH$^{+}$ with {\it second-order} extrapolation schemes (see
Eqs.~(\ref{eq:elim2_formula})
and (\ref{eq:EGICLIM2_formula})) are shown in
Fig.~\ref{fig:egiclim2_convergence}. As expected, ELIM2 and EGIC-LIM
decay similarly in the large $\mu$ limit. Nevertheless, the GIC still
ensures a faster convergence in $\mu$ towards the FCI result. Regarding
the GIC-LIM results, we observe a systematic improvement on
the excitation energies when adding higher-order extrapolation
corrections in the typical range $0.4\leq \mu\leq 1.0$. Note that EGIC-LIM2 reaches the FCI result for $\mu \sim 0.9a_{0}^{-1}$, which
is remarkable. This clearly demonstrates that GIC ensemble energies can
give very accurate excitation energies after extrapolation for
relatively small range-separation parameter values. Since EGIC-LIM is
already accurate for typical $\mu$ values, second-order extrapolation corrections will not
be considered in the rest of the discussion.

\begin{figure*}[h]
\begin{center}
\includegraphics[scale=0.95]{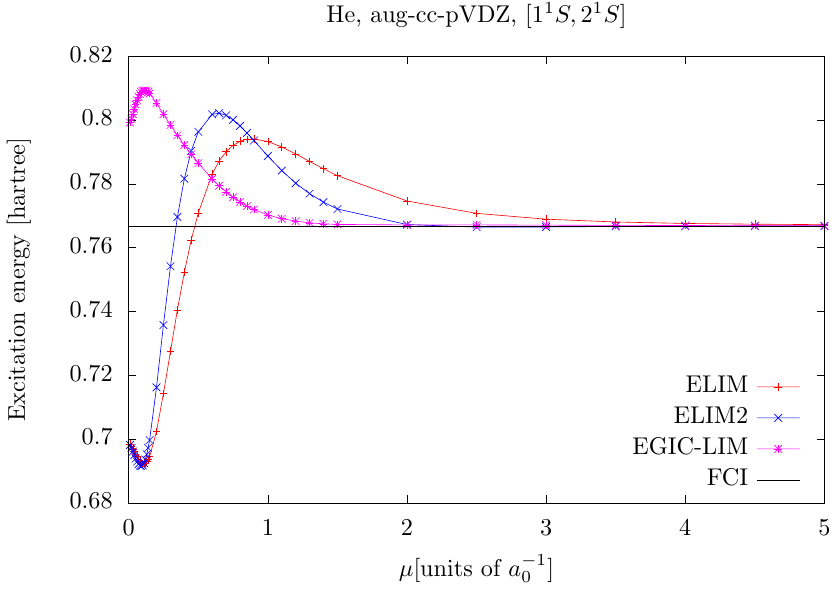}
\includegraphics[scale=0.95]{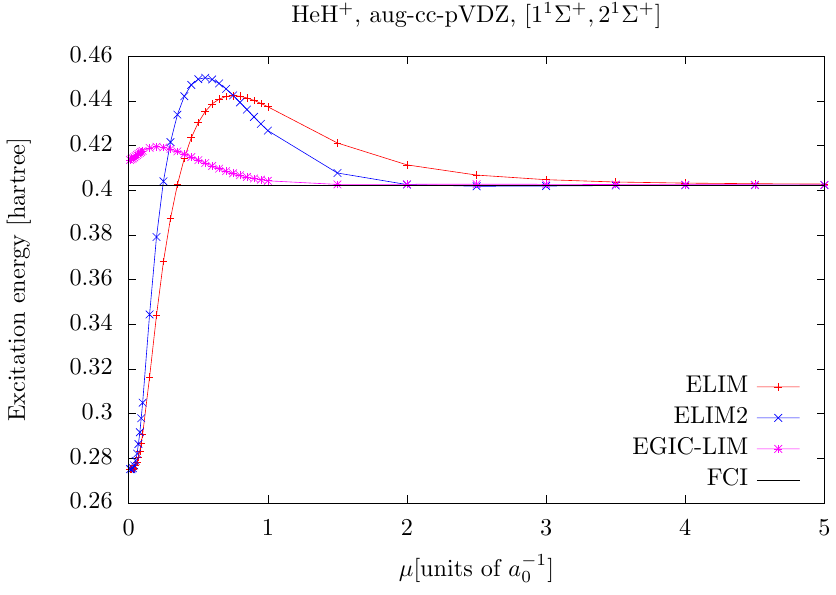}\\
\includegraphics[scale=0.95]{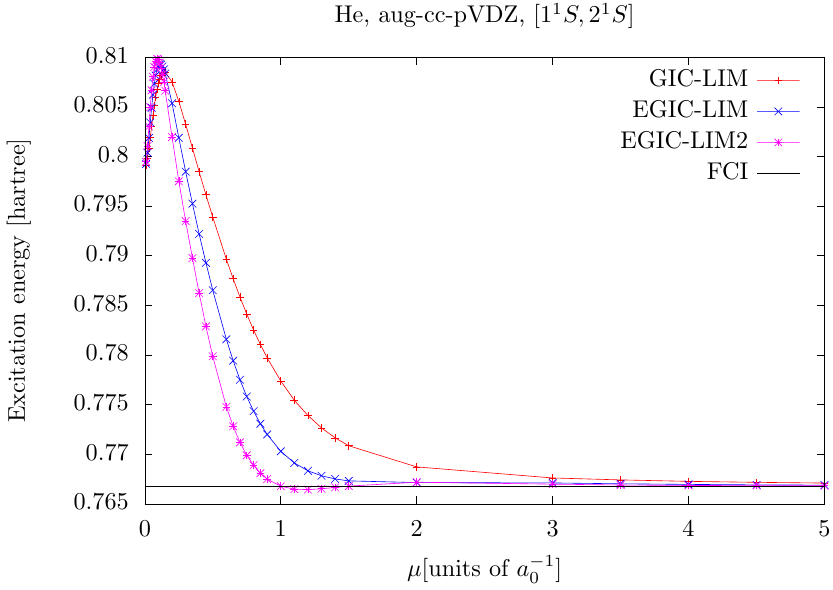}
\includegraphics[scale=0.95]{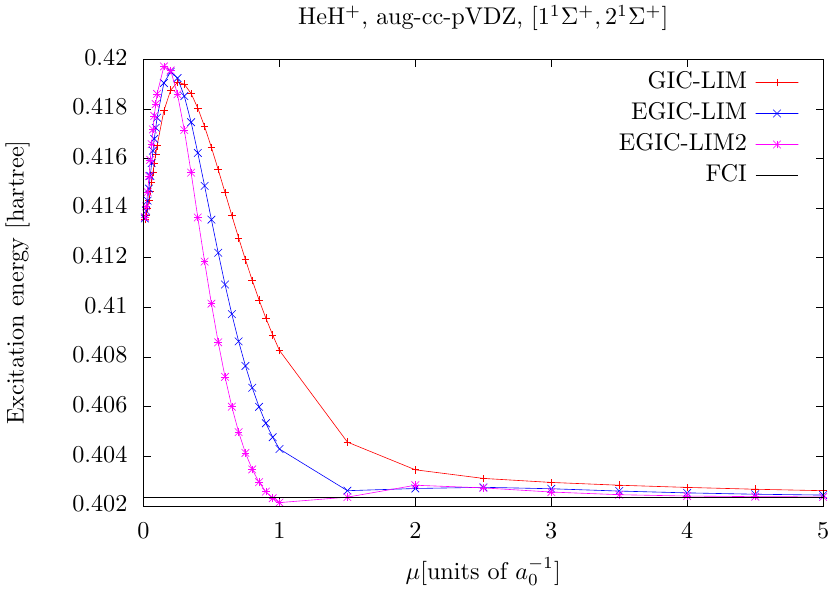}
\end{center}
\caption{
Variation of excitation energies (top panels: ELIM, ELIM2 and EGIC-LIM, bottom panels:
GIC-LIM, EGIC-LIM and EGIC-LIM2) with $\mu$ for He (left panels) and HeH$^{+}$ (right panels).
}\label{fig:egiclim2_convergence}
\end{figure*}

\subsection{H$_{2}$}\label{subsec:H2}
\begin{figure*}[h]
\begin{center}
\includegraphics[scale=0.65]{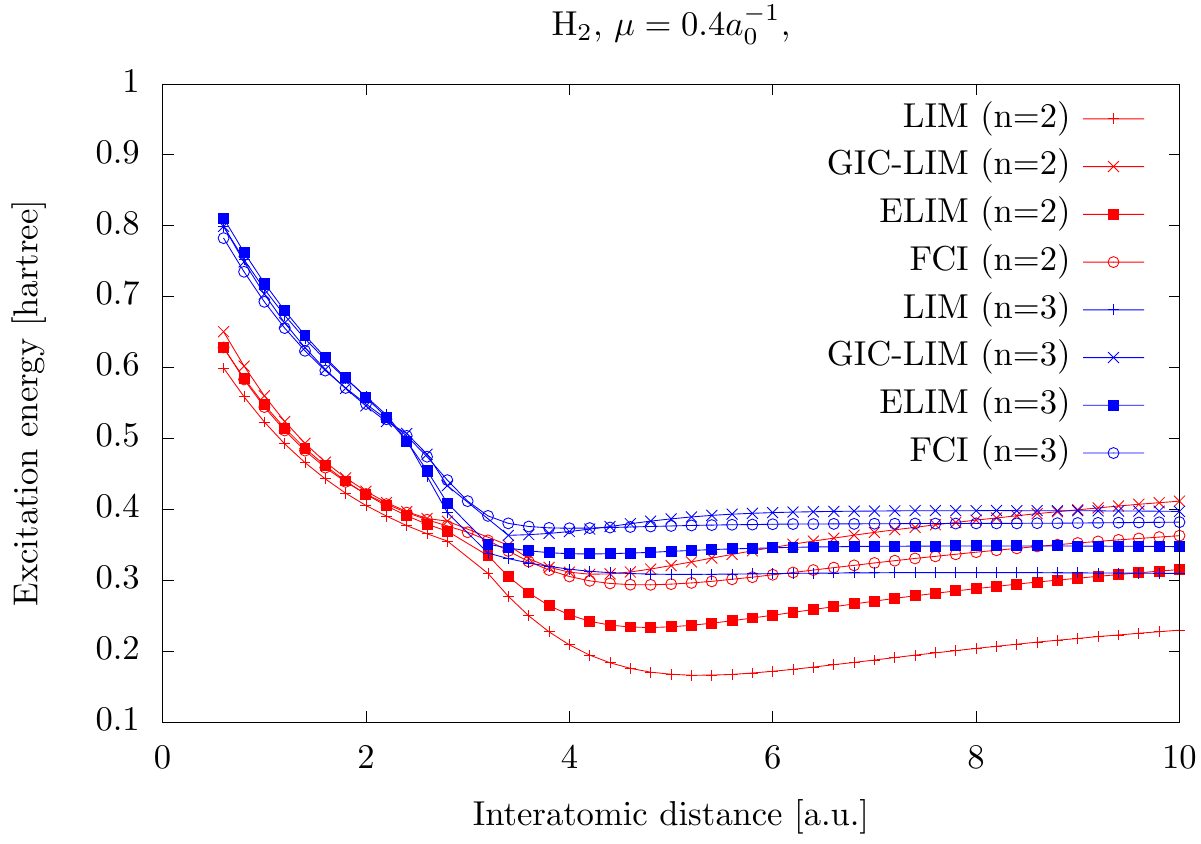}
\includegraphics[scale=0.65]{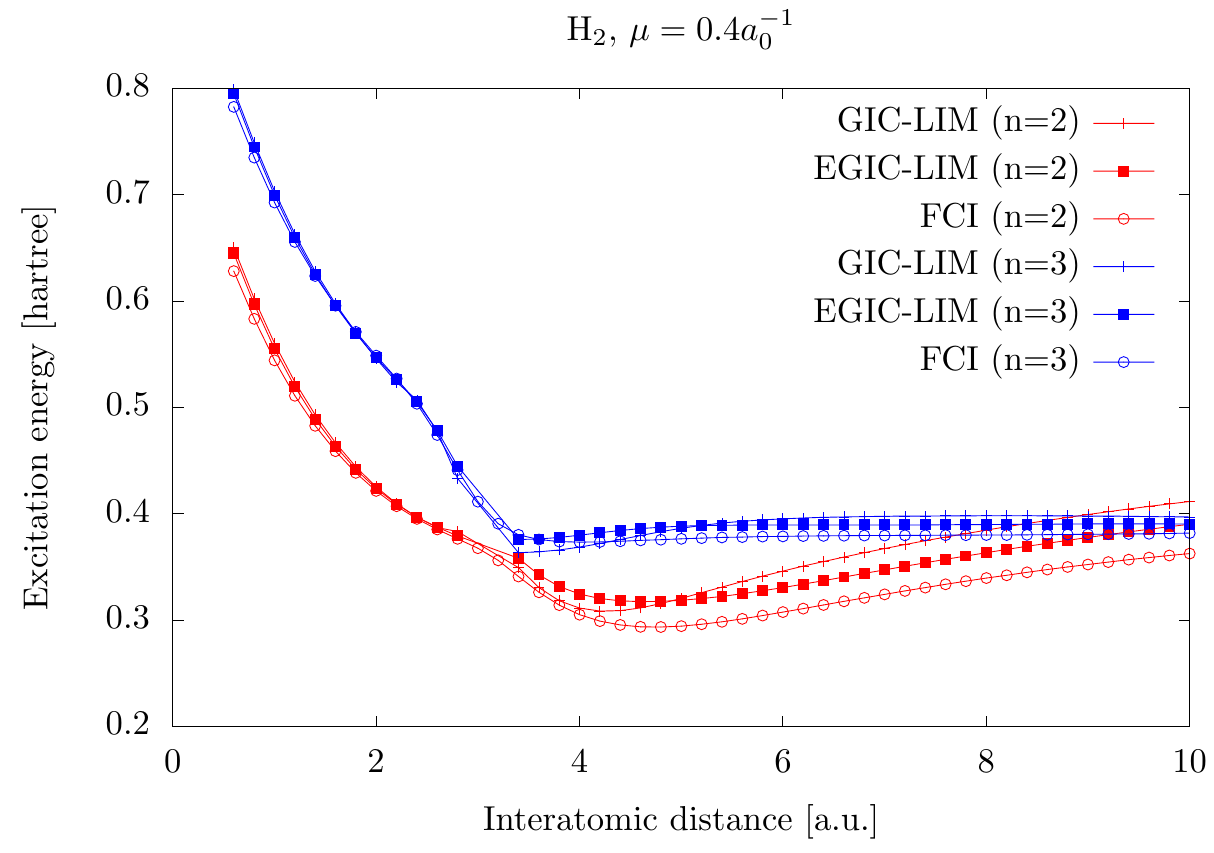}\\
\includegraphics[scale=0.65]{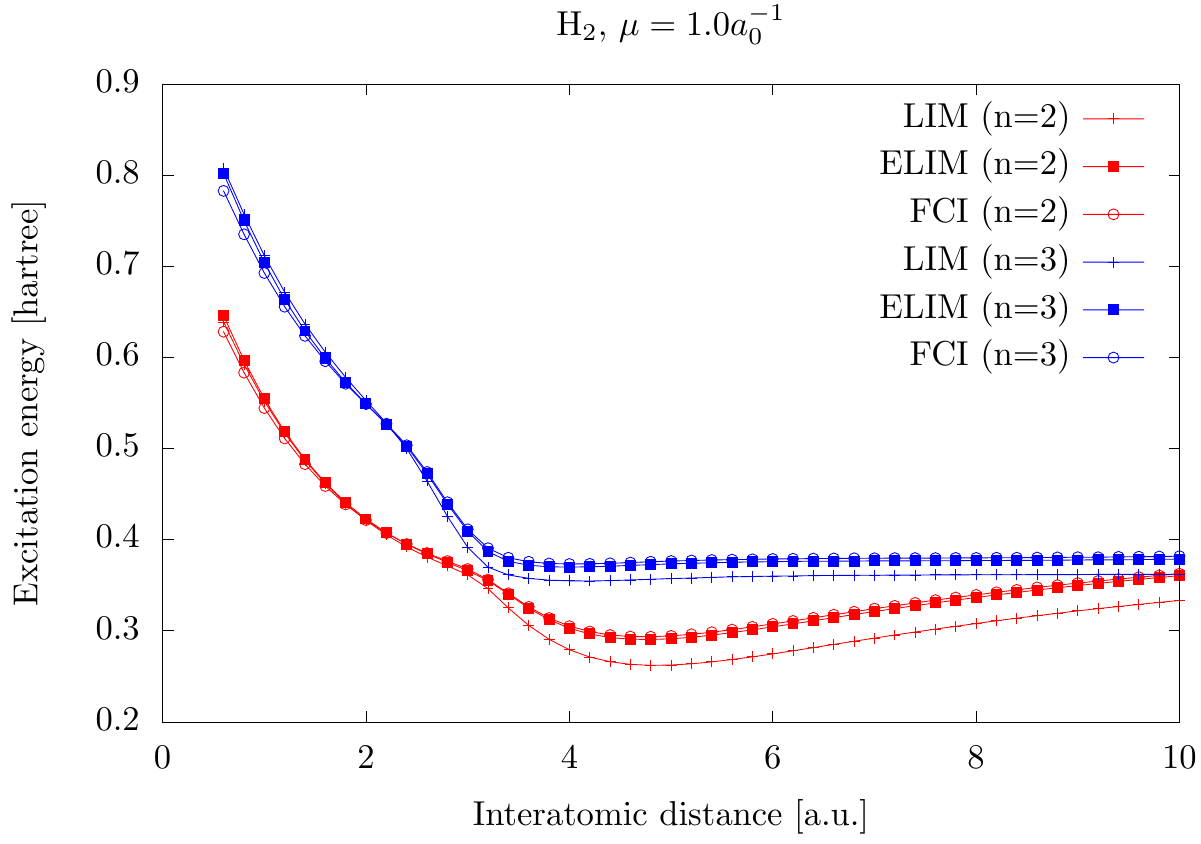}
\includegraphics[scale=0.65]{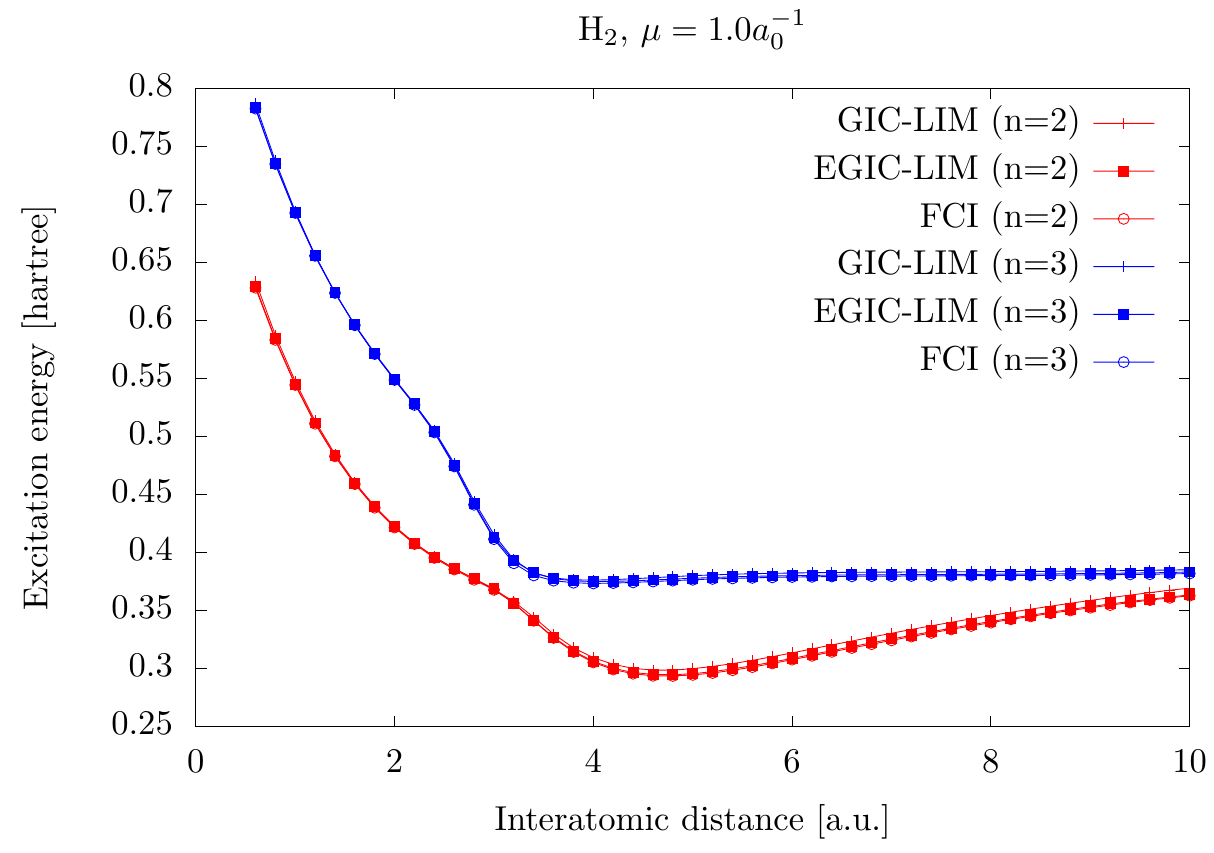}\\
\end{center}
\caption{
Variation of the $1^{1}\Sigma_{g}^{+} \rightarrow
n^{1}\Sigma_{g}^{+}$ excitation energy in H$_{2}$ for $n$=2 and 3 with
the interatomic distance. Results are shown for $\mu = 0.4a_{0}^{-1}$
(top panels) and $\mu = 1.0a_{0}^{-1}$ (bottom panels).
}\label{fig:firstandsecond_exenofh2}
\end{figure*}
\begin{figure*}[h]
\begin{center}
\includegraphics[scale=0.65]{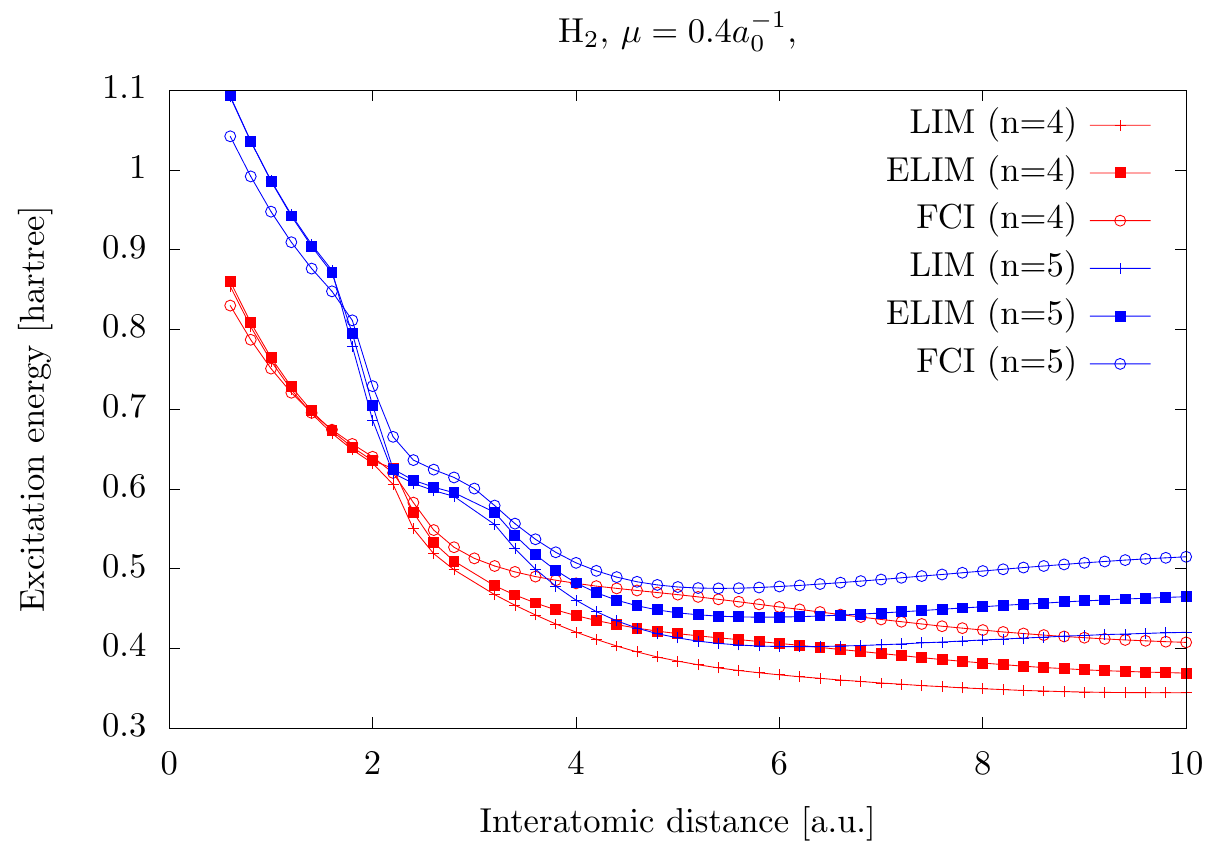}
\includegraphics[scale=0.65]{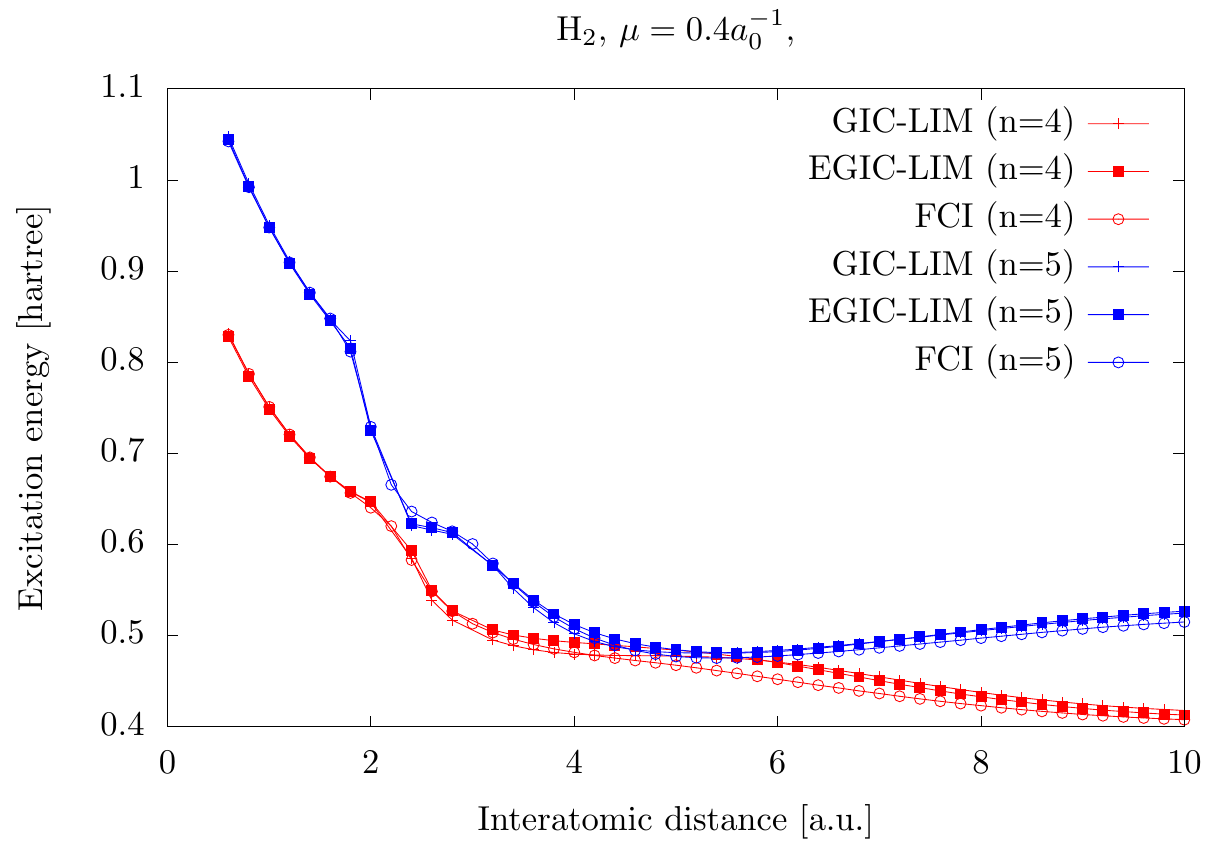}\\
\includegraphics[scale=0.65]{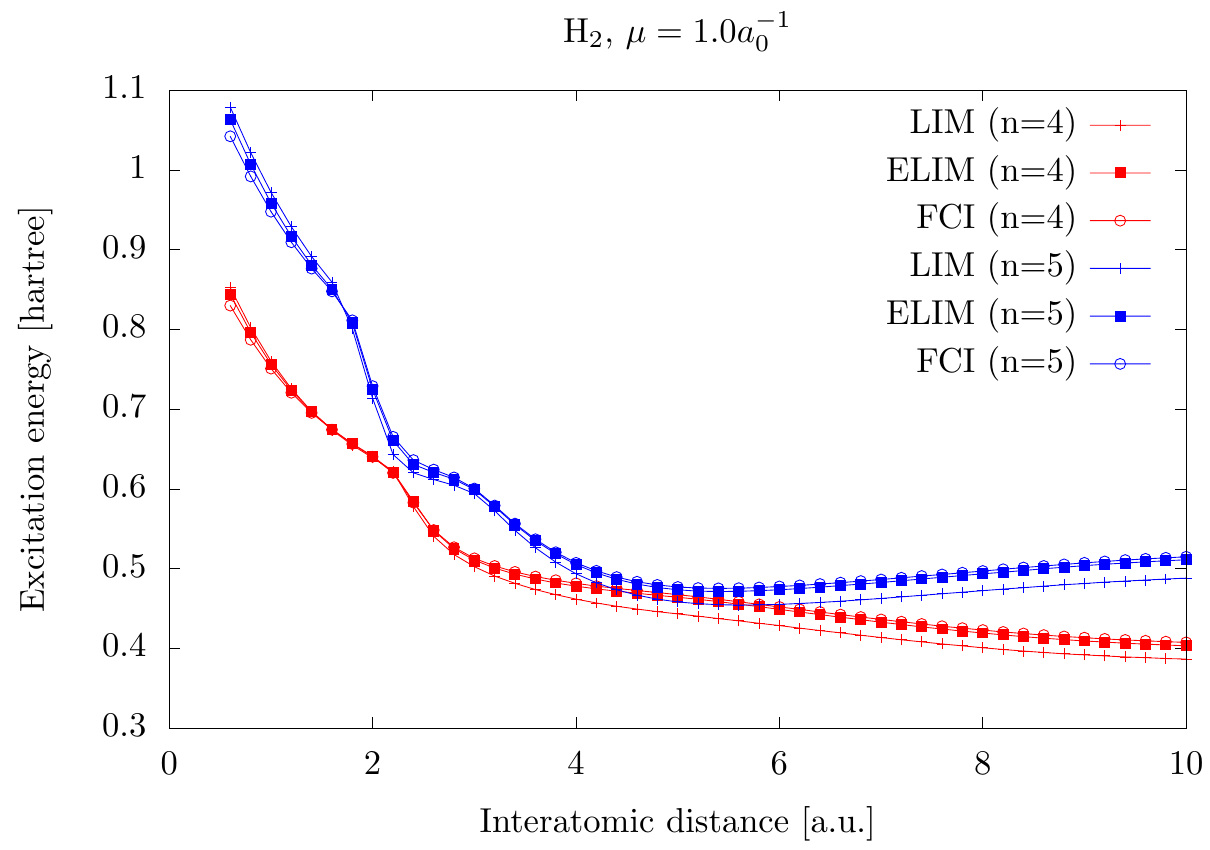}
\includegraphics[scale=0.65]{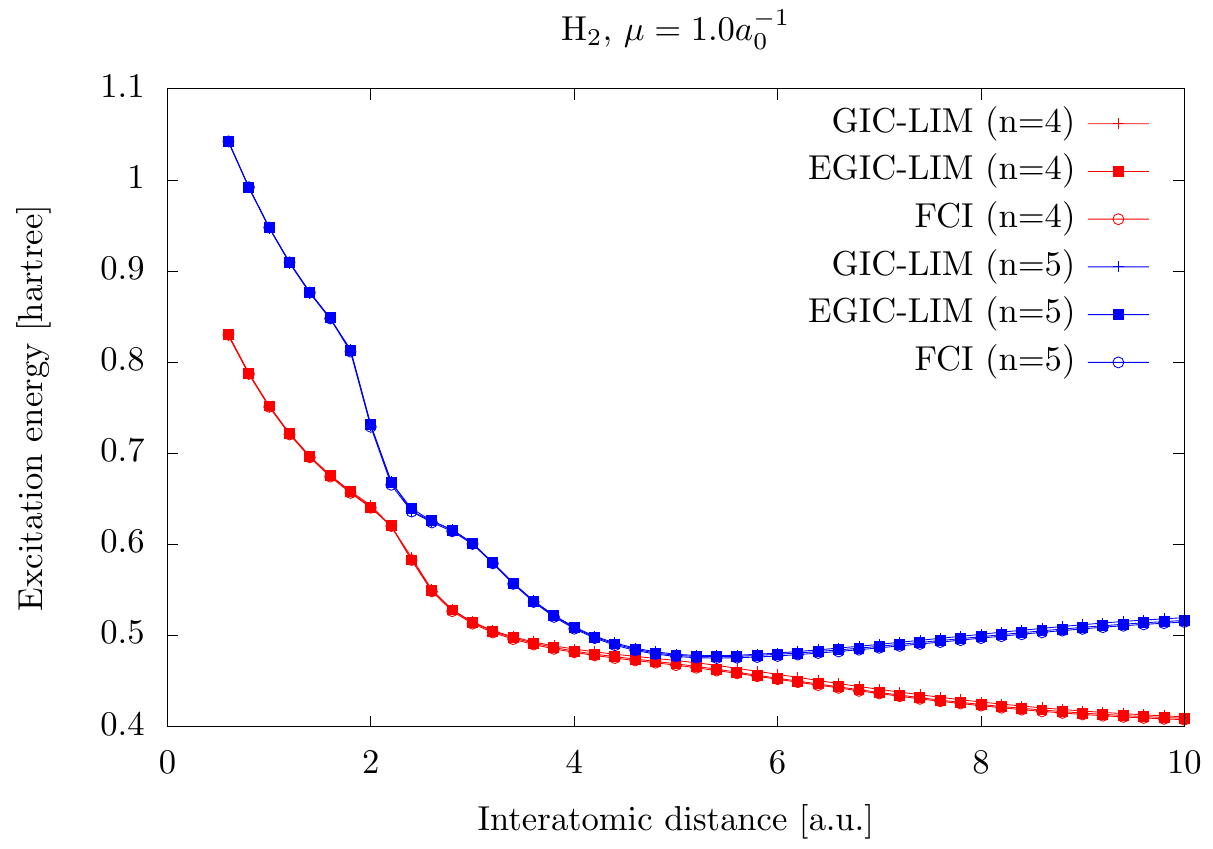}\\
\end{center}
\caption{
Variation of the $1^{1}\Sigma_{g}^{+} \rightarrow
n^{1}\Sigma_{g}^{+}$ excitation energy in H$_{2}$ for $n$=4 and 5 with
the interatomic distance. Results are shown for $\mu = 0.4a_{0}^{-1}$
(top panels) and $\mu = 1.0a_{0}^{-1}$ (bottom panels).
}\label{fig:thirdandfourth_exenofh2}
\end{figure*}
Excitation energies have been computed for the first and second (see
Fig.~\ref{fig:firstandsecond_exenofh2}) as well as third and fourth (see
Fig.~\ref{fig:thirdandfourth_exenofh2}) $^{1}\Sigma_{g}^{+}$ excited
states of H$_2$
along the bond breaking coordinate 
with $\mu=0.4$ and $1.0a_{0}^{-1}$.
Comparison is made with FCI. Recently, the first and second 
excitation energies obtained with the LIM and ELIM methods have 
been reported and discussed in detail by Senjean {\etal} in Ref.~\cite{extrapol_edft} Note that, around 
the equilibrium distance ($R=1.4a_{0}$), they are both relatively close
to the FCI value, especially when $\mu=1.0a_{0}^{-1}$. Substantial
differences appear when stretching the bond. 
LIM underestimates the four excitation energies when $R>3a_0$, in
particular for $\mu=0.4a_{0}^{-1}$. The extrapolation improves on the
results significantly but, for $\mu=0.4a_{0}^{-1}$, the ELIM excitation
energies are still too low. The overall performance of GIC-LIM, when
compared to LIM and ELIM, is far better. Still, 
for $\mu=0.4a_{0}^{-1}$, GIC-LIM
overestimates the excitation energies in the large-$R$ region. The
extrapolation slightly improves on the results. With the larger
$\mu=1.0a_{0}^{-1}$ value, GIC-LIM and FCI potential curves are almost on top of
each other. Small differences are visible at large distances where the
extrapolation correction actually brings some improvement.\\  
Let us now focus on the avoided crossing between the first and second excited
states at 
$\sim3.5a_{0}$. Beyond this distance the first excited state corresponds
to a double excitation. This is the reason why, in contrast to any
eDFT-based method, standard (adiabatic) TD-LDA does not
exhibit any avoided crossing~\cite{FromagerJCP2013}. We note that,  
for $\mu=0.4a_{0}^{-1}$, GIC-LIM improves on the individual
excitation energies when compared to LIM but the two states are then
too close in energy. The extrapolation slightly improves on GIC-LIM in this respect 
while making, when applied to LIM, the two states even closer in energy as
already shown in Ref.~\cite{extrapol_edft} 
For $\mu = 1.0a_{0}^{-1}$, the extrapolation brings 
larger improvement on the LIM values than the GIC-LIM ones.
Regarding the third and fourth excitation energies, two avoided crossings are found
at $\sim 2.4a_{0}$ and $\sim 5.0a_{0}$. Note that at the second 
avoided crossing the two FCI curves are closer to each
other than at the first avoided crossing. Noticeably, 
for $\mu=0.4a_{0}^{-1}$, this behavior is reproduced by GIC-LIM but not 
by LIM. At the second 
avoided crossing, the two GIC-LIM curves are closer to each other 
than the FCI curves. In contrast to EGIC-LIM results, the two ELIM 
curves cross at $R\sim 2.4a_{0}$. For $\mu=1.0a_{0}^{-1}$, 
both ELIM and EGIC-LIM are able to reproduce the two avoided crossings.
Note that, for the second
avoided crossing, EGIC-LIM is substantially improved by increasing $\mu$
from 0.4 to $1.0a_{0}^{-1}$. 

\section{Perspective: extracting individual state energies from 
range-separated 
ensemble energies 
}\label{sec:individual_energies}

Having access to state energies rather than ensemble energies or
excitation energies is important for modeling properties
like equilibrium structures in the excited states. While the extraction of individual energies from
the ensemble energy is trivial in pure wavefunction
theory~\cite{mcscf_pinkbook}, it is still unclear how this can be
achieved rigorously and efficiently (in terms of computational cost) in the context of eDFT. 
As readily seen from Eq.~(\ref{eq:XE_lim}), 
the individual state energy $E_I$ can be obtained (in principle
exactly) from two equiensemble energies, 
\begin{eqnarray}
E_I=\dfrac{1}{\mathrm{g}_{I}}\left(M_IE^{1/M_I}_I-M_{I-1}E^{1/M_{I-1}}_{I-1}\right),
\end{eqnarray}
the two equiensembles containing up to and including the multiplets with
energies $E_I$ and $E_{I-1}$, respectively. The disadvantage of such a
formulation is that it is not straightforward to calculate the energy of
any state belonging to the ensemble of interest. Following the idea of
Levy and Zahariev~\cite{levy2014ground}, we propose to rewrite the
exact range-separated density-functional ensemble energy in
Eq.~(\ref{eq:gok_rs_ens_en}) as a weighted sum of individual long-range
interacting energies. This can be achieved by introducing a 
density-functional shift    
\begin{eqnarray}
\label{constantshift}
C^{\rm{\mu,{\bf w}}}[n] 
= \dfrac{E^{\rm{sr,\mu,{\bf w}}}_{{\rm{Hxc}}}[n]
-\int d{\bf r}~\dfrac{\delta E_{\rm Hxc}^{\rm{sr,\mu,{\bf w}}}[n]}{\delta n({\bf
r})}n({\bf r})}
{\int d{\bf r}~n({\bf r})}
\end{eqnarray}
to the ensemble short-range potential in Eq.~(\ref{eq:gok_rs_scf}), thus
leading to the "shifted" eigenvalue equation 
\begin{eqnarray}
\Bigg( &&\hat{T} + \hat{W}_{\rm{ee}}^{\rm{lr,\mu}} 
+ \hat{V}_{\rm{ne}}
 + \int d{\bf r}
\Bigg[\frac{\delta E_{\rm{Hxc}}^{\rm{sr,\mu,{\bf w}}}[n_{\hat{\Gamma}^{\rm{\mu,{\bf w}}}}]}
{\delta n({\bf r})}
\nonumber\\
&&
+C^{\rm{\mu,{\bf w}}}[n_{\hat{\Gamma}^{\rm{\mu,{\bf w}}}}]\Bigg]\hat{n}({\bf r}) \Bigg) \vert \Psi_{k}^{\rm{\mu,{\bf w}}} \rangle 
 = \overline{\mathcal{E}}_{k}^{\rm{\mu,{\bf w}}} \vert \Psi_{k}^{\rm{\mu,{\bf w}}} \rangle,
\end{eqnarray}
where 
\begin{eqnarray}
\overline{\mathcal{E}}_{k}^{\rm{\mu,{\bf w}}} =
\mathcal{E}^{\rm{\mu,{\bf w}}}_{k} 
+ 
C^{\rm{\mu,{\bf
w}}}[n_{\hat{\Gamma}^{\rm{\mu,{\bf w}}}}]\int d{\bf
r}\;n_{\Psi_{k}^{\rm{\mu,{\bf w}}}}({\bf r}) 
,
\end{eqnarray}
so that, according to Eqs.~(\ref{eq:gok_rs_ens_en}),
(\ref{eq:gok_rs_scf}) and (\ref{constantshift}), the exact range-separated ensemble energy can be rewritten as
follows, 
\begin{eqnarray}
\label{ensembleenergyshifted}
E^{\rm{\bf w}} &=&
\sum\limits_{k = 0}^{M-1} w_{k}
{\mathcal{E}}^{\rm{\mu,{\bf w}}}_{k}
+
C^{\rm{\mu,{\bf
w}}}[n_{\hat{\Gamma}^{\rm{\mu,{\bf w}}}}]\int d{\bf
r}\;n_{\hat{\Gamma}^{\rm{\mu,{\bf w}}}}({\bf r}) 
\nonumber\\
&=& 
\sum\limits_{k = 0}^{M-1} w_{k}
\overline{\mathcal{E}}^{\rm{\mu,{\bf w}}}_{k}
.
\end{eqnarray}
It then becomes
natural to interpret each (weight- and $\mu$-dependent) individual shifted energy
$\overline{\mathcal{E}}^{\rm{\mu,{\bf w}}}_{k}$ as an approximation to
the exact individual state energy $E_k$ which is actually recovered for
any ensemble weight when
$\mu\rightarrow+\infty$. If we now expand the short-range Hxc energy 
and potential around each individual state density through first order
in $\delta n^{\rm{\mu,{\bf w}}}_{k}=n_{\hat{\Gamma}^{\rm{\mu,{\bf
w}}}}-n_{\Psi^{\rm{\mu,{\bf w}}}_{k}}$ it comes 
\begin{eqnarray}\label{eq:ind_ener_exp_1st_order}
\overline{\mathcal{E}}_{k}^{\rm{\mu,{\bf w}}}&=&
\left\langle \Psi_{k}^{\rm{\mu,{\bf w}}}\middle\vert \hat{T} + \hat{W}_{\rm{ee}}^{\rm{lr,\mu}} +
\hat{V}_{\rm{ne}} \middle\vert \Psi_{k}^{\rm{\mu,{\bf w}}}\right\rangle
+E_{\rm{Hxc}}^{\rm{sr,\mu,{\bf w}}}[n_{\hat{\Gamma}^{\rm{\mu,{\bf w}}}}]
\nonumber\\
&&
 - \int d{\bf r}
\frac{\delta E_{\rm{Hxc}}^{\rm{sr,\mu,{\bf w}}}[n_{\hat{\Gamma}^{\rm{\mu,{\bf w}}}}]}
{\delta n({\bf r})}\delta n^{\rm{\mu,{\bf w}}}_{k}({\bf r})
\nonumber\\
&=&
\left\langle \Psi_{k}^{\rm{\mu,{\bf w}}}\middle\vert \hat{T} + \hat{W}_{\rm{ee}}^{\rm{lr,\mu}} +
\hat{V}_{\rm{ne}} \middle\vert \Psi_{k}^{\rm{\mu,{\bf w}}}\right\rangle
+E_{\rm{Hxc}}^{\rm{sr,\mu,{\bf w}}}[n_{\Psi_{k}^{\rm{\mu,{\bf w}}}}]
\nonumber\\
&&
+\mathcal{O}\left[\left(\delta n^{\rm{\mu,{\bf w}}}_{k}\right)^2\right],
\end{eqnarray}
where we used the fact that both ensemble and individual densities
integrate to the number of electrons, according to the normalization
condition in Eq.~(\ref{eq:normalization_weights}). Interestingly, when
the WIDFA approximation is used, the
individual state energy expression proposed
by Pastorczak~{\etal}~\cite{PRA13_Pernal_srEDFT} for computing
approximate excitation energies is recovered from Eq.~(\ref{eq:ind_ener_exp_1st_order}) through first
order in $\delta n^{\rm{\mu,{\bf w}}}_{k}$. In the latter case, the use of individual
densities automatically removes ghost interaction
errors~\cite{pernal2016ghost}. The implementation and calibration of
the first line of Eq.~(\ref{eq:ind_ener_exp_1st_order}) within WIDFA is
currently in progress and will be presented in a separate paper.

\section{Conclusion}\label{sec:conclusion}

The extrapolation technique introduced by Savin~\cite{savin2014towards}
in the context of ground-state range-separated DFT has been extended to
ghost-interaction-corrected (GIC) ensemble energies of ground and excited
states. While the standard extrapolation correction relies on a Taylor
expansion of the range-separated energy that decays as $\mu^{-2}$ in the
$\mu\rightarrow+\infty$ limit, where $\mu$ is the range-separation parameter, the GIC
ensemble energy was shown to decay more rapidly as $\mu^{-3}$, thus
requiring a different extrapolation correction. The approach has been
combined with a linear interpolation (between equiensembles) method in
order to compute excitation energies. Promising results have been
obtained for singlet excitations (including charge transfer and double
excitations) on a small test set consisting of He, H$_{2}$, HeH$^{+}$ 
and LiH. In particular, avoided crossings could be described accurately
in H$_{2}$ by setting the range-separation parameter to
$\mu=1.0a_0^{-1}$, which is a typical value in range-separated eDFT
calculations~\cite{PRA13_Pernal_srEDFT,pernal2016ghost}. Interestingly,
convergence towards the pure wavefunction theory result
($\mu\rightarrow+\infty$ limit) is essentially reached for
$\mu=1.0a_0^{-1}$ thanks to both ghost-interaction and extrapolation
corrections. As expected, the results can be further improved for
smaller $\mu$ values with higher-order extrapolation corrections. The method is currently applied to the modeling of conical
intersections, which is still challenging for TD-DFT.  
Finally, the extraction of individual state energies from
range-separated ensemble energies has been discussed as a perspective.
Approximate energies have been constructed by introducing an
ensemble-density-functional shift in the exchange-correlation potential.
We could show that, by expanding these energies around the individual
densities, the ghost-interaction-free expressions of
Pastorczak~{\etal}~\cite{PRA13_Pernal_srEDFT} are recovered through
first order. The implementation and development of this approach, for
the calculation of excited-state molecular gradients, for example, is
left for future work.

\section{Acknowledgements}

The authors acknowledge financial support from the LABEX `Chemistry of complex 
systems' and the ANR [MCFUNEX project, Grant No. ANR-14-CE06- 0014-01].

\appendix
\section{Taylor expansion of the range-separated GIC ensemble
energy for large $\mu$ values.}


Let $\eta = 1/\mu$ so that the range-separated GIC ensemble energy can
be Taylor expanded as follows for large $\mu$ values,
\begin{eqnarray}
\label{eq:gic_ensen_expansion}
\tilde{E}_{\rm{GIC}}^{1/\eta,{\bf w}}=&& E^{\bf w} 
+ \tilde{E}_{\rm{GIC}}^{(-1),{\bf w}}\eta 
+ \frac{1}{2} \tilde{E}_{\rm{GIC}}^{(-2),{\bf w}}\eta^{2} 
+ \frac{1}{6} \tilde{E}_{\rm{GIC}}^{(-3),{\bf w}}\eta^{3} 
\nonumber\\
&&+ \mathcal{O}(\eta^{4}),
\end{eqnarray}
where
\begin{eqnarray}
\begin{array}{l}\label{eq:deriv_gic_ens_en}
\begin{split}
\tilde{E}_{\rm{GIC}}^{(-1),{\bf w}} & = 
\left.\frac{\partial \tilde{E}_{\rm{GIC}}^{\rm{1/\eta},{\bf w}}}
{\partial \eta} \right\vert_{\eta = 0},\\
\tilde{E}_{\rm{GIC}}^{(-2),{\bf w}} & = 
\left.\frac{\partial^{2} \tilde{E}_{\rm{GIC}}^{\rm{1/\eta},{\bf w}}}
{\partial \eta^{2}} \right\vert_{\eta = 0}.
\end{split}
\end{array}
\end{eqnarray}
We will show that these two derivatives vanish and hence the first 
$\eta$-dependence of $\tilde{E}_{\rm{GIC}}^{1/\eta,{\bf w}}$ appears at
third order. According to Eq.~(\ref{eq:gok_widfa_gic_ens_en}), 
we have (using real algebra)
\begin{eqnarray}
\begin{array}{l}\label{eq:deriv_of_ens_en_1}
\begin{split}
\frac{\partial \tilde{E}_{\rm{GIC}}^{1/\eta,{\bf w}}}{\partial \eta}
= & \sum\limits_{k=0}^{M-1}2w_{k} \left.\left.\left\langle \frac{\partial \tilde{\Psi}_{k}^{1/\eta,{\bf w}}}{\partial\eta}
\right\vert \hat{H} \right\vert \tilde{\Psi}_{k}^{1/\eta,{\bf w}} \right\rangle \\
& + \frac{\partial E_{\rm{c,md}}^{{\rm{sr}},1/\eta}[n_{\hat{\gamma}^{1/\eta,{\bf w}}}]}{\partial\eta}, \\
\frac{\partial^{2} \tilde{E}_{\rm{GIC}}^{1/\eta,{\bf w}}}{\partial \eta^{2}}
= & \sum\limits_{k=0}^{M-1}2w_{k}\left\{\left.\left.\left\langle \frac{\partial^{2} \tilde{\Psi}_{k}^{1/\eta,{\bf w}}}{\partial\eta^{2}}
\right\vert \hat{H} \right\vert \tilde{\Psi}_{k}^{1/\eta,{\bf w}} \right\rangle\right. \\
& + \left.\left.\left.\left\langle \frac{\partial \tilde{\Psi}_{k}^{1/\eta,{\bf w}}}{\partial\eta}
\right\vert \hat{H} \right\vert \frac{\partial \tilde{\Psi}_{k}^{1/\eta,{\bf w}}}{\partial\eta} \right\rangle \right\} \\
& + \frac{\partial^{2} E_{\rm{c,md}}^{{\rm{sr}},1/\eta}[n_{\hat{\gamma}^{1/\eta,{\bf w}}}]}{\partial\eta^{2}}.
\end{split}
\end{array}
\end{eqnarray}
Since~ \cite{Toulouse2005TCA}
\begin{eqnarray}
\begin{array}{l}\label{eq:toul_c_md}
E_{\rm{c,md}}^{\rm{sr,1/\eta}}[n] = 
E_{\rm{c,md}}^{(-3)}[n]\; \eta^{3} + \mathcal{O}(\eta^{4}),
\end{array}
\end{eqnarray}
the last term on the right-hand side of both equalities in 
Eq.~(\ref{eq:deriv_of_ens_en_1}) vanishes when $\eta=0$.
Furthermore, according to Eq.~(\ref{eq:widfa_scf}),
\begin{eqnarray}
\begin{array}{l}\label{eq:deriv_of_ens_en_2}
\begin{split}
\left.\left.\left.\left\langle \frac{\partial \tilde{\Psi}_{k}^{1/\eta,{\bf w}}}{\partial\eta}
\right\vert \hat{H} \right\vert \tilde{\Psi}_{k}^{1/\eta,{\bf w}} \right\rangle \right\vert_{\eta = 0}
= E_{k}\left.\left\langle \left.\frac{\partial \tilde{\Psi}_{k}^{1/\eta,{\bf w}}}{\partial \eta} \right\vert {\Psi}_{k} \right\rangle\right\vert_{\eta=0} 
\end{split}
\end{array}
\end{eqnarray}
and
\begin{eqnarray}
\begin{array}{l}\label{eq:deriv_of_ens_en_3}
\begin{split}
\left.\left.\left.\left\langle \frac{\partial^{2} \tilde{\Psi}_{k}^{1/\eta,{\bf w}}}{\partial\eta^{2}}
\right\vert \hat{H} \right\vert \tilde{\Psi}_{k}^{1/\eta,{\bf w}} \right\rangle \right\vert_{\eta = 0}
= E_{k} \left.\left\langle \left.\frac{\partial^{2} \tilde{\Psi}_{k}^{1/\eta,{\bf w}}}{\partial\eta^{2}}\right\vert {\Psi}_{k} \right\rangle \right\vert_{\eta=0}.
\end{split}
\end{array}
\end{eqnarray}
Since the long-range-interacting wavefunction is normalized for any
value of the range-separation parameter,
\begin{eqnarray}
\begin{array}{l}\label{eq:norm_lr_wf1}
\begin{split}
& \forall \eta\hspace{0.2cm}  \left.\left\langle \tilde{\Psi}_{k}^{1/\eta,{\bf w}} \right\vert  \tilde{\Psi}_{k}^{1/\eta,{\bf w}} \right\rangle = 1,
\end{split}
\end{array}
\end{eqnarray}
it comes
\begin{eqnarray}
\begin{array}{l}\label{eq:norm_lr_wf2}
\begin{split}
\left.\left.\left\langle \frac{\partial \tilde{\Psi}_{k}^{1/\eta,{\bf w}}}{\partial \eta} \right\vert  {\Psi}_{k} \right\rangle \right\vert_{\eta=0} = 0
\end{split}
\end{array}
\end{eqnarray}
and
\begin{eqnarray}
\begin{array}{l}\label{eq:norm_lr_wf3}
\begin{split}
\left.\left.\left\langle\frac{\partial^{2} \tilde{\Psi}_{k}^{1/\eta,{\bf w}}}{\partial \eta^{2}} \right\vert {\Psi}_{k} \right\rangle\right\vert_{\eta=0}
 = -\left.\left\langle \left.\frac{\partial \tilde{\Psi}_{k}^{1/\eta,{\bf w}}}{\partial \eta} \right\vert \frac{\partial \tilde{\Psi}_{k}^{1/\eta,{\bf w}}}{\partial \eta} \right\rangle \right\vert_{\eta=0}
.
\end{split}
\end{array}
\end{eqnarray}
Combining Eqs.~(\ref{eq:deriv_gic_ens_en}),
(\ref{eq:deriv_of_ens_en_1}),~(\ref{eq:deriv_of_ens_en_2}),~(\ref{eq:deriv_of_ens_en_3}),~(\ref{eq:norm_lr_wf2})
and~(\ref{eq:norm_lr_wf3}) leads to
\begin{eqnarray}
\begin{array}{l}\label{eq:deriv_of_ens_en_41}
\tilde{E}_{\rm{GIC}}^{(-1),{\bf w}}= 0
\end{array}
\end{eqnarray}
and
\begin{eqnarray}
\label{eq:deriv_of_ens_en_42}
&&\tilde{E}_{\rm{GIC}}^{(-2),{\bf w}}
=
 \sum\limits_{k=0}^{M-1} 2w_{k} \left.\left.\left\langle 
\left.\frac{\partial \tilde{\Psi}_{k}^{1/\eta,{\bf w}}}{\partial \eta} \right\vert \hat{H} - E_{k} \right\vert 
 \frac{\partial \tilde{\Psi}_{k}^{1/\eta,{\bf w}}}{\partial \eta} \right\rangle\right\vert_{\eta=0}. 
\end{eqnarray}
By applying first-order perturbation theory to Eq.~(\ref{eq:widfa_scf})
we obtain 
\begin{multline}\label{eq:first_derivs_lr_wf}
\displaystyle{
\dfrac{\partial \left\vert \tilde{\Psi}_{k}^{1/\eta,{\bf w}}\right\rangle}{\partial\eta} =
\sum^{+\infty}_{l\neq k} \dfrac{\left\langle \tilde{\Psi}_{l}^{1/\eta,{\bf w}} 
\vert \hat{\mathcal{W}}^{1/\eta}\vert
\tilde{\Psi}_{k}^{1/\eta,{\bf w}}\right\rangle}
{\tilde{\mathcal{E}}_{k}^{1/\eta,{\bf w}} - \tilde{\mathcal{E}}_{l}^{1/\eta,{\bf w}}} 
\left\vert \tilde{\Psi}_{l}^{1/\eta,{\bf w}} \right\rangle,
}
\end{multline}
where the perturbation operator reads
\begin{eqnarray}
\label{eq:pert_op}
\hspace{-0.7cm}\hat{\mathcal{W}}^{1/\eta} = \frac{\partial \hat{W}_{\rm{ee}}^{\rm{lr,1/\eta}}}
{\partial \eta} 
 + 
 \int d{\bf r}\frac{\partial}{\partial\eta}\frac{\delta E_{\rm{Hxc}}^{\rm{sr,1/\eta}}
[n_{\hat{\gamma}^{1/\eta,{\bf w}}}]}
{\delta n({\bf r})}\hat{n}({\bf r}) 
.
\end{eqnarray}
Note that the second term on the right-hand side of Eq.~(\ref{eq:pert_op}) 
vanishes when $\eta=0$ since~\cite{srDFT,paolacorr}
\begin{eqnarray}
\begin{array}{l}\label{eq:hxc_eta3}
E_{\rm{Hxc}}^{\rm{sr},1/\eta}[n] = E_{\rm{Hxc}}^{(-2)}[n]\,\eta^{2} 
+ \mathcal{O}(\eta^{3}),
\end{array}
\end{eqnarray}
thus leading to
\begin{eqnarray}
\begin{array}{l}\label{eq:deriv_of_ens_en_5}
\begin{split}
\displaystyle
& \left.\left.\left\langle 
\left.\frac{\partial \tilde{\Psi}_{k}^{1/\eta,{\bf w}}}{\partial \eta}
 \right\vert \hat{H} - E_{k} \right\vert 
 \frac{\partial \tilde{\Psi}_{k}^{1/\eta,{\bf w}}}{\partial \eta}
 \right\rangle \right\vert_{\eta=0} \\
& = - \sum^{+\infty}\limits_{l\neq k} \frac{1}{E_{k} - E_{l}}
\left.\left.\left. \left\vert \left\langle \Psi_{l} \left\vert 
\frac{\partial \hat{W}_{\rm{ee}}^{\rm{lr,1/\eta}}}{\partial\eta}
\right.\right\vert \Psi_{k} \right.\right\rangle \right\vert^{2} 
\right\vert_{\eta=0}
.
 \end{split}
\end{array}
\end{eqnarray}
Finally, using~\cite{extrapol_edft}
\begin{eqnarray}
\begin{array}{l}\label{eq:lr_intracule_2}
\displaystyle
\frac{\partial \hat{W}^{\rm{lr,1/\eta}}_{\rm{ee}}}{\partial\eta} 
=  -8\sqrt{\pi} \int^{+\infty}_0 d{{{r}}_{12}} \frac{r_{12}^{2}}{\eta^{2}} 
e^{-\frac{r_{12}^{2}}{\eta^{2}}}\hat{f}({{r}}_{12}),
\end{array}
\end{eqnarray}
where $r_{12} = \vert {\bf r}_{1} - {\bf r}_{2} \vert$ is the 
interelectronic distance and $\hat{f}({{r}}_{12})$ is the 
intracule density operator, we obtain 
\begin{eqnarray}
\label{eq:deriv_of_ens_en_6}
\left.\left. \left\langle \Psi_{l} \left\vert \frac{\partial \hat{W}_{\rm{ee}}^{\rm{lr,1/\eta}}}{\partial\eta} \right.\right\vert \Psi_{k} \right.\right\rangle = 
 -8\sqrt{\pi}&& \int^{+\infty}_0d{{{r}}_{12}}
\frac{r_{12}^{2}}{\eta^{2}} 
\text{exp}\left({-\frac{r_{12}^{2}}{\eta^{2}}}\right) 
\nonumber\\
&&\times\left\langle \Psi_{l} \left\vert \hat{f}({{r}}_{12})\right\vert \Psi_{k} \right\rangle 
.
\end{eqnarray}
Since $\frac{r_{12}^{2}}{\eta^{2}} 
\text{exp}\left({-\frac{r_{12}^{2}}{\eta^{2}}}\right) \rightarrow 0$
when $\eta \rightarrow 0$ for {\it all} values of $r_{12}$, we conclude
from Eqs.~(\ref{eq:deriv_of_ens_en_42}) and (\ref{eq:deriv_of_ens_en_5}) that 
\begin{eqnarray}
\begin{array}{l}
\tilde{E}_{\rm{GIC}}^{(-2),{\bf w}}= 0,
\end{array}
\end{eqnarray}
so that Eq.~(\ref{eq:gic_ensen_expansion}) can be simplified as follows,
\begin{eqnarray}
\begin{array}{l}\label{eq:gic_ensen_expansion5}
\tilde{E}_{\rm{GIC}}^{1/\eta,{\bf w}} =
 E^{\bf w} + \dfrac{1}{6} \tilde{E}_{\rm{GIC}}^{(-3),{\bf w}}\eta^{3} 
+ \mathcal{O}(\eta^{4}),
\end{array}
\end{eqnarray}
thus leading to the expansion in Eq.~(\ref{eq:gic_exp_largemu}). 
 


\begin{thebibliography}{64}%
\makeatletter
\providecommand \@ifxundefined [1]{%
 \@ifx{#1\undefined}
}%
\providecommand \@ifnum [1]{%
 \ifnum #1\expandafter \@firstoftwo
 \else \expandafter \@secondoftwo
 \fi
}%
\providecommand \@ifx [1]{%
 \ifx #1\expandafter \@firstoftwo
 \else \expandafter \@secondoftwo
 \fi
}%
\providecommand \natexlab [1]{#1}%
\providecommand \enquote  [1]{``#1''}%
\providecommand \bibnamefont  [1]{#1}%
\providecommand \bibfnamefont [1]{#1}%
\providecommand \citenamefont [1]{#1}%
\providecommand \href@noop [0]{\@secondoftwo}%
\providecommand \href [0]{\begingroup \@sanitize@url \@href}%
\providecommand \@href[1]{\@@startlink{#1}\@@href}%
\providecommand \@@href[1]{\endgroup#1\@@endlink}%
\providecommand \@sanitize@url [0]{\catcode `\\12\catcode `\$12\catcode
  `\&12\catcode `\#12\catcode `\^12\catcode `\_12\catcode `\%12\relax}%
\providecommand \@@startlink[1]{}%
\providecommand \@@endlink[0]{}%
\providecommand \url  [0]{\begingroup\@sanitize@url \@url }%
\providecommand \@url [1]{\endgroup\@href {#1}{\urlprefix }}%
\providecommand \urlprefix  [0]{URL }%
\providecommand \Eprint [0]{\href }%
\providecommand \doibase [0]{http://dx.doi.org/}%
\providecommand \selectlanguage [0]{\@gobble}%
\providecommand \bibinfo  [0]{\@secondoftwo}%
\providecommand \bibfield  [0]{\@secondoftwo}%
\providecommand \translation [1]{[#1]}%
\providecommand \BibitemOpen [0]{}%
\providecommand \bibitemStop [0]{}%
\providecommand \bibitemNoStop [0]{.\EOS\space}%
\providecommand \EOS [0]{\spacefactor3000\relax}%
\providecommand \BibitemShut  [1]{\csname bibitem#1\endcsname}%
\let\auto@bib@innerbib\@empty
\bibitem [{\citenamefont {Hohenberg}\ and\ \citenamefont
  {Kohn}(1964)}]{hktheo}%
  \BibitemOpen
  \bibfield  {author} {\bibinfo {author} {\bibfnamefont {P.}~\bibnamefont
  {Hohenberg}}\ and\ \bibinfo {author} {\bibfnamefont {W.}~\bibnamefont
  {Kohn}},\ }\href@noop {} {\bibfield  {journal} {\bibinfo  {journal} {Phys.
  Rev.}\ }\textbf {\bibinfo {volume} {136}},\ \bibinfo {pages} {B864} (\bibinfo
  {year} {1964})}\BibitemShut {NoStop}%
\bibitem [{\citenamefont {Kohn}\ and\ \citenamefont {Sham}(1965)}]{KS}%
  \BibitemOpen
  \bibfield  {author} {\bibinfo {author} {\bibfnamefont {W.}~\bibnamefont
  {Kohn}}\ and\ \bibinfo {author} {\bibfnamefont {L.}~\bibnamefont {Sham}},\
  }\href {\doibase 10.1103/PhysRev.140.A1133} {\bibfield  {journal} {\bibinfo
  {journal} {Phys. Rev.}\ }\textbf {\bibinfo {volume} {140}},\ \bibinfo {pages}
  {A1133} (\bibinfo {year} {1965})}\BibitemShut {NoStop}%
\bibitem [{\citenamefont {Runge}\ and\ \citenamefont
  {Gross}(1984)}]{TDDFT_Gross}%
  \BibitemOpen
  \bibfield  {author} {\bibinfo {author} {\bibfnamefont {E.}~\bibnamefont
  {Runge}}\ and\ \bibinfo {author} {\bibfnamefont {E.~K.~U.}\ \bibnamefont
  {Gross}},\ }\href {\doibase 10.1103/PhysRevLett.52.997} {\bibfield  {journal}
  {\bibinfo  {journal} {Phys. Rev. Lett.}\ }\textbf {\bibinfo {volume} {52}},\
  \bibinfo {pages} {997} (\bibinfo {year} {1984})}\BibitemShut {NoStop}%
\bibitem [{\citenamefont {Casida}\ and\ \citenamefont
  {Huix-Rotllant}(2012)}]{Casida_tddft_review_2012}%
  \BibitemOpen
  \bibfield  {author} {\bibinfo {author} {\bibfnamefont {M.}~\bibnamefont
  {Casida}}\ and\ \bibinfo {author} {\bibfnamefont {M.}~\bibnamefont
  {Huix-Rotllant}},\ }\href@noop {} {\bibfield  {journal} {\bibinfo  {journal}
  {Annu. Rev. Phys. Chem.}\ }\textbf {\bibinfo {volume} {63}},\ \bibinfo
  {pages} {287} (\bibinfo {year} {2012})}\BibitemShut {NoStop}%
\bibitem [{\citenamefont {Casida}(1995)}]{Casida_eqs}%
  \BibitemOpen
  \bibfield  {author} {\bibinfo {author} {\bibfnamefont {M.}~\bibnamefont
  {Casida}},\ }in\ \href@noop {} {\emph {\bibinfo {booktitle} {Recent Advances
  in Density Functional Methods}}},\ \bibinfo {editor} {edited by\ \bibinfo
  {editor} {\bibfnamefont {D.~P.}\ \bibnamefont {Chong}}}\ (\bibinfo
  {publisher} {World Scientific},\ \bibinfo {address} {Singapore},\ \bibinfo
  {year} {1995})\BibitemShut {NoStop}%
\bibitem [{\citenamefont {Marques}\ and\ \citenamefont
  {Gross}(2004)}]{marques2004time}%
  \BibitemOpen
  \bibfield  {author} {\bibinfo {author} {\bibfnamefont {M.}~\bibnamefont
  {Marques}}\ and\ \bibinfo {author} {\bibfnamefont {E.}~\bibnamefont
  {Gross}},\ }\href@noop {} {\bibfield  {journal} {\bibinfo  {journal} {Annu.
  Rev. Phys. Chem.}\ }\textbf {\bibinfo {volume} {55}},\ \bibinfo {pages} {427}
  (\bibinfo {year} {2004})}\BibitemShut {NoStop}%
\bibitem [{\citenamefont {Gunnarsson}\ and\ \citenamefont
  {Lundqvist}(1976)}]{GUNNARSSON:1976p1781}%
  \BibitemOpen
  \bibfield  {author} {\bibinfo {author} {\bibfnamefont {O.}~\bibnamefont
  {Gunnarsson}}\ and\ \bibinfo {author} {\bibfnamefont {B.~I.}\ \bibnamefont
  {Lundqvist}},\ }\href@noop {} {\bibfield  {journal} {\bibinfo  {journal}
  {Phys. Rev. B}\ }\textbf {\bibinfo {volume} {13}},\ \bibinfo {pages} {4274}
  (\bibinfo {year} {1976})}\BibitemShut {NoStop}%
\bibitem [{\citenamefont {Dederichs}\ \emph {et~al.}(1984)\citenamefont
  {Dederichs}, \citenamefont {Bl\"ugel}, \citenamefont {Zeller},\ and\
  \citenamefont {Akai}}]{DederichsPRL1984}%
  \BibitemOpen
  \bibfield  {author} {\bibinfo {author} {\bibfnamefont {P.~H.}\ \bibnamefont
  {Dederichs}}, \bibinfo {author} {\bibfnamefont {S.}~\bibnamefont {Bl\"ugel}},
  \bibinfo {author} {\bibfnamefont {R.}~\bibnamefont {Zeller}}, \ and\ \bibinfo
  {author} {\bibfnamefont {H.}~\bibnamefont {Akai}},\ }\href {\doibase
  10.1103/PhysRevLett.53.2512} {\bibfield  {journal} {\bibinfo  {journal}
  {Phys. Rev. Lett.}\ }\textbf {\bibinfo {volume} {53}},\ \bibinfo {pages}
  {2512} (\bibinfo {year} {1984})}\BibitemShut {NoStop}%
\bibitem [{\citenamefont {Ziegler}, \citenamefont {Rauk},\ and\ \citenamefont
  {Baerends}(1977)}]{Ziegler1977}%
  \BibitemOpen
  \bibfield  {author} {\bibinfo {author} {\bibfnamefont {T.}~\bibnamefont
  {Ziegler}}, \bibinfo {author} {\bibfnamefont {A.}~\bibnamefont {Rauk}}, \
  and\ \bibinfo {author} {\bibfnamefont {E.~J.}\ \bibnamefont {Baerends}},\
  }\href@noop {} {\bibfield  {journal} {\bibinfo  {journal} {Theor. Chim.
  Acta}\ }\textbf {\bibinfo {volume} {43}},\ \bibinfo {pages} {261} (\bibinfo
  {year} {1977})}\BibitemShut {NoStop}%
\bibitem [{\citenamefont {von Barth}(1979)}]{BarthPRA1979}%
  \BibitemOpen
  \bibfield  {author} {\bibinfo {author} {\bibfnamefont {U.}~\bibnamefont {von
  Barth}},\ }\href {\doibase 10.1103/PhysRevA.20.1693} {\bibfield  {journal}
  {\bibinfo  {journal} {Phys. Rev. A}\ }\textbf {\bibinfo {volume} {20}},\
  \bibinfo {pages} {1693} (\bibinfo {year} {1979})}\BibitemShut {NoStop}%
\bibitem [{\citenamefont {Theophilou}(1979)}]{JPC79_Theophilou_equi-ensembles}%
  \BibitemOpen
  \bibfield  {author} {\bibinfo {author} {\bibfnamefont {A.~K.}\ \bibnamefont
  {Theophilou}},\ }\href@noop {} {\bibfield  {journal} {\bibinfo  {journal} {J.
  Phys. C (Solid State Phys.)}\ }\textbf {\bibinfo {volume} {12}},\ \bibinfo
  {pages} {5419} (\bibinfo {year} {1979})}\BibitemShut {NoStop}%
\bibitem [{\citenamefont {Gross}, \citenamefont {Oliveira},\ and\ \citenamefont
  {Kohn}(1988{\natexlab{a}})}]{PRA_GOK_RRprinc}%
  \BibitemOpen
  \bibfield  {author} {\bibinfo {author} {\bibfnamefont {E.~K.~U.}\
  \bibnamefont {Gross}}, \bibinfo {author} {\bibfnamefont {L.~N.}\ \bibnamefont
  {Oliveira}}, \ and\ \bibinfo {author} {\bibfnamefont {W.}~\bibnamefont
  {Kohn}},\ }\href {\doibase 10.1103/PhysRevA.37.2805} {\bibfield  {journal}
  {\bibinfo  {journal} {Phys. Rev. A}\ }\textbf {\bibinfo {volume} {37}},\
  \bibinfo {pages} {2805} (\bibinfo {year} {1988}{\natexlab{a}})}\BibitemShut
  {NoStop}%
\bibitem [{\citenamefont {Gross}, \citenamefont {Oliveira},\ and\ \citenamefont
  {Kohn}(1988{\natexlab{b}})}]{PRA_GOK_EKSDFT}%
  \BibitemOpen
  \bibfield  {author} {\bibinfo {author} {\bibfnamefont {E.~K.~U.}\
  \bibnamefont {Gross}}, \bibinfo {author} {\bibfnamefont {L.~N.}\ \bibnamefont
  {Oliveira}}, \ and\ \bibinfo {author} {\bibfnamefont {W.}~\bibnamefont
  {Kohn}},\ }\href {\doibase 10.1103/PhysRevA.37.2809} {\bibfield  {journal}
  {\bibinfo  {journal} {Phys. Rev. A}\ }\textbf {\bibinfo {volume} {37}},\
  \bibinfo {pages} {2809} (\bibinfo {year} {1988}{\natexlab{b}})}\BibitemShut
  {NoStop}%
\bibitem [{\citenamefont {Gross}, \citenamefont {Oliveira},\ and\ \citenamefont
  {Kohn}(1988{\natexlab{c}})}]{GOK3}%
  \BibitemOpen
  \bibfield  {author} {\bibinfo {author} {\bibfnamefont {E.~K.~U.}\
  \bibnamefont {Gross}}, \bibinfo {author} {\bibfnamefont {L.~N.}\ \bibnamefont
  {Oliveira}}, \ and\ \bibinfo {author} {\bibfnamefont {W.}~\bibnamefont
  {Kohn}},\ }\href {\doibase 10.1103/PhysRevA.37.2821} {\bibfield  {journal}
  {\bibinfo  {journal} {Phys. Rev. A}\ }\textbf {\bibinfo {volume} {37}},\
  \bibinfo {pages} {2821} (\bibinfo {year} {1988}{\natexlab{c}})}\BibitemShut
  {NoStop}%
\bibitem [{\citenamefont {Nagy}(1996)}]{Nagy_enseXpot}%
  \BibitemOpen
  \bibfield  {author} {\bibinfo {author} {\bibfnamefont {{\'A}.}~\bibnamefont
  {Nagy}},\ }\href@noop {} {\bibfield  {journal} {\bibinfo  {journal} {J. Phys.
  B: At. Mol. Opt. Phys.}\ }\textbf {\bibinfo {volume} {29}},\ \bibinfo {pages}
  {389} (\bibinfo {year} {1996})}\BibitemShut {NoStop}%
\bibitem [{\citenamefont {Tasn\'{a}di}\ and\ \citenamefont
  {Nagy}(2003)}]{TasnadiJPB2003}%
  \BibitemOpen
  \bibfield  {author} {\bibinfo {author} {\bibfnamefont {F.}~\bibnamefont
  {Tasn\'{a}di}}\ and\ \bibinfo {author} {\bibfnamefont {{\'A}.}~\bibnamefont
  {Nagy}},\ }\href {http://stacks.iop.org/0953-4075/36/i=20/a=002} {\bibfield
  {journal} {\bibinfo  {journal} {Journal of Physics B: Atomic, Molecular and
  Optical Physics}\ }\textbf {\bibinfo {volume} {36}},\ \bibinfo {pages} {4073}
  (\bibinfo {year} {2003})}\BibitemShut {NoStop}%
\bibitem [{\citenamefont {Gritsenko}\ \emph {et~al.}(2000)\citenamefont
  {Gritsenko}, \citenamefont {Van~Gisbergen}, \citenamefont {Gorling},\ and\
  \citenamefont {Baerends}}]{gritsenko2000excitation}%
  \BibitemOpen
  \bibfield  {author} {\bibinfo {author} {\bibfnamefont {O.}~\bibnamefont
  {Gritsenko}}, \bibinfo {author} {\bibfnamefont {S.}~\bibnamefont
  {Van~Gisbergen}}, \bibinfo {author} {\bibfnamefont {A.}~\bibnamefont
  {Gorling}}, \ and\ \bibinfo {author} {\bibfnamefont {E.}~\bibnamefont
  {Baerends}},\ }\href@noop {} {\bibfield  {journal} {\bibinfo  {journal} {J.
  Chem. Phys.}\ }\textbf {\bibinfo {volume} {113}},\ \bibinfo {pages} {8478}
  (\bibinfo {year} {2000})}\BibitemShut {NoStop}%
\bibitem [{\citenamefont {Casida}\ \emph {et~al.}(1998)\citenamefont {Casida},
  \citenamefont {Jamorski}, \citenamefont {Casida},\ and\ \citenamefont
  {Salahub}}]{CasidaJCP1998chargetransfer}%
  \BibitemOpen
  \bibfield  {author} {\bibinfo {author} {\bibfnamefont {M.~E.}\ \bibnamefont
  {Casida}}, \bibinfo {author} {\bibfnamefont {C.}~\bibnamefont {Jamorski}},
  \bibinfo {author} {\bibfnamefont {K.~C.}\ \bibnamefont {Casida}}, \ and\
  \bibinfo {author} {\bibfnamefont {D.~R.}\ \bibnamefont {Salahub}},\ }\href
  {\doibase http://dx.doi.org/10.1063/1.475855} {\bibfield  {journal} {\bibinfo
   {journal} {J. Chem. Phys.}\ }\textbf {\bibinfo {volume} {108}},\ \bibinfo
  {pages} {4439} (\bibinfo {year} {1998})}\BibitemShut {NoStop}%
\bibitem [{\citenamefont {Dreuw}, \citenamefont {Weisman},\ and\ \citenamefont
  {Head-Gordon}(2003)}]{DreuwJCP2003chargetransfer}%
  \BibitemOpen
  \bibfield  {author} {\bibinfo {author} {\bibfnamefont {A.}~\bibnamefont
  {Dreuw}}, \bibinfo {author} {\bibfnamefont {J.~L.}\ \bibnamefont {Weisman}},
  \ and\ \bibinfo {author} {\bibfnamefont {M.}~\bibnamefont {Head-Gordon}},\
  }\href {\doibase http://dx.doi.org/10.1063/1.1590951} {\bibfield  {journal}
  {\bibinfo  {journal} {J. Chem. Phys.}\ }\textbf {\bibinfo {volume} {119}},\
  \bibinfo {pages} {2943} (\bibinfo {year} {2003})}\BibitemShut {NoStop}%
\bibitem [{\citenamefont {Maitra}\ \emph {et~al.}(2004)\citenamefont {Maitra},
  \citenamefont {Zhang}, \citenamefont {Cave},\ and\ \citenamefont
  {Burke}}]{maitra2004double}%
  \BibitemOpen
  \bibfield  {author} {\bibinfo {author} {\bibfnamefont {N.~T.}\ \bibnamefont
  {Maitra}}, \bibinfo {author} {\bibfnamefont {F.}~\bibnamefont {Zhang}},
  \bibinfo {author} {\bibfnamefont {R.~J.}\ \bibnamefont {Cave}}, \ and\
  \bibinfo {author} {\bibfnamefont {K.}~\bibnamefont {Burke}},\ }\href@noop {}
  {\bibfield  {journal} {\bibinfo  {journal} {J. Chem. Phys.}\ }\textbf
  {\bibinfo {volume} {120}},\ \bibinfo {pages} {5932} (\bibinfo {year}
  {2004})}\BibitemShut {NoStop}%
\bibitem [{\citenamefont {Savin}(1988)}]{savin1988combined}%
  \BibitemOpen
  \bibfield  {author} {\bibinfo {author} {\bibfnamefont {A.}~\bibnamefont
  {Savin}},\ }\href@noop {} {\bibfield  {journal} {\bibinfo  {journal} {Int. J.
  Quantum Chem.}\ }\textbf {\bibinfo {volume} {34}},\ \bibinfo {pages} {59}
  (\bibinfo {year} {1988})}\BibitemShut {NoStop}%
\bibitem [{\citenamefont {Stoll}\ and\ \citenamefont
  {Savin}(1985)}]{savinstoll}%
  \BibitemOpen
  \bibfield  {author} {\bibinfo {author} {\bibfnamefont {H.}~\bibnamefont
  {Stoll}}\ and\ \bibinfo {author} {\bibfnamefont {A.}~\bibnamefont {Savin}},\
  }in\ \href@noop {} {\emph {\bibinfo {booktitle} {Density Functional Methods
  in Physics}}},\ \bibinfo {editor} {edited by\ \bibinfo {editor}
  {\bibfnamefont {R.~M.}\ \bibnamefont {Dreizler}}\ and\ \bibinfo {editor}
  {\bibfnamefont {J.}~\bibnamefont {da~Providencia}}}\ (\bibinfo  {publisher}
  {Plenum},\ \bibinfo {address} {New York},\ \bibinfo {year}
  {1985})\BibitemShut {NoStop}%
\bibitem [{\citenamefont {Savin}(1996)}]{savinbook}%
  \BibitemOpen
  \bibfield  {author} {\bibinfo {author} {\bibfnamefont {A.}~\bibnamefont
  {Savin}},\ }\href@noop {} {\emph {\bibinfo {title} {Recent Developments and
  Applications of Modern Density Functional Theory}}}\ (\bibinfo  {publisher}
  {Elsevier},\ \bibinfo {address} {Amsterdam},\ \bibinfo {year} {1996})\ p.\
  \bibinfo {pages} {327}\BibitemShut {NoStop}%
\bibitem [{\citenamefont {Huix-Rotllant}\ \emph {et~al.}(2011)\citenamefont
  {Huix-Rotllant}, \citenamefont {Ipatov}, \citenamefont {Rubio},\ and\
  \citenamefont {Casida}}]{HuixRotllantCP2011hybridkernel}%
  \BibitemOpen
  \bibfield  {author} {\bibinfo {author} {\bibfnamefont {M.}~\bibnamefont
  {Huix-Rotllant}}, \bibinfo {author} {\bibfnamefont {A.}~\bibnamefont
  {Ipatov}}, \bibinfo {author} {\bibfnamefont {A.}~\bibnamefont {Rubio}}, \
  and\ \bibinfo {author} {\bibfnamefont {M.~E.}\ \bibnamefont {Casida}},\
  }\href {\doibase http://dx.doi.org/10.1016/j.chemphys.2011.03.019} {\bibfield
   {journal} {\bibinfo  {journal} {Chem. Phys.}\ }\textbf {\bibinfo {volume}
  {391}},\ \bibinfo {pages} {120 } (\bibinfo {year} {2011})}\BibitemShut
  {NoStop}%
\bibitem [{\citenamefont {Filatov}, \citenamefont {Huix-Rotllant},\ and\
  \citenamefont {Burghardt}(2015)}]{filatov2015ensemble}%
  \BibitemOpen
  \bibfield  {author} {\bibinfo {author} {\bibfnamefont {M.}~\bibnamefont
  {Filatov}}, \bibinfo {author} {\bibfnamefont {M.}~\bibnamefont
  {Huix-Rotllant}}, \ and\ \bibinfo {author} {\bibfnamefont {I.}~\bibnamefont
  {Burghardt}},\ }\href@noop {} {\bibfield  {journal} {\bibinfo  {journal} {J.
  Chem. Phys.}\ }\textbf {\bibinfo {volume} {142}},\ \bibinfo {pages} {184104}
  (\bibinfo {year} {2015})}\BibitemShut {NoStop}%
\bibitem [{\citenamefont {Krykunov}\ and\ \citenamefont
  {Ziegler}(2013)}]{krykunov2013jctc}%
  \BibitemOpen
  \bibfield  {author} {\bibinfo {author} {\bibfnamefont {M.}~\bibnamefont
  {Krykunov}}\ and\ \bibinfo {author} {\bibfnamefont {T.}~\bibnamefont
  {Ziegler}},\ }\href {\doibase 10.1021/ct300891k} {\bibfield  {journal}
  {\bibinfo  {journal} {Journal of Chemical Theory and Computation}\ }\textbf
  {\bibinfo {volume} {9}},\ \bibinfo {pages} {2761} (\bibinfo {year} {2013})},\
  \bibinfo {note} {pMID: 26583867},\ \Eprint
  {http://arxiv.org/abs/http://dx.doi.org/10.1021/ct300891k}
  {http://dx.doi.org/10.1021/ct300891k} \BibitemShut {NoStop}%
\bibitem [{\citenamefont {Ziegler}\ \emph {et~al.}(2009)\citenamefont
  {Ziegler}, \citenamefont {Seth}, \citenamefont {Krykunov}, \citenamefont
  {Autschbach},\ and\ \citenamefont {Wang}}]{ziegler2009jcp}%
  \BibitemOpen
  \bibfield  {author} {\bibinfo {author} {\bibfnamefont {T.}~\bibnamefont
  {Ziegler}}, \bibinfo {author} {\bibfnamefont {M.}~\bibnamefont {Seth}},
  \bibinfo {author} {\bibfnamefont {M.}~\bibnamefont {Krykunov}}, \bibinfo
  {author} {\bibfnamefont {J.}~\bibnamefont {Autschbach}}, \ and\ \bibinfo
  {author} {\bibfnamefont {F.}~\bibnamefont {Wang}},\ }\href {\doibase
  10.1063/1.3114988} {\bibfield  {journal} {\bibinfo  {journal} {The Journal of
  Chemical Physics}\ }\textbf {\bibinfo {volume} {130}},\ \bibinfo {pages}
  {154102} (\bibinfo {year} {2009})},\ \Eprint
  {http://arxiv.org/abs/http://dx.doi.org/10.1063/1.3114988}
  {http://dx.doi.org/10.1063/1.3114988} \BibitemShut {NoStop}%
\bibitem [{\citenamefont {Glushkov}\ and\ \citenamefont
  {Levy}(2016)}]{levy2016computation}%
  \BibitemOpen
  \bibfield  {author} {\bibinfo {author} {\bibfnamefont {V.}~\bibnamefont
  {Glushkov}}\ and\ \bibinfo {author} {\bibfnamefont {M.}~\bibnamefont
  {Levy}},\ }\href {\doibase 10.3390/computation4030028} {\bibfield  {journal}
  {\bibinfo  {journal} {Computation}\ }\textbf {\bibinfo {volume} {4}}
  (\bibinfo {year} {2016}),\ 10.3390/computation4030028}\BibitemShut {NoStop}%
\bibitem [{\citenamefont {Ayers}\ and\ \citenamefont
  {Levy}(2009)}]{ayers2009pra}%
  \BibitemOpen
  \bibfield  {author} {\bibinfo {author} {\bibfnamefont {P.~W.}\ \bibnamefont
  {Ayers}}\ and\ \bibinfo {author} {\bibfnamefont {M.}~\bibnamefont {Levy}},\
  }\href {\doibase 10.1103/PhysRevA.80.012508} {\bibfield  {journal} {\bibinfo
  {journal} {Phys. Rev. A}\ }\textbf {\bibinfo {volume} {80}},\ \bibinfo
  {pages} {012508} (\bibinfo {year} {2009})}\BibitemShut {NoStop}%
\bibitem [{\citenamefont {Yang}\ \emph {et~al.}(2017)\citenamefont {Yang},
  \citenamefont {Pribram-Jones}, \citenamefont {Burke},\ and\ \citenamefont
  {Ullrich}}]{yang2017prl}%
  \BibitemOpen
  \bibfield  {author} {\bibinfo {author} {\bibfnamefont {Z.-h.}\ \bibnamefont
  {Yang}}, \bibinfo {author} {\bibfnamefont {A.}~\bibnamefont {Pribram-Jones}},
  \bibinfo {author} {\bibfnamefont {K.}~\bibnamefont {Burke}}, \ and\ \bibinfo
  {author} {\bibfnamefont {C.~A.}\ \bibnamefont {Ullrich}},\ }\href {\doibase
  10.1103/PhysRevLett.119.033003} {\bibfield  {journal} {\bibinfo  {journal}
  {Phys. Rev. Lett.}\ }\textbf {\bibinfo {volume} {119}},\ \bibinfo {pages}
  {033003} (\bibinfo {year} {2017})}\BibitemShut {NoStop}%
\bibitem [{\citenamefont {Pastorczak}, \citenamefont {Gidopoulos},\ and\
  \citenamefont {Pernal}(2013)}]{PRA13_Pernal_srEDFT}%
  \BibitemOpen
  \bibfield  {author} {\bibinfo {author} {\bibfnamefont {E.}~\bibnamefont
  {Pastorczak}}, \bibinfo {author} {\bibfnamefont {N.~I.}\ \bibnamefont
  {Gidopoulos}}, \ and\ \bibinfo {author} {\bibfnamefont {K.}~\bibnamefont
  {Pernal}},\ }\href@noop {} {\bibfield  {journal} {\bibinfo  {journal} {Phys.
  Rev. A}\ }\textbf {\bibinfo {volume} {87}},\ \bibinfo {pages} {062501}
  (\bibinfo {year} {2013})}\BibitemShut {NoStop}%
\bibitem [{\citenamefont {Franck}\ and\ \citenamefont
  {Fromager}(2014)}]{franck2014generalised}%
  \BibitemOpen
  \bibfield  {author} {\bibinfo {author} {\bibfnamefont {O.}~\bibnamefont
  {Franck}}\ and\ \bibinfo {author} {\bibfnamefont {E.}~\bibnamefont
  {Fromager}},\ }\href@noop {} {\bibfield  {journal} {\bibinfo  {journal} {Mol.
  Phys.}\ }\textbf {\bibinfo {volume} {112}},\ \bibinfo {pages} {1684}
  (\bibinfo {year} {2014})}\BibitemShut {NoStop}%
\bibitem [{\citenamefont {Deur}, \citenamefont {Mazouin},\ and\ \citenamefont
  {Fromager}(2017)}]{PRB17_Deur_eDFT_Hubbard_dimer}%
  \BibitemOpen
  \bibfield  {author} {\bibinfo {author} {\bibfnamefont {K.}~\bibnamefont
  {Deur}}, \bibinfo {author} {\bibfnamefont {L.}~\bibnamefont {Mazouin}}, \
  and\ \bibinfo {author} {\bibfnamefont {E.}~\bibnamefont {Fromager}},\ }\href
  {\doibase 10.1103/PhysRevB.95.035120} {\bibfield  {journal} {\bibinfo
  {journal} {Phys. Rev. B}\ }\textbf {\bibinfo {volume} {95}},\ \bibinfo
  {pages} {035120} (\bibinfo {year} {2017})}\BibitemShut {NoStop}%
\bibitem [{\citenamefont {Theophilou}(1987)}]{theophilou_book}%
  \BibitemOpen
  \bibfield  {author} {\bibinfo {author} {\bibfnamefont {A.~K.}\ \bibnamefont
  {Theophilou}},\ }\href@noop {} {\emph {\bibinfo {title} {The single particle
  density in physics and chemistry}}},\ edited by\ \bibinfo {editor}
  {\bibfnamefont {N.~H.}\ \bibnamefont {March}}\ and\ \bibinfo {editor}
  {\bibfnamefont {B.~M.}\ \bibnamefont {Deb}}\ (\bibinfo  {publisher} {Academic
  Press},\ \bibinfo {year} {1987})\ pp.\ \bibinfo {pages}
  {210--212}\BibitemShut {NoStop}%
\bibitem [{\citenamefont {Kohn}(1986)}]{KohnPRA1986}%
  \BibitemOpen
  \bibfield  {author} {\bibinfo {author} {\bibfnamefont {W.}~\bibnamefont
  {Kohn}},\ }\href {\doibase 10.1103/PhysRevA.34.737} {\bibfield  {journal}
  {\bibinfo  {journal} {Phys. Rev. A}\ }\textbf {\bibinfo {volume} {34}},\
  \bibinfo {pages} {737} (\bibinfo {year} {1986})}\BibitemShut {NoStop}%
\bibitem [{\citenamefont {Nagy}(1998)}]{Nagy_functional}%
  \BibitemOpen
  \bibfield  {author} {\bibinfo {author} {\bibfnamefont {A.}~\bibnamefont
  {Nagy}},\ }\href@noop {} {\bibfield  {journal} {\bibinfo  {journal} {Int. J.
  Quantum Chem.}\ }\textbf {\bibinfo {volume} {69}},\ \bibinfo {pages} {247}
  (\bibinfo {year} {1998})}\BibitemShut {NoStop}%
\bibitem [{\citenamefont {Nagy}(2001)}]{nagy2001}%
  \BibitemOpen
  \bibfield  {author} {\bibinfo {author} {\bibfnamefont {A.}~\bibnamefont
  {Nagy}},\ }\href@noop {} {\bibfield  {journal} {\bibinfo  {journal} {J. Phys.
  B: At. Mol. Opt. Phys.}\ }\textbf {\bibinfo {volume} {34}},\ \bibinfo {pages}
  {2363} (\bibinfo {year} {2001})}\BibitemShut {NoStop}%
\bibitem [{\citenamefont {Yang}\ \emph {et~al.}(2014)\citenamefont {Yang},
  \citenamefont {Trail}, \citenamefont {Pribram-Jones}, \citenamefont {Burke},
  \citenamefont {Needs},\ and\ \citenamefont {Ullrich}}]{yang2014exact}%
  \BibitemOpen
  \bibfield  {author} {\bibinfo {author} {\bibfnamefont {Z.-h.}\ \bibnamefont
  {Yang}}, \bibinfo {author} {\bibfnamefont {J.~R.}\ \bibnamefont {Trail}},
  \bibinfo {author} {\bibfnamefont {A.}~\bibnamefont {Pribram-Jones}}, \bibinfo
  {author} {\bibfnamefont {K.}~\bibnamefont {Burke}}, \bibinfo {author}
  {\bibfnamefont {R.~J.}\ \bibnamefont {Needs}}, \ and\ \bibinfo {author}
  {\bibfnamefont {C.~A.}\ \bibnamefont {Ullrich}},\ }\href@noop {} {\bibfield
  {journal} {\bibinfo  {journal} {Phys. Rev. A}\ }\textbf {\bibinfo {volume}
  {90}},\ \bibinfo {pages} {042501} (\bibinfo {year} {2014})}\BibitemShut
  {NoStop}%
\bibitem [{\citenamefont {Pribram-Jones}\ \emph {et~al.}(2014)\citenamefont
  {Pribram-Jones}, \citenamefont {hui Yang}, \citenamefont {R.Trail},
  \citenamefont {Burke}, \citenamefont {J.Needs},\ and\ \citenamefont
  {A.Ullrich}}]{Burke_ensemble}%
  \BibitemOpen
  \bibfield  {author} {\bibinfo {author} {\bibfnamefont {A.}~\bibnamefont
  {Pribram-Jones}}, \bibinfo {author} {\bibfnamefont {Z.}~\bibnamefont {hui
  Yang}}, \bibinfo {author} {\bibfnamefont {J.}~\bibnamefont {R.Trail}},
  \bibinfo {author} {\bibfnamefont {K.}~\bibnamefont {Burke}}, \bibinfo
  {author} {\bibfnamefont {R.}~\bibnamefont {J.Needs}}, \ and\ \bibinfo
  {author} {\bibfnamefont {C.}~\bibnamefont {A.Ullrich}},\ }\href {\doibase
  10.1063/1.4872255} {\bibfield  {journal} {\bibinfo  {journal} {J. Chem.
  Phys.}\ }\textbf {\bibinfo {volume} {140}},\ \bibinfo {pages} {18A541}
  (\bibinfo {year} {2014})}\BibitemShut {NoStop}%
\bibitem [{\citenamefont {Senjean}\ \emph {et~al.}(2015)\citenamefont
  {Senjean}, \citenamefont {Knecht}, \citenamefont {Jensen},\ and\
  \citenamefont {Fromager}}]{SenjeanPRA2015}%
  \BibitemOpen
  \bibfield  {author} {\bibinfo {author} {\bibfnamefont {B.}~\bibnamefont
  {Senjean}}, \bibinfo {author} {\bibfnamefont {S.}~\bibnamefont {Knecht}},
  \bibinfo {author} {\bibfnamefont {H.~J.~A.}\ \bibnamefont {Jensen}}, \ and\
  \bibinfo {author} {\bibfnamefont {E.}~\bibnamefont {Fromager}},\ }\href
  {http://link.aps.org/doi/10.1103/PhysRevA.92.012518} {\bibfield  {journal}
  {\bibinfo  {journal} {Phys. Rev. A}\ }\textbf {\bibinfo {volume} {92}},\
  \bibinfo {pages} {012518} (\bibinfo {year} {2015})}\BibitemShut {NoStop}%
\bibitem [{\citenamefont {Gidopoulos}, \citenamefont {Papaconstantinou},\ and\
  \citenamefont {Gross}(2002)}]{ensemble_ghost_interaction}%
  \BibitemOpen
  \bibfield  {author} {\bibinfo {author} {\bibfnamefont {N.}~\bibnamefont
  {Gidopoulos}}, \bibinfo {author} {\bibfnamefont {P.}~\bibnamefont
  {Papaconstantinou}}, \ and\ \bibinfo {author} {\bibfnamefont
  {E.}~\bibnamefont {Gross}},\ }\href@noop {} {\bibfield  {journal} {\bibinfo
  {journal} {Phys. Rev. Lett.}\ }\textbf {\bibinfo {volume} {88}},\ \bibinfo
  {pages} {033003} (\bibinfo {year} {2002})}\BibitemShut {NoStop}%
\bibitem [{\citenamefont {Pastorczak}\ and\ \citenamefont
  {Pernal}(2014)}]{pastorczak2014ensemble}%
  \BibitemOpen
  \bibfield  {author} {\bibinfo {author} {\bibfnamefont {E.}~\bibnamefont
  {Pastorczak}}\ and\ \bibinfo {author} {\bibfnamefont {K.}~\bibnamefont
  {Pernal}},\ }\href@noop {} {\bibfield  {journal} {\bibinfo  {journal} {J.
  Chem. Phys.}\ }\textbf {\bibinfo {volume} {140}},\ \bibinfo {pages} {18A514}
  (\bibinfo {year} {2014})}\BibitemShut {NoStop}%
\bibitem [{\citenamefont {Pastorczak}\ and\ \citenamefont
  {Pernal}(2016)}]{pernal2016ghost}%
  \BibitemOpen
  \bibfield  {author} {\bibinfo {author} {\bibfnamefont {E.}~\bibnamefont
  {Pastorczak}}\ and\ \bibinfo {author} {\bibfnamefont {K.}~\bibnamefont
  {Pernal}},\ }\href {\doibase http://dx.doi.org/10.1002/qua.25107} {\bibfield
  {journal} {\bibinfo  {journal} {Int. J. Quantum Chem.}\ } (\bibinfo {year}
  {2016}),\ http://dx.doi.org/10.1002/qua.25107}\BibitemShut {NoStop}%
\bibitem [{\citenamefont {Alam}, \citenamefont {Knecht},\ and\ \citenamefont
  {Fromager}(2016)}]{alam_gic}%
  \BibitemOpen
  \bibfield  {author} {\bibinfo {author} {\bibfnamefont {M.~M.}\ \bibnamefont
  {Alam}}, \bibinfo {author} {\bibfnamefont {S.}~\bibnamefont {Knecht}}, \ and\
  \bibinfo {author} {\bibfnamefont {E.}~\bibnamefont {Fromager}},\ }\href@noop
  {} {\bibfield  {journal} {\bibinfo  {journal} {Physical Review A}\ }\textbf
  {\bibinfo {volume} {94}},\ \bibinfo {pages} {012511} (\bibinfo {year}
  {2016})}\BibitemShut {NoStop}%
\bibitem [{\citenamefont {Fromager}, \citenamefont {Toulouse},\ and\
  \citenamefont {Jensen}(2007)}]{FromagerJCP2007}%
  \BibitemOpen
  \bibfield  {author} {\bibinfo {author} {\bibfnamefont {E.}~\bibnamefont
  {Fromager}}, \bibinfo {author} {\bibfnamefont {J.}~\bibnamefont {Toulouse}},
  \ and\ \bibinfo {author} {\bibfnamefont {H.~J.~A.}\ \bibnamefont {Jensen}},\
  }\href
  {http://scitation.aip.org/content/aip/journal/jcp/126/7/10.1063/1.2566459}
  {\bibfield  {journal} {\bibinfo  {journal} {J. Chem. Phys.}\ }\textbf
  {\bibinfo {volume} {126}},\ \bibinfo {eid} {074111} (\bibinfo {year}
  {2007})}\BibitemShut {NoStop}%
\bibitem [{\citenamefont {Gerber}\ and\ \citenamefont
  {\'{A}ngy\'{a}n}(2005)}]{nancycalib}%
  \BibitemOpen
  \bibfield  {author} {\bibinfo {author} {\bibfnamefont {I.~C.}\ \bibnamefont
  {Gerber}}\ and\ \bibinfo {author} {\bibfnamefont {J.~G.}\ \bibnamefont
  {\'{A}ngy\'{a}n}},\ }\href@noop {} {\bibfield  {journal} {\bibinfo  {journal}
  {Chem. Phys. Lett.}\ }\textbf {\bibinfo {volume} {415}},\ \bibinfo {pages}
  {100} (\bibinfo {year} {2005})}\BibitemShut {NoStop}%
\bibitem [{\citenamefont {Franck}\ \emph {et~al.}(2015)\citenamefont {Franck},
  \citenamefont {Mussard}, \citenamefont {Luppi},\ and\ \citenamefont
  {Toulouse}}]{JCP15_Odile_basis_convergence_srDFT}%
  \BibitemOpen
  \bibfield  {author} {\bibinfo {author} {\bibfnamefont {O.}~\bibnamefont
  {Franck}}, \bibinfo {author} {\bibfnamefont {B.}~\bibnamefont {Mussard}},
  \bibinfo {author} {\bibfnamefont {E.}~\bibnamefont {Luppi}}, \ and\ \bibinfo
  {author} {\bibfnamefont {J.}~\bibnamefont {Toulouse}},\ }\href {\doibase
  http://dx.doi.org/10.1063/1.4907920} {\bibfield  {journal} {\bibinfo
  {journal} {J. Chem. Phys.}\ }\textbf {\bibinfo {volume} {142}},\ \bibinfo
  {eid} {074107} (\bibinfo {year} {2015})}\BibitemShut {NoStop}%
\bibitem [{\citenamefont {Senjean}\ \emph {et~al.}(2016)\citenamefont
  {Senjean}, \citenamefont {Hedeg{\aa}rd}, \citenamefont {Alam}, \citenamefont
  {Knecht},\ and\ \citenamefont {Fromager}}]{extrapol_edft}%
  \BibitemOpen
  \bibfield  {author} {\bibinfo {author} {\bibfnamefont {B.}~\bibnamefont
  {Senjean}}, \bibinfo {author} {\bibfnamefont {E.~D.}\ \bibnamefont
  {Hedeg{\aa}rd}}, \bibinfo {author} {\bibfnamefont {M.~M.}\ \bibnamefont
  {Alam}}, \bibinfo {author} {\bibfnamefont {S.}~\bibnamefont {Knecht}}, \ and\
  \bibinfo {author} {\bibfnamefont {E.}~\bibnamefont {Fromager}},\ }\href@noop
  {} {\bibfield  {journal} {\bibinfo  {journal} {Mol. Phys.}\ }\textbf
  {\bibinfo {volume} {114}},\ \bibinfo {pages} {968} (\bibinfo {year}
  {2016})}\BibitemShut {NoStop}%
\bibitem [{\citenamefont {Savin}(2014)}]{savin2014towards}%
  \BibitemOpen
  \bibfield  {author} {\bibinfo {author} {\bibfnamefont {A.}~\bibnamefont
  {Savin}},\ }\href@noop {} {\bibfield  {journal} {\bibinfo  {journal} {J.
  Chem. Phys.}\ }\textbf {\bibinfo {volume} {140}},\ \bibinfo {pages} {18A509}
  (\bibinfo {year} {2014})}\BibitemShut {NoStop}%
\bibitem [{\citenamefont {Rebolini}\ \emph
  {et~al.}(2015{\natexlab{a}})\citenamefont {Rebolini}, \citenamefont
  {Toulouse}, \citenamefont {Teale}, \citenamefont {Helgaker},\ and\
  \citenamefont {Savin}}]{rebolini2015pra}%
  \BibitemOpen
  \bibfield  {author} {\bibinfo {author} {\bibfnamefont {E.}~\bibnamefont
  {Rebolini}}, \bibinfo {author} {\bibfnamefont {J.}~\bibnamefont {Toulouse}},
  \bibinfo {author} {\bibfnamefont {A.~M.}\ \bibnamefont {Teale}}, \bibinfo
  {author} {\bibfnamefont {T.}~\bibnamefont {Helgaker}}, \ and\ \bibinfo
  {author} {\bibfnamefont {A.}~\bibnamefont {Savin}},\ }\href {\doibase
  10.1103/PhysRevA.91.032519} {\bibfield  {journal} {\bibinfo  {journal} {Phys.
  Rev. A}\ }\textbf {\bibinfo {volume} {91}},\ \bibinfo {pages} {032519}
  (\bibinfo {year} {2015}{\natexlab{a}})}\BibitemShut {NoStop}%
\bibitem [{\citenamefont {Toulouse}, \citenamefont {Colonna},\ and\
  \citenamefont {Savin}(2004)}]{srDFT}%
  \BibitemOpen
  \bibfield  {author} {\bibinfo {author} {\bibfnamefont {J.}~\bibnamefont
  {Toulouse}}, \bibinfo {author} {\bibfnamefont {F.}~\bibnamefont {Colonna}}, \
  and\ \bibinfo {author} {\bibfnamefont {A.}~\bibnamefont {Savin}},\ }\href
  {\doibase 10.1103/PhysRevA.70.062505} {\bibfield  {journal} {\bibinfo
  {journal} {Phys. Rev. A}\ }\textbf {\bibinfo {volume} {70}},\ \bibinfo
  {pages} {062505} (\bibinfo {year} {2004})}\BibitemShut {NoStop}%
\bibitem [{\citenamefont {Goll}, \citenamefont {Werner},\ and\ \citenamefont
  {Stoll}(2005)}]{Goll2005PCCP}%
  \BibitemOpen
  \bibfield  {author} {\bibinfo {author} {\bibfnamefont {E.}~\bibnamefont
  {Goll}}, \bibinfo {author} {\bibfnamefont {H.~J.}\ \bibnamefont {Werner}}, \
  and\ \bibinfo {author} {\bibfnamefont {H.}~\bibnamefont {Stoll}},\
  }\href@noop {} {\bibfield  {journal} {\bibinfo  {journal} {Phys. Chem. Chem.
  Phys.}\ }\textbf {\bibinfo {volume} {7}},\ \bibinfo {pages} {3917} (\bibinfo
  {year} {2005})}\BibitemShut {NoStop}%
\bibitem [{\citenamefont {Toulouse}, \citenamefont {Gori-Giorgi},\ and\
  \citenamefont {Savin}(2005)}]{Toulouse2005TCA}%
  \BibitemOpen
  \bibfield  {author} {\bibinfo {author} {\bibfnamefont {J.}~\bibnamefont
  {Toulouse}}, \bibinfo {author} {\bibfnamefont {P.}~\bibnamefont
  {Gori-Giorgi}}, \ and\ \bibinfo {author} {\bibfnamefont {A.}~\bibnamefont
  {Savin}},\ }\href@noop {} {\bibfield  {journal} {\bibinfo  {journal} {Theor.
  Chem. Acc.}\ }\textbf {\bibinfo {volume} {114}},\ \bibinfo {pages} {305}
  (\bibinfo {year} {2005})}\BibitemShut {NoStop}%
\bibitem [{\citenamefont {Paziani}\ \emph {et~al.}(2006)\citenamefont
  {Paziani}, \citenamefont {Moroni}, \citenamefont {Gori-Giorgi},\ and\
  \citenamefont {Bachelet}}]{Paziani2006PRB}%
  \BibitemOpen
  \bibfield  {author} {\bibinfo {author} {\bibfnamefont {S.}~\bibnamefont
  {Paziani}}, \bibinfo {author} {\bibfnamefont {S.}~\bibnamefont {Moroni}},
  \bibinfo {author} {\bibfnamefont {P.}~\bibnamefont {Gori-Giorgi}}, \ and\
  \bibinfo {author} {\bibfnamefont {G.~B.}\ \bibnamefont {Bachelet}},\
  }\href@noop {} {\bibfield  {journal} {\bibinfo  {journal} {Phys. Rev. B}\
  }\textbf {\bibinfo {volume} {73}},\ \bibinfo {pages} {155111} (\bibinfo
  {year} {2006})}\BibitemShut {NoStop}%
\bibitem [{\citenamefont {Rebolini}\ \emph
  {et~al.}(2015{\natexlab{b}})\citenamefont {Rebolini}, \citenamefont
  {Toulouse}, \citenamefont {Teale}, \citenamefont {Helgaker},\ and\
  \citenamefont {Savin}}]{rebolini2015calculating}%
  \BibitemOpen
  \bibfield  {author} {\bibinfo {author} {\bibfnamefont {E.}~\bibnamefont
  {Rebolini}}, \bibinfo {author} {\bibfnamefont {J.}~\bibnamefont {Toulouse}},
  \bibinfo {author} {\bibfnamefont {A.~M.}\ \bibnamefont {Teale}}, \bibinfo
  {author} {\bibfnamefont {T.}~\bibnamefont {Helgaker}}, \ and\ \bibinfo
  {author} {\bibfnamefont {A.}~\bibnamefont {Savin}},\ }\href@noop {}
  {\bibfield  {journal} {\bibinfo  {journal} {Phys. Rev. A}\ }\textbf {\bibinfo
  {volume} {91}},\ \bibinfo {pages} {032519} (\bibinfo {year}
  {2015}{\natexlab{b}})}\BibitemShut {NoStop}%
\bibitem [{DAL()}]{DALTON}%
  \BibitemOpen
  \href@noop {} {\enquote {\bibinfo {title} {Dalton, a molecular electronic
  structure program, release dalton2015 (2015), see
  http://daltonprogram.org/},}\ }\BibitemShut {NoStop}%
\bibitem [{\citenamefont {Aidas}\ \emph {et~al.}(2015)\citenamefont {Aidas},
  \citenamefont {Angeli}, \citenamefont {Bak}, \citenamefont {Bakken},
  \citenamefont {Bast}, \citenamefont {Boman}, \citenamefont {Christiansen},
  \citenamefont {Cimiraglia}, \citenamefont {Coriani}, \citenamefont {Dahle},
  \citenamefont {Dalskov}, \citenamefont {Ekstr\"{o}m}, \citenamefont
  {Enevoldsen}, \citenamefont {Eriksen}, \citenamefont {Ettenhuber},
  \citenamefont {Fern\'{a}ndez}, \citenamefont {Ferrighi}, \citenamefont
  {Fliegl}, \citenamefont {Frediani}, \citenamefont {Hald}, \citenamefont
  {Halkier}, \citenamefont {H\"{a}ttig}, \citenamefont {Heiberg}, \citenamefont
  {Helgaker}, \citenamefont {Hennum}, \citenamefont {Hettema}, \citenamefont
  {Hjerten\ae{}s}, \citenamefont {H\o{}st}, \citenamefont {H\o{}yvik},
  \citenamefont {Iozzi}, \citenamefont {Jans\'{i}k}, \citenamefont {Jensen},
  \citenamefont {Jonsson}, \citenamefont {J\o{}rgensen}, \citenamefont
  {Kauczor}, \citenamefont {Kirpekar}, \citenamefont {Kj\ae{}rgaard},
  \citenamefont {Klopper}, \citenamefont {Knecht}, \citenamefont {Kobayashi},
  \citenamefont {Koch}, \citenamefont {Kongsted}, \citenamefont {Krapp},
  \citenamefont {Kristensen}, \citenamefont {Ligabue}, \citenamefont
  {Lutn\ae{}s}, \citenamefont {Melo}, \citenamefont {Mikkelsen}, \citenamefont
  {Myhre}, \citenamefont {Neiss}, \citenamefont {Nielsen}, \citenamefont
  {Norman}, \citenamefont {Olsen}, \citenamefont {Olsen}, \citenamefont
  {Osted}, \citenamefont {Packer}, \citenamefont {Pawlowski}, \citenamefont
  {Pedersen}, \citenamefont {Provasi}, \citenamefont {Reine}, \citenamefont
  {Rinkevicius}, \citenamefont {Ruden}, \citenamefont {Ruud}, \citenamefont
  {Rybkin}, \citenamefont {Sa\l{}ek}, \citenamefont {Samson}, \citenamefont
  {de~Mer\'{a}s}, \citenamefont {Saue}, \citenamefont {Sauer}, \citenamefont
  {Schimmelpfennig}, \citenamefont {Sneskov}, \citenamefont {Steindal},
  \citenamefont {Sylvester-Hvid}, \citenamefont {Taylor}, \citenamefont
  {Teale}, \citenamefont {Tellgren}, \citenamefont {Tew}, \citenamefont
  {Thorvaldsen}, \citenamefont {Th\o{}gersen}, \citenamefont {Vahtras},
  \citenamefont {Watson}, \citenamefont {Wilson}, \citenamefont {Ziolkowski},\
  and\ \citenamefont {\AA{}gren}}]{DALTON2}%
  \BibitemOpen
  \bibfield  {author} {\bibinfo {author} {\bibfnamefont {K.}~\bibnamefont
  {Aidas}}, \bibinfo {author} {\bibfnamefont {C.}~\bibnamefont {Angeli}},
  \bibinfo {author} {\bibfnamefont {K.~L.}\ \bibnamefont {Bak}}, \bibinfo
  {author} {\bibfnamefont {V.}~\bibnamefont {Bakken}}, \bibinfo {author}
  {\bibfnamefont {R.}~\bibnamefont {Bast}}, \bibinfo {author} {\bibfnamefont
  {L.}~\bibnamefont {Boman}}, \bibinfo {author} {\bibfnamefont
  {O.}~\bibnamefont {Christiansen}}, \bibinfo {author} {\bibfnamefont
  {R.}~\bibnamefont {Cimiraglia}}, \bibinfo {author} {\bibfnamefont
  {S.}~\bibnamefont {Coriani}}, \bibinfo {author} {\bibfnamefont
  {P.}~\bibnamefont {Dahle}}, \bibinfo {author} {\bibfnamefont {E.~K.}\
  \bibnamefont {Dalskov}}, \bibinfo {author} {\bibfnamefont {U.}~\bibnamefont
  {Ekstr\"{o}m}}, \bibinfo {author} {\bibfnamefont {T.}~\bibnamefont
  {Enevoldsen}}, \bibinfo {author} {\bibfnamefont {J.~J.}\ \bibnamefont
  {Eriksen}}, \bibinfo {author} {\bibfnamefont {P.}~\bibnamefont {Ettenhuber}},
  \bibinfo {author} {\bibfnamefont {B.}~\bibnamefont {Fern\'{a}ndez}}, \bibinfo
  {author} {\bibfnamefont {L.}~\bibnamefont {Ferrighi}}, \bibinfo {author}
  {\bibfnamefont {H.}~\bibnamefont {Fliegl}}, \bibinfo {author} {\bibfnamefont
  {L.}~\bibnamefont {Frediani}}, \bibinfo {author} {\bibfnamefont
  {K.}~\bibnamefont {Hald}}, \bibinfo {author} {\bibfnamefont {A.}~\bibnamefont
  {Halkier}}, \bibinfo {author} {\bibfnamefont {C.}~\bibnamefont {H\"{a}ttig}},
  \bibinfo {author} {\bibfnamefont {H.}~\bibnamefont {Heiberg}}, \bibinfo
  {author} {\bibfnamefont {T.}~\bibnamefont {Helgaker}}, \bibinfo {author}
  {\bibfnamefont {A.~C.}\ \bibnamefont {Hennum}}, \bibinfo {author}
  {\bibfnamefont {H.}~\bibnamefont {Hettema}}, \bibinfo {author} {\bibfnamefont
  {E.}~\bibnamefont {Hjerten\ae{}s}}, \bibinfo {author} {\bibfnamefont
  {S.}~\bibnamefont {H\o{}st}}, \bibinfo {author} {\bibfnamefont {I.-M.}\
  \bibnamefont {H\o{}yvik}}, \bibinfo {author} {\bibfnamefont {M.~F.}\
  \bibnamefont {Iozzi}}, \bibinfo {author} {\bibfnamefont {B.}~\bibnamefont
  {Jans\'{i}k}}, \bibinfo {author} {\bibfnamefont {H.~J.~{\relax Aa}.}\
  \bibnamefont {Jensen}}, \bibinfo {author} {\bibfnamefont {D.}~\bibnamefont
  {Jonsson}}, \bibinfo {author} {\bibfnamefont {P.}~\bibnamefont
  {J\o{}rgensen}}, \bibinfo {author} {\bibfnamefont {J.}~\bibnamefont
  {Kauczor}}, \bibinfo {author} {\bibfnamefont {S.}~\bibnamefont {Kirpekar}},
  \bibinfo {author} {\bibfnamefont {T.}~\bibnamefont {Kj\ae{}rgaard}}, \bibinfo
  {author} {\bibfnamefont {W.}~\bibnamefont {Klopper}}, \bibinfo {author}
  {\bibfnamefont {S.}~\bibnamefont {Knecht}}, \bibinfo {author} {\bibfnamefont
  {R.}~\bibnamefont {Kobayashi}}, \bibinfo {author} {\bibfnamefont
  {H.}~\bibnamefont {Koch}}, \bibinfo {author} {\bibfnamefont {J.}~\bibnamefont
  {Kongsted}}, \bibinfo {author} {\bibfnamefont {A.}~\bibnamefont {Krapp}},
  \bibinfo {author} {\bibfnamefont {K.}~\bibnamefont {Kristensen}}, \bibinfo
  {author} {\bibfnamefont {A.}~\bibnamefont {Ligabue}}, \bibinfo {author}
  {\bibfnamefont {O.~B.}\ \bibnamefont {Lutn\ae{}s}}, \bibinfo {author}
  {\bibfnamefont {J.~I.}\ \bibnamefont {Melo}}, \bibinfo {author}
  {\bibfnamefont {K.~V.}\ \bibnamefont {Mikkelsen}}, \bibinfo {author}
  {\bibfnamefont {R.~H.}\ \bibnamefont {Myhre}}, \bibinfo {author}
  {\bibfnamefont {C.}~\bibnamefont {Neiss}}, \bibinfo {author} {\bibfnamefont
  {C.~B.}\ \bibnamefont {Nielsen}}, \bibinfo {author} {\bibfnamefont
  {P.}~\bibnamefont {Norman}}, \bibinfo {author} {\bibfnamefont
  {J.}~\bibnamefont {Olsen}}, \bibinfo {author} {\bibfnamefont {J.~M.~H.}\
  \bibnamefont {Olsen}}, \bibinfo {author} {\bibfnamefont {A.}~\bibnamefont
  {Osted}}, \bibinfo {author} {\bibfnamefont {M.~J.}\ \bibnamefont {Packer}},
  \bibinfo {author} {\bibfnamefont {F.}~\bibnamefont {Pawlowski}}, \bibinfo
  {author} {\bibfnamefont {T.~B.}\ \bibnamefont {Pedersen}}, \bibinfo {author}
  {\bibfnamefont {P.~F.}\ \bibnamefont {Provasi}}, \bibinfo {author}
  {\bibfnamefont {S.}~\bibnamefont {Reine}}, \bibinfo {author} {\bibfnamefont
  {Z.}~\bibnamefont {Rinkevicius}}, \bibinfo {author} {\bibfnamefont {T.~A.}\
  \bibnamefont {Ruden}}, \bibinfo {author} {\bibfnamefont {K.}~\bibnamefont
  {Ruud}}, \bibinfo {author} {\bibfnamefont {V.~V.}\ \bibnamefont {Rybkin}},
  \bibinfo {author} {\bibfnamefont {P.}~\bibnamefont {Sa\l{}ek}}, \bibinfo
  {author} {\bibfnamefont {C.~C.~M.}\ \bibnamefont {Samson}}, \bibinfo {author}
  {\bibfnamefont {A.~S.}\ \bibnamefont {de~Mer\'{a}s}}, \bibinfo {author}
  {\bibfnamefont {T.}~\bibnamefont {Saue}}, \bibinfo {author} {\bibfnamefont
  {S.~P.~A.}\ \bibnamefont {Sauer}}, \bibinfo {author} {\bibfnamefont
  {B.}~\bibnamefont {Schimmelpfennig}}, \bibinfo {author} {\bibfnamefont
  {K.}~\bibnamefont {Sneskov}}, \bibinfo {author} {\bibfnamefont {A.~H.}\
  \bibnamefont {Steindal}}, \bibinfo {author} {\bibfnamefont {K.~O.}\
  \bibnamefont {Sylvester-Hvid}}, \bibinfo {author} {\bibfnamefont {P.~R.}\
  \bibnamefont {Taylor}}, \bibinfo {author} {\bibfnamefont {A.~M.}\
  \bibnamefont {Teale}}, \bibinfo {author} {\bibfnamefont {E.~I.}\ \bibnamefont
  {Tellgren}}, \bibinfo {author} {\bibfnamefont {D.~P.}\ \bibnamefont {Tew}},
  \bibinfo {author} {\bibfnamefont {A.~J.}\ \bibnamefont {Thorvaldsen}},
  \bibinfo {author} {\bibfnamefont {L.}~\bibnamefont {Th\o{}gersen}}, \bibinfo
  {author} {\bibfnamefont {O.}~\bibnamefont {Vahtras}}, \bibinfo {author}
  {\bibfnamefont {M.~A.}\ \bibnamefont {Watson}}, \bibinfo {author}
  {\bibfnamefont {D.~J.~D.}\ \bibnamefont {Wilson}}, \bibinfo {author}
  {\bibfnamefont {M.}~\bibnamefont {Ziolkowski}}, \ and\ \bibinfo {author}
  {\bibfnamefont {H.}~\bibnamefont {\AA{}gren}},\ }\href {\doibase
  10.1002/wcms.1172} {\bibfield  {journal} {\bibinfo  {journal} {WIREs
  Comput.~Mol.~Sci.}\ }\textbf {\bibinfo {volume} {4}},\ \bibinfo {pages} {269}
  (\bibinfo {year} {2015})}\BibitemShut {NoStop}%
\bibitem [{\citenamefont {Toulouse}, \citenamefont {Savin},\ and\ \citenamefont
  {Flad}(2004)}]{toulda}%
  \BibitemOpen
  \bibfield  {author} {\bibinfo {author} {\bibfnamefont {J.}~\bibnamefont
  {Toulouse}}, \bibinfo {author} {\bibfnamefont {A.}~\bibnamefont {Savin}}, \
  and\ \bibinfo {author} {\bibfnamefont {H.~J.}\ \bibnamefont {Flad}},\
  }\href@noop {} {\bibfield  {journal} {\bibinfo  {journal} {Int. J. Quantum
  Chem.}\ }\textbf {\bibinfo {volume} {100}},\ \bibinfo {pages} {1047}
  (\bibinfo {year} {2004})}\BibitemShut {NoStop}%
\bibitem [{\citenamefont {Dunning~Jr}(1989)}]{dunning1989gaussian}%
  \BibitemOpen
  \bibfield  {author} {\bibinfo {author} {\bibfnamefont {T.~H.}\ \bibnamefont
  {Dunning~Jr}},\ }\href@noop {} {\bibfield  {journal} {\bibinfo  {journal} {J.
  Chem. Phys.}\ }\textbf {\bibinfo {volume} {90}},\ \bibinfo {pages} {1007}
  (\bibinfo {year} {1989})}\BibitemShut {NoStop}%
\bibitem [{\citenamefont {Woon}\ and\ \citenamefont
  {Dunning~Jr}(1994)}]{woon1994gaussian}%
  \BibitemOpen
  \bibfield  {author} {\bibinfo {author} {\bibfnamefont {D.~E.}\ \bibnamefont
  {Woon}}\ and\ \bibinfo {author} {\bibfnamefont {T.~H.}\ \bibnamefont
  {Dunning~Jr}},\ }\href@noop {} {\bibfield  {journal} {\bibinfo  {journal} {J.
  Chem. Phys.}\ }\textbf {\bibinfo {volume} {100}},\ \bibinfo {pages} {2975}
  (\bibinfo {year} {1994})}\BibitemShut {NoStop}%
\bibitem [{\citenamefont {Fromager}, \citenamefont {Knecht},\ and\
  \citenamefont {{Aa. Jensen}}(2013)}]{FromagerJCP2013}%
  \BibitemOpen
  \bibfield  {author} {\bibinfo {author} {\bibfnamefont {E.}~\bibnamefont
  {Fromager}}, \bibinfo {author} {\bibfnamefont {S.}~\bibnamefont {Knecht}}, \
  and\ \bibinfo {author} {\bibfnamefont {H.~J.}\ \bibnamefont {{Aa. Jensen}}},\
  }\href@noop {} {\bibfield  {journal} {\bibinfo  {journal} {J. Chem. Phys.}\
  }\textbf {\bibinfo {volume} {138}},\ \bibinfo {pages} {084101} (\bibinfo
  {year} {2013})}\BibitemShut {NoStop}%
\bibitem [{\citenamefont {Helgaker}, \citenamefont {J{\o}rgensen},\ and\
  \citenamefont {Olsen}(2004)}]{mcscf_pinkbook}%
  \BibitemOpen
  \bibfield  {author} {\bibinfo {author} {\bibfnamefont {T.}~\bibnamefont
  {Helgaker}}, \bibinfo {author} {\bibfnamefont {P.}~\bibnamefont
  {J{\o}rgensen}}, \ and\ \bibinfo {author} {\bibfnamefont {J.}~\bibnamefont
  {Olsen}},\ }\enquote {\bibinfo {title} {Molecular electronic-structure
  theory},}\ \ (\bibinfo  {publisher} {Wiley},\ \bibinfo {address}
  {Chichester},\ \bibinfo {year} {2004})\ pp.\ \bibinfo {pages}
  {598--647}\BibitemShut {NoStop}%
\bibitem [{\citenamefont {Levy}\ and\ \citenamefont
  {Zahariev}(2014)}]{levy2014ground}%
  \BibitemOpen
  \bibfield  {author} {\bibinfo {author} {\bibfnamefont {M.}~\bibnamefont
  {Levy}}\ and\ \bibinfo {author} {\bibfnamefont {F.}~\bibnamefont
  {Zahariev}},\ }\href@noop {} {\bibfield  {journal} {\bibinfo  {journal}
  {Phys. Rev. Lett.}\ }\textbf {\bibinfo {volume} {113}},\ \bibinfo {pages}
  {113002} (\bibinfo {year} {2014})}\BibitemShut {NoStop}%
\bibitem [{\citenamefont {Gori-Giorgi}\ and\ \citenamefont
  {Savin}(2006)}]{paolacorr}%
  \BibitemOpen
  \bibfield  {author} {\bibinfo {author} {\bibfnamefont {P.}~\bibnamefont
  {Gori-Giorgi}}\ and\ \bibinfo {author} {\bibfnamefont {A.}~\bibnamefont
  {Savin}},\ }\href@noop {} {\bibfield  {journal} {\bibinfo  {journal} {Phys.
  Rev. A}\ }\textbf {\bibinfo {volume} {73}},\ \bibinfo {pages} {032506}
  (\bibinfo {year} {2006})}\BibitemShut {NoStop}%
\end{thebibliography}
\newcommand{\Aa}[0]{Aa}

\end{document}